\documentclass[12pt]{iopart}


\usepackage{iopams}
\usepackage{cite}

\makeatletter
\expandafter\let\csname equation*\endcsname\relax					
\expandafter\let\csname endequation*\endcsname\relax			
\let\original@equation\equation												
\let\original@endequation\endequation									
\usepackage{amsmath}
\let\equation\original@equation												
\let\endequation\original@endequation									
\makeatother

\usepackage{mathtools}
\usepackage{graphicx}
\usepackage{color}
\usepackage{hyperref}

\newcommand{\vek}[1]{\bi{#1}}		
\newcommand{\veknull}{\boldsymbol{0}}
\newcommand{\vnabla}{\boldsymbol{\nabla}}
\renewcommand{\mat}[1]{{#1}}

\newcommand{\bra}[1]{\langle#1|}
\newcommand{\ket}[1]{|#1\rangle}

\newcommand{\bigbraOket}[3]{\big\langle#1\big|#2\big|#3\big\rangle}

\newcommand{\rmf}{{\rm f}}
\newcommand{\rmT}{{\rm T}}

\newcommand{\Ord}{\Or}					

\newcommand{\pr}[1]{\tilde{#1}}

\newcommand{\tE}{t_{\mathrm{E}}}
\newcommand{\qp}{$qp$}

\newcommand{\Mstab}{\mathbb{M}_t}
\newcommand{\Mcan}{\tilde{\mathbb{M}}_t}
\newcommand{\McanT}{\tilde{\mathbb{M}}_t^\rmT}

\newcommand{\Ralt}{\mathcal{R}}

\newcommand{\Wi}{W_\rmi}
\newcommand{\Wf}{W_\rmf}
\newcommand{\Ow}{\Omega_{\mathrm{W}}}

\newcommand{\bigzero}{\mbox{\normalfont\Large 0}}
\newcommand{\lL}{\lambda_{\mathrm{L}}}

\newcommand{\Sq}{{\mathcal{S}}}
\newcommand{\SP}{\mathcal{S}_\mathcal{P}}
\newcommand{\sap}[1]{{\overline{#1}}}

\newcommand{\para}{\parallel}

\newcommand{\vq}{{\vek{q}}}
\newcommand{\vp}{{\vek{p}}}
\newcommand{\vx}{{\vek{x}}}
\newcommand{\vz}{{\vek{z}}}
\newcommand{\vk}{{\vek{k}}}
\newcommand{\vR}{{\vek{R}}}
\newcommand{\vP}{{\vek{P}}}
\newcommand{\vX}{{\vek{X}}}

\newcommand{\vRi}{\vR^\rmi}
\newcommand{\vRf}{\vR^\rmf}
\newcommand{\vPi}{\vP^\rmi}

\newcommand{\vXi}{\vX^\rmi}
\newcommand{\vXf}{\vX^\rmf}

\newcommand{\vRin}{\vR^\rmi_0}
\newcommand{\vRfn}{\vR^\rmf_0}
\newcommand{\vPin}{\vP^\rmi_0}

\newcommand{\vXin}{\vX^\rmi_0}
\newcommand{\vXfn}{\vX^\rmf_0}

\newcommand{\vRiperp}{\vR^\rmi_\perp}
\newcommand{\vRfperp}{\vR^\rmf_\perp}
\newcommand{\vPiperp}{\vP^\rmi_\perp}

\newcommand{\vXiperp}{\vX^\rmi_\perp}
\newcommand{\vXfperp}{\vX^\rmf_\perp}

\newcommand{\vRipara}{\vR^\rmi_\para}
\newcommand{\vRfpara}{\vR^\rmf_\para}
\newcommand{\vPipara}{\vP^\rmi_\para}

\newcommand{\vXipara}{\vX^\rmi_\para}
\newcommand{\vXfpara}{\vX^\rmf_\para}

\newcommand{\vqpara}{\vq_\para}
\newcommand{\vqperp}{\vq_\perp}

\newcommand{\vppara}{\vp_\para}
\newcommand{\vpperp}{\vp_\perp}

\newcommand{\vpig}{\vp^\rmi_\gamma}
\newcommand{\vpfg}{\vp^\rmf_\gamma}
\newcommand{\vpigpara}{\vp^\rmi_{\gamma, \para}}
\newcommand{\vpfgpara}{\vp^\rmf_{\gamma, \para}}
\newcommand{\vpigperp}{\vp^\rmi_{\gamma, \perp}}
\newcommand{\vpfgperp}{\vp^\rmf_{\gamma, \perp}}

\newcommand{\vqsap}{\sap{\vq}}
\newcommand{\vqsapn}{\sap{\vq}_0}
\newcommand{\vqsappara}{\sap{\vq}_\para}
\newcommand{\vqsapperp}{\sap{\vq}_\perp}

\newcommand{\vpsap}{\sap{\vp}}

\newcommand{\vpsapig}{\sap{\vp}^\rmi_\gamma}
\newcommand{\vpsapign}{\sap{\vp}^\rmi_{\gamma, 0}}
\newcommand{\vpsapigpara}{\sap{\vp}^\rmi_{\gamma, \para}}
\newcommand{\vpsapigperp}{\sap{\vp}^\rmi_{\gamma, \perp}}

\newcommand{\vRbi}{\overline{\vR}^\rmi}
\newcommand{\vPbi}{\overline{\vP}^\rmi}
\newcommand{\vXbi}{\overline{\vX}^\rmi}

\newcommand{\vRbin}{\overline{\vR}^\rmi_0}

\newcommand{\vXbin}{\overline{\vX}^\rmi_0}

\newcommand{\vekphi}{\boldsymbol{\varphi}}
\newcommand{\vekPhi}{\boldsymbol{\Phi}}
\newcommand{\vekpsi}{\boldsymbol{\psi}}
\newcommand{\vekPsi}{\boldsymbol{\Psi}}

\newcommand{\Tstrut}[1]{\rule{0pt}{#1}}         
\newcommand{\Bstrut}[1]{\rule[-#1]{0pt}{0pt}}   
\newcommand{\psep}{\,}					

\newcommand{{\eq}}{~}
\newcommand{\eqs}{~}
\newcommand{\rf}{~}

\begin{document}
\title[Symmetry-induced many-body interference: an augmented Truncated Wigner method]{Symmetry-induced many-body quantum interference in chaotic bosonic systems: an augmented Truncated Wigner method}

\author{Q Hummel$^{1,2}$, P Schlagheck$^1$}

\address{$^1$ CESAM research unit, University of Li\`ege (ULiege), 4000 Li\`ege, Belgium}
\address{$^2$ Institut f\"ur Theoretische Physik, Universit\"at Regensburg, D-93040 Regensburg, Germany}
\ead{quirin.hummel@uliege.be}

\vspace{10pt}
\begin{indented}
	\item[]January 2022
\end{indented}

\begin{abstract}
Although highly successful, the Truncated Wigner Approximation (TWA) does not account for genuine many-body quantum interference between different solutions of the mean-field equations of a bosonic many-body (MB) system.
This renders the TWA essentially classical, where a large number of particles formally takes the role of the inverse of Planck's constant $\hbar$.
The failure to describe genuine interference phenomena, such as localization and scarring in Fock space, can be seen as a virtue of this quasiclassical method, which thereby allows one to identify genuine quantum effects when being compared with ``exact'' quantum calculations that do not involve any a priori approximation.
A rather prominent cause for such quantum effects that are not accounted for by the TWA is the constructive interference between the contributions of symmetry-related trajectories, which would occur in the presence of discrete symmetries provided the phase-space distribution of the initial state and the observable to be evaluated feature a strong localization about the corresponding symmetry subspaces.
Here we show how one can conceive an augmented version of the TWA which can account for this particular effect.
This augmented TWA effectively amounts to complementing conventional TWA calculations by separate Truncated Wigner simulations that are restricted to symmetric subspaces and involve weight factors that account for the dynamical stability of sampling trajectories with respect to perpendicular deviations from those subspaces.
We illustrate the validity of this method at pre- as well as post-Ehrenfest time scales in prototypical Bose-Hubbard systems displaying chaotic classical dynamics, where it also reveals the existence of additional MB interference effects.
\end{abstract}

\vspace{2pc}


\maketitle


\section{Introduction}
\label{sec:Intro}
Questions related to quantum chaos, i.e., to characteristic signatures of quantum systems whose classical counterpart exhibits chaotic dynamics \cite{Haake2010,Stoeckmann1999}, have recently attracted renewed attention in the context of many-body systems in general and of ultracold quantum gases in particular.
As illustrated by recent milestone experiments on thermalization and localization properties \cite{Choi2016,Kaufman2016,Lukin2019}, those particular physical systems allow one to achieve an unprecedented degree of precision and control in the preparation and readout stages of a quantum many-body experiment, together with a wide flexibility to tune system parameters and change fundamental properties of the involved particles.
A particularly intriguing object of study in this context, which gained a lot of attention most recently, are signatures of ergodicity breaking in quantum many-body systems that are expected to exhibit thermalization from a classical point of view, traced back, most prominently, to many-body scars \cite{Bernien2017,Turner2018,Zhao2020,Serbyn2021}.
Those particular objects can be understood as many-body eigenstates that are localized on unstable periodic orbits of the system's classical dynamics \cite{Heller1984,Bogomolny1988}, thereby defying the eigenstate thermalization hypothesis \cite{Deutsch1991,Srednicki1994}.

The theoretical investigation of many-body scars, and also of other nonclassical transport phenomena related, e.g., to (dynamical) localization \cite{Anderson1958,Fishman1982,Shepelyansky1983,Bohigas1993,%
Altshuler1997,Gornyi2005,Basko2006,Oganesyan2007} or tunneling \cite{Davis1981,Bohigas1993,Tomsovic1994,Hensinger2001,Vanhaele2021}, can certainly be carried out via a spectral analysis of the system under consideration \cite{Hummel2022unpub}.
An alternative approach, which is more closely simulating state-of-the-art experiments in this context \cite{Choi2016,Kaufman2016,Lukin2019,Bernien2017}, consists in studying time evolution processes that would result from a quantum quench, i.e., a sudden change of system parameters at initial time, after having prepared the system in a given initial state.
A particularly interesting choice for that initial state, especially for the purpose of exploring the impact of scars, is a coherent state, i.e., a minimum-uncertainty wave packet which mimics most closely the motion of a classical trajectory, at least during the initial stage of the evolution process.
In the context of ultracold bosonic atoms, such a coherent state would correspond to a perfect Bose-Einstein condensate which can most straightforwardly be prepared in optical lattices for suitable system parameters%
\footnote{%
	This consideration neglects the presence of quantum depletion, i.e., of a minority of bosonic atoms that are not in the condensate state, which is an inevitable consequence of atom-atom interaction even at zero temperature.
	A quench in the atom-atom interaction strength, induced by Feshbach tuning, can be used to overcome this limitation if needed \cite{Chin2010}.}%
.
A particularly useful numerical tool for the purpose of describing such a quench theoretically is the Truncated Wigner method \cite{Steel1998,Sinatra2002,Polkovnikov2010}.
This method consists in a quasiclassical simulation of the wave packet's time evolution in terms of classical trajectories, chosen such that their initial phase-space points properly sample the initial Wigner function of the quantum wave packet.
A perfect quantum-classical correspondence in the time evolution of a system's observable, given in terms of the expectation value of a one- or many-body operator, implies that the Truncated Wigner Approximation (TWA) quantitatively reproduces this time evolution, while deviations between the TWA and the exact time evolution of the observable are indicative of genuine quantum effects typically related to many-body interference.
If we simply consider the return probability to the system's initial state as observable under study, then the presence of scars and/or dynamical localization would most characteristically manifest in terms of an enhancement of that observable as compared to TWA simulations.

Besides those quantum dynamical effects, enhancements of return probabilities to the initial state as compared to quasiclassical predictions can also occur due to the presence of one or several \emph{discrete symmetries} of the system under consideration, provided the phase-space distribution of that initial state is tightly localized about the corresponding symmetric submanifolds.
In that case, each trajectory belonging to the quasiclassical sampling of the initial distribution will have one or several symmetry-related partner trajectories belonging to that same sampling, which will contribute to the TWA-based simulation with equal weight.
For long evolution times and chaotic classical dynamics, those partner trajectories generically belong to different trajectory \emph{families}, which, from a semiclassical perspective, implies that their respective contributions to the quantum time evolution give rise to constructive interference.
These are totally neglected in the quasiclassical TWA, which incorporates only interference due to quantum fluctuations in the \emph{immediate} vicinity of classical trajectories.
While there are extensions that improve on the TWA's inherent second-order description of these fluctuations, e.g., in terms of quantum Brownian motion \cite{Polkovnikov2010}, the innate inability to account for interference of \emph{classically well-separated} trajectories, i.e., different trajectory families, remains.
An augmented version of the TWA is therefore required to correctly account for those interference effects, in order to discriminate them from dynamical interference effects related to localization and scars.
In\rf\cite{Schlagheck2019} such an augmented TWA was proposed and successfully implemented for ultracold bosonic atoms in optical lattices of finite extent.
It crucially relies on an approximate identification of trajectories belonging to symmetric or nonsymmetric families, using a distance threshold criterion with respect to the symmetry submanifold under consideration.
While this particular approach can be straightforwardly implemented without significantly increasing the numerical effort of a TWA simulation, it suffers from an intrinsic ambiguity related to the definition of appropriate distance thresholds in phase space.

The purpose of the present paper is to provide a more solid theoretical foundation for the applicability of such a method.
We shall, to this end, employ a complementary approach to discriminate between the respective contributions of symmetric and nonsymmetric trajectory families to the TWA simulation, which does not involve any adjustable parameter.
The key idea of this approach is to perform TWA simulations that are restricted to the symmetric subspaces or submanifolds about which the wave packet is initially localized, together with an additional weight that accounts for the stability of the involved trajectories with respect to deviations from symmetry.
As detailed below, such a subspace-restricted sampling can yield the specific contribution of the symmetric trajectory families to the quasiclassical simulation of the wave packet's time evolution, and hence, by subtraction, also the complementary contribution of nonsymmetric trajectories.
Properly re-weighting the latter with respect to the former then yields an augmented version of the TWA that quantitatively accounts for the presence of discrete symmetries as a function of the evolution time.
We shall demonstrate below that this method agrees very well with the more heuristic implementation of the augmented TWA described in\rf\cite{Schlagheck2019} provided the distance threshold parameter employed in the latter is chosen sufficiently small to enclose only the immediate dynamical vicinity of the symmetry subspace, but not so small that it would cut into the Wigner function of the wave packet's initial state.
Both methods reproduce fairly well the exact time evolution of the many-body wave packet, in stark contrast to the ordinary TWA which grossly underestimates the return probability to the initial state at long evolution times.

To lay down proper foundations, we start in \sref{sec:reviewTWA} with a review of the semiclassical derivation of the TWA via the van Vleck-Gutzwiller propagator \cite{Gutzwiller1990}.
This particular derivation, which was already provided in various ways \cite{Sun1998,Dittrich2006,Dujardin2015,Schlagheck2019}, is complementary to the usual justification of the TWA based on the truncation of the time evolution equation of the system's Wigner function \cite{Steel1998,Sinatra2002,Polkovnikov2010} and essentially explains why, in practice, the TWA can, for sufficiently simple one-body observables, yield reliable predictions even for long evolution times \cite{Dujardin2015,Schlagheck2019}.
\Sref{sec:discrsym} is devoted to a discussion of discrete symmetries and explains the notion of symmetric and nonsymmetric trajectory families. In \sref{sec:augTWA} we discuss how one can formally derive the augmented TWA, using a number of intermediate calculation steps the details of which are described in the appendices of this paper.
\Sref{sec:applicMB} contains the particularization of this method to bosonic many-body systems.
Numerical results, obtained within Bose-Hubbard pla\-quettes that are populated by a mesoscopic number of particles, are presented and discussed in \sref{sec:BHapplication}.

\section{The Truncated Wigner approach from a semiclassical perspective}
\label{sec:reviewTWA}
Let us first review how the TWA can be derived from the point of view of semiclassics \cite{Sun1998,Dittrich2006}.
We consider a closed quantum system consisting of $L$ (continuous) degrees of freedom, characterized by position and conjugate momentum operators 
$\hat{\vq} \equiv ( \hat{q}_1,\ldots,\hat{q}_L )$ and
$\hat{\vp} \equiv ( \hat{p}_1,\ldots,\hat{p}_L )$ with continuous spectrum $\vq, \vp \in \mathbb{R}^L$.
In the many-body context these can, e.g., be chosen to be the quadrature operators of lattice sites.
At time $t=0$ the system is prepared in an initial state described by a density matrix $\hat{\rho}_\rmi$.
After a time $t$ the system has evolved into a state
\begin{equation}
	\hat{\rho}_\rmi(t) = \hat{U}(t) \hat{\rho}_\rmi(0) \hat{U}^\dagger(t) \psep,
\end{equation}
where $ \hat{U}(t) $ is the time evolution operator of the system.
We investigate the dynamics by means of the time evolved expectation value
\begin{equation} \label{eq:Omegatdef}
	\langle \hat{\Omega} \rangle_t = \tr \!\left[ \hat{\rho}_\rmi(t) \hat{\Omega} \right]
\end{equation}
of some observable $\hat{\Omega}$.
The goal is to describe and understand the expectation value~\eref{eq:Omegatdef} in terms of trajectories of the underlying classical system.
By expanding\eq\eref{eq:Omegatdef} in the basis of position coordinates,
\begin{equation} \label{eq:OmegatUU}
	\fl
	\langle \hat{\Omega} \rangle_t = \int \!\rmd^L \vq^\rmi \int \!\rmd^L \tilde{\vq}^\rmi \int \!\rmd^L \vq^\rmf \int \!\rmd^L \tilde{\vq}^\rmf \;
		\bigbraOket {\vq^\rmf} {\hat{U}(t)} {\vq^\rmi}
		\bigbraOket {\vq^\rmi} {\hat{\rho}_\rmi} {\tilde{\vq}^\rmi} 
		\bigbraOket {\tilde{\vq}^\rmi} {\hat{U}^\dagger(t)} {\tilde{\vq}^\rmf}
		\bigbraOket {\tilde{\vq}^\rmf} {\hat{\Omega}} {\vq^\rmf} \psep,
\end{equation}
the dynamical information is reduced to the propagator $K(\vq^\rmf,\vq^\rmi,t) = \bigbraOket{\vq^\rmf}{\hat{U}(t)}{\vq^\rmi}$, i.e., the position matrix element of the time evolution operator.
The semiclassical approximation consists in representing this latter
quantity as a sum over all classical trajectories, indexed by $\gamma$, 
that start at $t=0$ at positions $\vq^\rmi = (q_1^\rmi, \ldots, q_L^\rmi)$ 
and end at time $t$ at positions $\vq^\rmf = (q_1^\rmf, \ldots, q_L^\rmf)$%
\footnote{%
	Depending on the context we may identify vectors $\vek{v}$ with either their column representation $\vek{v} = (v_1, \ldots, v_n)^\rmT$ or their row representation $\vek{v} = (v_1, \ldots, v_n)$ without writing the transpose symbol ${(\cdot)}^\rmT$ explicitly.%
}%
.
This is unambiguously implemented by the \textit{van Vleck-Gutzwiller propagator} that can be rigorously derived from Feynman's path integral by means of saddle point approximation \cite{Gutzwiller1990}.
It approximates the full quantum propagator by
\begin{equation} \label{eq:K}
  K(\vq^\rmf,\vq^\rmi,t) \simeq \sum_\gamma A_{\gamma}(\vq^\rmf,\vq^\rmi,t)
  e^{\rmi R_{\gamma}(\vq^\rmf,\vq^\rmi,t)/\hbar} \psep,
\end{equation}
where the phases correspond to Hamilton's principal function associated
with the trajectories $\gamma$, i.e., we have
\begin{equation} \label{eq:Rgamma}
  R_\gamma(\vq^\rmf,\vq^\rmi,t) = \int_0^t 
  	L\left[ \vq_\gamma(t'),\dot{\vq}_\gamma(t') \right] \rmd t^\prime
\end{equation}
with $L$ the Lagrangian of the underlying classical system.
The amplitude prefactors can be expressed as
\begin{equation} \label{eq:Ag}
  A_{\gamma}(\vq^\rmf,\vq^\rmi,t) = \left|\frac{1}{(2\pi\hbar)^L}
  \det\left(-\frac{\partial^2 R_\gamma}{\partial q_l^\rmi \partial q_{l^\prime}^\rmf}
  (\vq^\rmf,\vq^\rmi,t)\right)_{l, l^\prime}\right|^{1/2} e^{i \kappa_\gamma \pi/4} \psep,
\end{equation}
where the \textit{Morse index} $\kappa_\gamma$ counts the number of conjugated
points that are encountered along the trajectory.
Throughout we use the notation $ \left( a_{ij} \right)_{ij} $ to denote the matrix with elements $a_{ij}$ in row $i$ and column $j$.
We do not denote explicitly the ranges of indexes, which in~\eref{eq:Ag} are $ l, l^\prime \in \{ 1,  \ldots, L \} $.

The time evolved expectation value~\eref{eq:OmegatUU} is then semiclassically expressed as a double sum over classical trajectories $\gamma, \tilde{\gamma}$ according to
\begin{eqnarray} \label{eq:Omegatsumgg}
	\eqalign{
		\fl
			\langle \hat{\Omega} \rangle_t \simeq  \int \!\rmd^L \vq^\rmi \int \!\rmd^L \tilde{\vq}^\rmi \int \!\rmd^L \vq^\rmf \int \!\rmd^L \tilde{\vq}^\rmf \;
			\bigbraOket {\vq^\rmi} {\hat{\rho}_\rmi} {\tilde{\vq}^\rmi}
			\bigbraOket {\tilde{\vq}^\rmf} {\hat{\Omega}} {\vq^\rmf}
		\\ \times
			\sum_{\gamma, \tilde{\gamma}}
			A_\gamma(\vq^\rmf,\vq^\rmi,t) A_{\tilde{\gamma}}^\ast(\tilde{\vq}^\rmf,\tilde{\vq}^\rmi,t)
			\rme^{ \rmi \left[ R_\gamma(\vq^\rmf, \vq^\rmi, t) - R_{\tilde{\gamma}}(\tilde{\vq}^\rmf, \tilde{\vq}^\rmi, t) \right]/\hbar} \psep.
	}
\end{eqnarray}
We may refer to the coordinates $\tilde{\vq}^{\rmi,\rmf}$ that appear together with the time-reversed evolution $\hat{U}^\dagger$ as \textit{backward} coordinates and to $\tilde{\gamma}$ and as \textit{backward} trajectory as opposed to the \textit{forward} coordinates and trajectory,  $\vq^{\rmi,\rmf}$ and $\gamma$, respectively.

This double sum gives rise to a number of rapidly oscillating contributions as a classical action difference is compared to Planck's quantum of action $\hbar$ in the exponential in\eq\eref{eq:Omegatsumgg}.
More precisely, these oscillations occur when some external parameter is varied by values parametrically small in $\hbar$, as long as this parameter, e.g., the propagation time $t$, is of classical nature.
Thus, under an additional average (e.g., over a time window large compared to $\hbar$ divided by a typical energy of the classical counterpart of the system), these oscillations will get washed out.

Non-vanishing contributions in the presence of such an average are expected to arise only from those trajectory pairs $(\gamma,\tilde{\gamma})$ for which the associated principal functions $R_\gamma$ and $R_{\pr{\gamma}}$ are systematically correlated with each other.
In the absence of any discrete symmetries, such systematic correlations do generically not occur if $\gamma$ is different from $\tilde{\gamma}$, which means that only the pairing of trajectories with themselves, i.e., the \textit{diagonal approximation} $\tilde{\gamma} = \gamma$, contributes significantly%
\footnote{%
While for systems with time reversal symmetry, it is well-known \cite{Sieber2001} that additional correlations exist between non-identical trajectory families (involving self-crossings), such \textit{loop corrections} noticeable, e.g., in spectral correlations at energy differences comparable to the mean level spacing, are subdominant (when one formally sends $\hbar \to 0$) and are neglected here.%
}%
.

More precisely, with $\tilde{\gamma} = \gamma$ one denotes equal trajectory \textit{families}, where initial and final points of the two paired trajectories can be different, $ \tilde{\vq}^{\rmi,\rmf} \neq \vq^{\rmi,\rmf} $, and hence the two trajectories are not literally equal.
Instead, trajectories, as representatives of trajectory families, are functions of the initial and final position and the transition time $t$.
They are represented by the solutions $ \vp^\rmi = \vpig( \vq^\rmf, \vq^\rmi, t ) $ for the initial momentum that solve the boundary value problem $ \vq^\rmf = \vq(\vq^\rmi, \vp^\rmi, t) $, where $\vq(\vq^\rmi, \vp^\rmi, t)$ is the unique classical time evolution of the position, initialized at $( \vq^\rmi, \vp^\rmi )$ in phase space.
The enumeration of different solutions $\gamma$ is what defines the set of trajectory families $\gamma$, whereas the individual trajectories within one family relate to each other by smooth deformation with variation of the boundary values $ \vq^{\rmi,\rmf} $ and $ t $ without crossing a point for which the stability amplitude~\eref{eq:Ag} diverges.
For example, two trajectories where one passes through a caustic and the other doesn't must be representatives of two different families.
In other words, the number of conjugate points $ \kappa_\gamma $, as an integer, cannot change smoothly and is the same for all trajectories within one family.

In diagonal approximation, as justified in presence of some average, the above expression~\eref{eq:Omegatsumgg} reads
\begin{eqnarray} \label{eq:Omegatdiag}
	\eqalign{
		\fl
			\langle \hat{\Omega} \rangle_t^{\mathrm{diag}} = \int \!\rmd^L \vq^\rmi \int \!\rmd^L \tilde{\vq}^\rmi \int \!\rmd^L \vq^\rmf \int \!\rmd^L \tilde{\vq}^\rmf \;
			\bigbraOket {\vq^\rmi} {\hat{\rho}_\rmi} {\tilde{\vq}^\rmi}
			\bigbraOket {\tilde{\vq}^\rmf} {\hat{\Omega}} {\vq^\rmf}
		\\ \times
			\sum_{\gamma}
			A_\gamma(\vq^\rmf,\vq^\rmi,t) A_{\gamma}^\ast(\tilde{\vq}^\rmf,\tilde{\vq}^\rmi,t)
			\rme^{ \rmi \left[ R_\gamma(\vq^\rmf, \vq^\rmi, t) - R_{\gamma}(\tilde{\vq}^\rmf, \tilde{\vq}^\rmi, t) \right]/\hbar} \psep.
	}
\end{eqnarray}
In particular, at the point $\tilde{\vq}^{\rmi,\rmf} = \vq^{\rmi,\rmf}$, the equivalence $\tilde{\gamma} = \gamma$ of two trajectory families yields the equivalence of the two trajectories (as representatives of the same family) for which the action difference \textit{vanishes} exactly, resulting in a robust contribution that survives the average.
With increasing separation of backward and forward coordinates, the action difference becomes larger, such that eventually the contribution gets suppressed in the average.
The scale of this crossover is parametrically small in $\hbar$ (divided by typical classical momenta of the system), which justifies an expansion around identical forward and backward coordinates to further simplify~\eref{eq:Omegatdiag}.
We perform a variable transformation from forward and backward coordinates to their center-of-mass $\vek{R}^\alpha \equiv (R_1^\alpha,\ldots,R_L^\alpha)$ and relative coordinates (also called \textit{chords}) $\vek{r}^\alpha \equiv (r_1^\alpha,\ldots,r_L^\alpha)$, defined through
\begin{eqnarray}
  \vq^\alpha &{}\equiv{}& \vek{R}^\alpha + \vek{r}^\alpha / 2 \psep, \label{eq:qRr}\\
	\tilde{\vq}^\alpha &{}\equiv{}& \vek{R}^\alpha - \vek{r}^\alpha / 2 \psep, \label{eq:qtRr}
\end{eqnarray}
where $\alpha = \rmi, \rmf$.
We treat $\vek{r}^\alpha$ as small variables, i.e., positive powers of $\hbar$, as is induced by the average.
Using the well-known identity
\begin{equation} \label{eq:palphaR}
	p_{\gamma,l}^{\alpha}( \vq^\rmf, \vq^\rmi, t ) = \sigma_\alpha \frac{\partial R_\gamma}{\partial q^\alpha_l}(\vq^\rmf, \vq^\rmi,t) \qquad
			\sigma_\alpha \equiv \left\{
				\begin{array}{rcl}
					-1 &:& \alpha = \rmi \psep, \\
					1  &:& \alpha = \rmf \psep,
		    \end{array} \right.
\end{equation}
that relates the principal function $R_\gamma$ to the initial and final momenta of the trajectory $\gamma$, denoted by $\vp_\gamma^\rmi(\vq^\rmf,\vq^\rmi,t)$ and $\vp_\gamma^\rmf(\vq^\rmf,\vq^\rmi,t)$, respectively, we expand the action difference in the diagonal approximation~\eref{eq:Omegatdiag} up to linear order in $\vek{r}^{\alpha}$,
\begin{equation} \label{eq:RminusR}
	R_\gamma(\vq^\rmf, \vq^\rmi, t) - R_{\gamma}(\tilde{\vq}^\rmf, \tilde{\vq}^\rmi, t)
		\simeq \vek{r}^\rmf \cdot \vpfg \big\rvert_R
		- \vek{r}^\rmi \cdot \vpig \big\rvert_R \psep,
\end{equation}
where we introduced the short-hand notation $ \vp^\alpha_\gamma \big\rvert_R \equiv \vp^\alpha_\gamma(\vRf, \vRi, t) $.
Owing to the choice for the coordinate transformation given by\eqs\eref{eq:qRr}, \eref{eq:qtRr}, the linear expansion~\eref{eq:RminusR} is actually valid up to corrections of cubic order in the coordinates $\vek{r}^\alpha$.
Keeping only the dominant power in $\hbar$ (henceforth indicated by $\cdot \simeq \cdot$) in the smoothly varying prefactors~\eref{eq:Ag}, we evaluate them at the center-of-mass coordinates,
\begin{eqnarray}
	A_\gamma(\vq^\rmf, \vq^\rmi, t) \simeq  A_\gamma(\vRf, \vRi, t) \psep, \\
	A_\gamma^\ast(\tilde{\vq}^\rmf, \tilde{\vq}^\rmi, t) \simeq{} A_\gamma^\ast(\vRf, \vRi, t) \psep,
\end{eqnarray}
where those approximations are actually valid up to quadratic corrections in $\vek{r}^\alpha$ owing to\eqs\eref{eq:qRr}, \eref{eq:qtRr}.
The leading order of the diagonal approximation~\eref{eq:Omegatdiag} is then written as
\begin{equation} \label{eq:Omegatdiaglinr}
	\eqalign{
		\fl
		\langle \hat{\Omega} \rangle_t^{\rm diag} \simeq{} \int \!\rmd^L \vRi \int \!\rmd^L \vek{r}^\rmi \int \!\rmd^L \vRf \int \!\rmd^L \vek{r}^\rmf \;
			\sum_{\gamma} (2 \pi \hbar)^{-L}
				\left| \det \left(
					\frac{\partial p_{\gamma, l}^{\rmi}}{\partial R_{l^\prime}^\rmf} \biggr\rvert_R
				\right)_{l,l^\prime} \right|
		\nonumber \\ \times
			\bigbraOket {\vRi + \vek{r}^\rmi/2} {\hat{\rho}_\rmi} {\vRi - \vek{r}^\rmi/2} \,
			\bigbraOket {\vRf - \vek{r}^\rmf/2} {\hat{\Omega}} {\vRf + \vek{r}^\rmf/2}
		\nonumber \\ \times
			\exp \left[ \frac{\rmi}{\hbar} \left( \vek{r}^\rmf \cdot \vpfg \big\rvert_R   - \vek{r}^\rmi \cdot \vpig \big\rvert_R \right) \right] \psep.
	}
\end{equation}
Throughout, we denote evaluation at $(\vRf,\vRi,t)$ with $(\cdot)\vert_R$.
We recognize the Weyl symbol of $\hat{\rho}_\rmi$, i.e., the Wigner function (up to normalization) of the initial state,
\begin{equation} \label{eq:Wi}
	\Wi(\vek{R}, \vp) = (2\pi\hbar)^{-L} \int \rmd^L \vek{r} \;
		\bigbraOket {\vek{R} + \vek{r}/2} {\hat{\rho}_\rmi} {\vek{R} - \vek{r}/2}
		\rme^{-\rmi \vek{r} \cdot \vp/\hbar} \psep,
\end{equation}
and the Weyl symbol of the observable
\begin{equation} \label{eq:Ow}
	\Ow(\vek{R}, \vp) = \int \rmd^L \vek{r} \;
		\bigbraOket {\vek{R} - \vek{r}/2} {\hat{\Omega}} {\vek{R} + \vek{r}/2} 
		\rme^{\rmi \vek{r} \cdot \vp/\hbar} \psep,
\end{equation}
to identify
\begin{equation} \label{eq:OmegatdiagWO}
		\fl
		\langle \hat{\Omega} \rangle_t^{\rm diag} \simeq{} \int \!\rmd^L \vRi \int \!\rmd^L \vRf \;
		\sum_{\gamma}
		\left| \det \!\left(
		\frac{\partial p_{\gamma, l}^{\rmi}}{\partial R_{l^\prime}^\rmf} \biggr\rvert_R
		\right)_{l,l^\prime} \right|
		\Wi\!\left( \vRi, \vp_{\gamma}^\rmi \big\rvert_R \right)
		\Ow\!\left( \vRf, \vp_{\gamma}^\rmf \big\rvert_R \right) \psep.
\end{equation}

In order to make the above expression much more amenable to implementations one can transform the double integral over \textit{boundary values} $\vRi, \vRf$ into an integral over \textit{initial values} $\vRi, \vPi$ by applying a \textit{sum rule} argument.
To this end, we make the generic assumption that all trajectories $\gamma$ going from $\vRi$ to $\vRf$ in time $t$ are well isolated from each other (which neglects the occasional occurrence 
of trajectory bifurcations that could arise for specific choices of $\gamma$, $\vRi$ and $\vRf$).
We can then use the composition property of the Dirac delta distribution to identify
\begin{equation} \label{eq:deltaRdeltaP}
	\fl
	\prod_l \delta\!\left[ R^{\rmf}_l - q_l(\vRi, \vPi, t) \right] =
		\sum_{\gamma}
			\left| \det \!\left(
				\frac{\partial p_{\gamma, l}^{\rmi}}{\partial R_{l^\prime}^\rmf} \biggr\rvert_R
			\right)_{l,l^\prime} \right|
		\prod_l \delta\!\left[ P^{\rmi}_l - p^\rmi_{\gamma,l}(\vRf, \vRi, t) \right] \psep.
\end{equation}
As above, the enumeration of solutions $\vPi$ that fulfill $\vR^{\rmf} = \vq(\vRi, \vPi, t)$ is exactly given by the different trajectories $\gamma$ from $\vRi$ to $\vRf$ in time $t$ and $\vq(\vRi,\vPi,t)$ is the position that results from the classical time evolution starting with 
the initial position $\vRi$ and momentum $\vPi$.
The Dirac delta identity~\eref{eq:deltaRdeltaP} implies the integral form
\begin{equation} \label{eq:sumrule}
	\eqalign{
		\fl
	  	\sum_\gamma\left| \det\left(
	  		\frac{\partial p_{\gamma, l}^{\rmi}}{\partial R_{l^\prime}^\rmf} \biggr\rvert_R
	  	\right)_{l, l^\prime}  \right| 
		 	f\!\left( \vp_{\gamma}^\rmi \big\rvert_R, \vp_{\gamma}^\rmf \big\rvert_R \right) 
	\\
	=
	  \int \!\rmd^L \vPi \; f\!\left[ \vPi,\vp(\vRi,\vPi,t) \right] 
	  \prod_{l=1}^L \delta\!\left[ R_l^\rmf - q_l(\vRi,\vPi,t) \right]
	}
\end{equation}
for arbitrary (smooth) functions $f$ of the initial and final momenta, with $\vp(\vRi,\vPi,t)$ the unique classical time evolution of momentum coordinates starting with the initial positions $\vRi$ and momenta $\vPi$.
Note that $ \vp\!\left( \vRi, \vpig( \vRf, \vRi, t ), t \right) = \vpfg( \vRf, \vRi, t ) $.
Using the Dirac delta identity~\eref{eq:deltaRdeltaP}, respectively~\eref{eq:sumrule}, in~\eref{eq:OmegatdiagWO} finally yields the TWA
\begin{equation} \label{eq:TWOmega}
	\langle \hat{\Omega} \rangle_t^{\rm diag} \simeq \int \!\rmd^L \vRi \int \!\rmd^L \vPi \;
		\Wi\!\left( \vRi, \vPi \right)
		\Ow\!\left[ \vq(\vRi, \vPi, t), \vp(\vRi, \vPi, t) \right] \psep.
\end{equation}
Evaluating this double integral numerically through a Monte-Carlo method amounts then in practice to
computing the time evolution of the expectation value of an observable $\hat{\Omega}$ (with Weyl symbol $\Ow$) given an initial state Wigner distribution $\Wi$.
The application to transition probabilities $ P(t) = \tr \!\left[ \hat{\rho}_\rmi(t) \hat{\rho}_\rmf \right] $ from a state $ \hat{\rho}_\rmi $ to a state $ \hat{\rho}_\rmf $ in time $ t $, which are of particular interest here, is obtained by identifying $ \hat{\Omega} = \hat{\rho}_\rmf $ (see \sref{sec:transprobcs}).

One significant property of the standard TWA~\eref{eq:TWOmega} that gains particular clarity from the semiclassical derivation is that \textit{interference effects} are neglected.
This is the essence behind the diagonal approximation~\eref{eq:Omegatdiag}, i.e, $ \tilde{\gamma} = \gamma $, under which contributions of different classical paths are only summed up \textit{incoherently}.
This renders the standard TWA essentially classical, unable to describe genuine interference effects, e.g., in the many-body context.
This is one of its major draw-backs, which cannot be overcome easily, since the diagonal approximation is what allows the transformation from boundary value to initial value sampling in the first place, which in turn is one of the major strengths of the standard TWA.
This can partially be resolved by adding sub-leading corrections in $\hbar$, e.g., in form of stochastic quantum jumps \cite{Polkovnikov2010}, but the main deficiency to incorporate interference between well-separated classical paths (like in a double-slit scenario) remains.

While generically some external average (e.g., temporal or configurational) will destroy such coherence and thus validate the TWA, we will argue in the following that systems with discrete symmetries constitute a special class of counter examples in which the coherence of certain interference effects survives, resulting in significant deviation from TWA.
Most notably, in this case it is possible to correctly account for these robust interference effects in an \textit{augmented Truncated Wigner approach}.

\section{Discrete symmetries}
\label{sec:discrsym}
\subsection{Preliminaries}
\label{sec:preliminaries}
The above approach is generally expected to yield satisfactory agreement 
with exact quantum calculations in the presence of some configurational or 
temporal average. 
It systematically fails, however, as soon as discrete symmetries play a role \cite{Schlagheck2019}.
More precisely, the diagonal approximation~\eref{eq:Omegatdiag} may be too simplistic an approximation if the Hamiltonian $\hat{H}$ exhibits a discrete symmetry, depending on the observable under consideration.
We consider symmetries under positional point transformations
\begin{equation} \label{eq:f}
	\vek{f}: \mathbb{R}^L \to \mathbb{R}^L \qquad \vq \mapsto \vek{f}(\vq) \psep,
\end{equation}
naturally assumed to restore the identity after a finite number of iterations, $\vek{f} \circ \vek{f} \circ \dots \circ \vek{f} = {\rm id}_{\mathbb{R}^L}$, which implies that it is volume preserving.
As a consequence, the induced symmetry operation acting on Hilbert space, given by
\begin{eqnarray} \label{eq:Uf}
	\eqalign{
		\hat{U}_{\vek{f}} \ket{ \vq } \equiv \ket{ \vek{f}(\vq) } \psep,\\
		\!\bigl[ \hat{U}_{\vek{f}} , \hat{H} \bigr] = 0 \psep,
	}
\end{eqnarray}
is unitary, i.e., $ \hat{U}_{\vek{f}}^\dagger \hat{U}_{\vek{f}} = 1 $.
For instance, the symmetry could be a parity with respect to a mirror operation in the case of a one-dimensional chain of oscillators with identical properties.
We further presume that both the initial state and the observable are symmetric, $ \bigl[ \hat{U}_{\vek{f}} , \hat{\rho}_\rmi \bigr] = \bigl[ \hat{U}_{\vek{f}} , \hat{\Omega} \bigr] = 0 $, equivalent to
\begin{equation} \label{eq:rhoiOsym}
	\eqalign{
		\bigbraOket {\vq} {\hat{\rho}_\rmi} {\vq^\prime}
			= \bigbraOket {\vek{f}(\vq)} {\hat{\rho}_\rmi} {\vek{f}(\vq^\prime)} \psep,
	\\
		\bigbraOket {\vq} {\hat{\Omega}} {\vq^\prime}
			= \bigbraOket {\vek{f}(\vq)}  {\hat{\Omega}} {\vek{f}(\vq^\prime)} \psep.
	}
\end{equation}
in position representation.
Equivalently, if the system is prepared in a pure state $\hat{\rho}_\rmi = \ket{\phi_\rmi} \bra{\phi_\rmi}$, then this has to be an eigenstate of the symmetry operation, $ \hat{U}_{\vek{f}} \ket{ \phi_\rmi} = \rme^{\rmi \theta} \ket{ \phi_\rmi} $.
A mixed initial state is symmetric if all populated pure states are eigenstates (e.g., in the eigenvalue decomposition).
If one is interested in transition probabilities, this holds analogously for the final state $\hat{\Omega} = \hat{\rho}_\rmf $.

At the level of the Wigner phase-space representations~\eref{eq:Wi} and~\eref{eq:Ow} this implies the approximate symmetry condition
\begin{equation} \label{eq:WOsymApprox}
	\eqalign{
		\Wi(\vq, \vp) \simeq \Wi(\vq^\prime, \vp^\prime) \psep,
	\\
		\Ow(\vq, \vp) \simeq \Ow(\vq^\prime, \vp^\prime) \psep,
	}
\end{equation}
where $(\vq^\prime, \vp^\prime)$ is the canonical symmetry transform of $(\vq, \vp)$ in phase space, i.e.,
\begin{equation} \label{eq:qptrafo}
	\eqalign{
		\vq^\prime = \vek{f}(\vq) \psep,
	\\
		\vp^\prime = \tilde{\vek{f}}_\vq(\vp) \equiv [D(\vq)]^{-\rmT} \cdot \vp \qquad \text{with} \qquad D(\vq) = \left( \frac{\partial f_{l}}{\partial q_{l^\prime}} \right)_{l, l^\prime} \psep,
	}
\end{equation}
and $(\cdot)^{-\rmT}$ denotes transposition of the inverse.
If the symmetry operation $\vek{f}$ is a \textit{linear} coordinate transformation---which happens to be the case for all practical applications we are considering later---then the symmetry condition~\eref{eq:WOsymApprox} is \textit{exact}, otherwise it holds in the vicinity of the symmetric manifold in \textit{phase space} that is invariant under~\eref{eq:qptrafo},
\begin{equation}
	{\SP} \equiv \bigl\{ (\vq,\vp) \in \mathbb{R}^{2L}
		\,\bigm\vert\,
		( \vek{f}(\vq) = \vq) \wedge ( \tilde{\vek{f}}_\vq(\vp) = \vp) \bigr\} \psep.
\end{equation}
Similarly, we define the symmetric manifold in \textit{position space} as
\begin{equation}
	\Sq \equiv \bigl\{ \vq \in \mathbb{R}^{L}
		\,\bigm\vert\,
		\vek{f}(\vq) = \vq \bigr\} \psep.
\end{equation}

\subsection{Degeneracy of trajectories and the role of symmetric families}

\begin{figure}
	\includegraphics[width=0.9\textwidth]{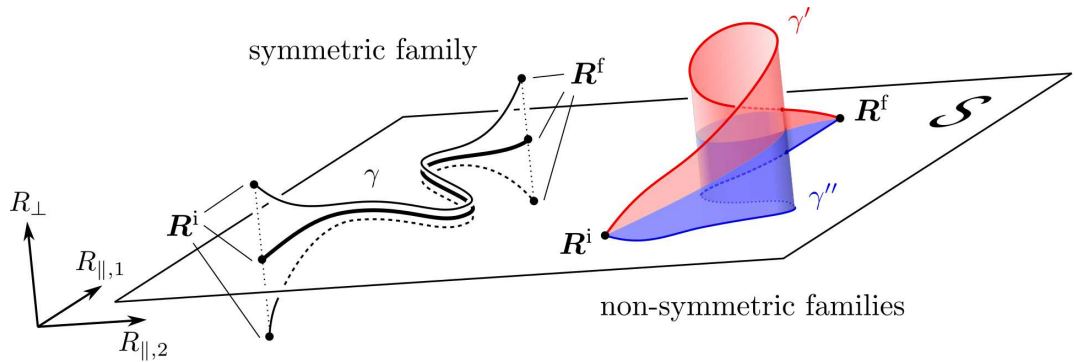}
	\caption{
		\label{fig:symnonsymgamma}
		Sketch of a symmetric trajectory family $\gamma$ (left) which comprises symmetry partners of all contained trajectories (thin lines).
		Consequently its trajectories for boundary positions $\vRi$ and $\vRf$ (dots) chosen in the symmetric manifold $\Sq$ lie fully within $\Sq$, and stay close to $\Sq$ otherwise (generic scenario for chaotic dynamics).
		In contrast, nonsymmetric families $\gamma^\prime$ (right) do not map onto themselves under the symmetry operation but to \textit{distinct}, equivalent families, $\gamma^\prime \mapsto \gamma^{\prime \prime} \neq \gamma^\prime$ (and vice-versa).
		They comprise trajectories (red and blue lines) that do not lie inside of $\Sq$ regardless of the initial and final positions.
		As a guide for the eye, their projections onto $\Sq$ are indicated by shaded surfaces (red and blue).
	}
\end{figure}

Under these prerequisites, a trajectory going from the initial state to the final state will have one or several partner trajectories that can be explicitly obtained by applying the symmetry operation to it.
Those partner trajectories exhibit exactly the same principal functions~\eref{eq:Rgamma}, amplitude prefactors~\eref{eq:Ag} and weights~\eref{eq:rhoiOsym} from the initial state and observable.
Their existence therefore gives rise to non-vanishing non-diagonal terms in the double sum arising within the semiclassical expression~\eref{eq:Omegatsumgg}, whenever they belong to \textit{different} families $\tilde{\gamma} \neq \gamma$.
As a consequence, the effective contribution from this trajectory and its symmetry-related partners in semiclassical approximation is underestimated by some integer degeneracy factor $g$ within the diagonal approximation~\eref{eq:Omegatdiag}.
In the case of a simple parity (as depicted in \fref{fig:symnonsymgamma}) giving rise to ``even'' and ``odd'' eigenstates of the Hamiltonian, this degeneracy factor would equal $g=2$.

It would be tempting to therefore simply multiply the final expression~\eref{eq:TWOmega} for the approximated expectation value by the degeneracy factor $g$ in order to account for this constructive interference effect.
While this strategy is expected to work out very well in the presence of chaotic dynamics and in the case of long evolution times $t$ well beyond the Ehrenfest time $\tE$, it systematically fails for shorter times and/or in the absence of chaos (not mentioning additional quantum effects such as (dynamical) localization \cite{Anderson1958,Fishman1982,Shepelyansky1983,Bohigas1993}, dynamical tunneling \cite{Davis1981,Bohigas1993,Tomsovic1994}, or scarring \cite{Heller1984,Bogomolny1988}).
There, a finite fraction of trajectory families $\gamma$ that significantly contribute to the expectation value, according to~\eref{eq:Omegatdiag}, are \textit{symmetric}.
By symmetric we denote those families that also include the symmetry-related counterparts of any contained trajectory.
As these families are mapped onto themselves under the discrete symmetry operation, they are identical with their symmetry-partners, and thus already accounted for in the diagonal approximation $\tilde{\gamma} = \gamma$.
Therefore, their contributions to $ \langle \hat{\Omega} \rangle_t $ or to the transition probability $P(t)$ must not be multiplied with the degeneracy factor $g$.
Similarly, if the overall symmetry contains sub-symmetries, the degeneracy factor $g$ might get reduced to an integer factor if $\gamma$ is symmetric with respect to a subset of symmetries.

The generic coexistence of symmetric and nonsymmetric families is illustrated in \fref{fig:symnonsymgamma}.
Note that, when both the initial and final positions $\vRi, \vRf$ lie exactly within the symmetric subspace $\Sq$, then the corresponding trajectory from a \textit{symmetric} family is itself invariant under the symmetry transformation and lies fully within ${\SP}$.
In contrast, this is not the case for nonsymmetric families.
Put around, any phase-space point in ${\SP}$ belongs to a symmetric trajectory.
Also note that individual trajectories from symmetric families are \textit{not} invariant under the symmetry transformation as soon as $\vRi$ or $\vRf$ lie outside of $\Sq$.
Nevertheless, assuming chaotic dynamics, they will stay very close to ${\SP}$ during the entire time evolution.
In other words, moving $\vRi, \vRf$ away from $\Sq$ (while keeping the \textit{family} $\gamma$ fixed) corresponds to a variation in the \textit{stable directions} of a reference trajectory that lies fully inside ${\SP}$.
The reason is that otherwise exponential separation from the reference trajectory would lead to an orbit that explores phase space in a way that is completely unrelated to the reference trajectory, e.g., passing through unrelated conjugate points, i.e., it would belong to a \textit{different} family $\gamma$.
This generic behaviour for chaotic motion was used in\rf\cite{Schlagheck2019} to introduce a heuristic distance criterion in order to discriminate symmetric from nonsymmetric families.
While such a criterion gives satisfactory results in many cases, it nonetheless suffers from innate ambiguity and moreover seems to fail in phase-space regions with regular or mixed dynamics (see applications in \sref{sec:BHapplication}).

\subsection{The idea of an augmented Truncated Wigner Approximation}
\label{sec:augTWAidea}

Here, we pursue a viable alternative strategy to account for this complication in the framework of a Truncated Wigner approach.
We promote the idea to separately calculate the particular contribution of symmetric trajectory families to~\eref{eq:OmegatdiagWO} through a Monte Carlo sampling that is restricted to the symmetric subspace ${\SP}$.
Multiplying those contributions by $g-1$ and then subtracting them from the numerically determined TWA~\eref{eq:TWOmega}, multiplied by $g$, will yield a more precise semiclassical prediction of the expectation value $ \langle \hat{\Omega} \rangle_t $ or transition probability $ P(t) $ under consideration.
Analogously, if sub-symmetries are present, an appropriate combination of such individual TWA-like samplings, one for each relevant sub-symmetry, has to be taken (see application in \sref{sec:BHapplication}).

A sampling within ${\SP}$ will automatically yield all the symmetric trajectories.
This selects all symmetric families $\gamma$, each by the particular representatives that are invariant under~\eref{eq:qptrafo}.
The influence from these families as a whole will then be incorporated by the \textit{local behaviour} of their trajectories in the vicinity of ${\SP}$.
To make this approach quantitatively accurate, $\Wi$ and $\Ow$ have to be sufficiently \textit{localized} around ${\SP}$, such that they drop off in magnitude fast enough with increasing separation from ${\SP}$.
In particular this is the case for coherent states centered around $\vek{z} \in {\SP}$ (e.g., corresponding to a quantum state within a one-dimensional symmetric chain of an odd number of coupled oscillators where only the central oscillator is excited).
A more precise statement will be given later.

\section{The augmented Truncated Wigner Approximation}
\label{sec:augTWA}
\subsection{Coordinates}
\label{sec:coordinates}
We suppose that the phase-space variables chosen to represent the $L$ degrees of freedom that constitute the system have been adapted to the presence of the discrete symmetry under consideration.
Denoting by $S$ the dimension of the symmetric subspace $\mathcal{S}$ in position space, we assume that the first $S$ coordinates of the position and momentum vectors $\vq$ and $\vp$, referred to as ``parallel'' components in the following, represent the degrees of freedom that lie within the symmetry subspace $ {\SP} $, while the remaining $L-S$ coordinates, henceforth referred to as ``perpendicular'' components, comprise the degrees of freedom that describe the motion out of that subspace.
By convention, we will use the index symbols
\begin{equation} \label{eq:siglam}
	\sigma \in \{ 1, \ldots, S \} \quad \text{ and } \quad
	\lambda \in \{ S + 1, \ldots, L \}
\end{equation}
to index the parallel and perpendicular components, respectively.
To be more specific, we may write
\begin{eqnarray}
	\vq = ( \vqpara, \vqperp ) = (
		\underbrace{q_1, \ldots, q_S}_{\vqpara},
		\underbrace{q_{S+1}, \ldots, q_L}_{\vqperp}
		) \psep, \\
	\vp = ( \vppara, \vpperp ) = (
		\underbrace{p_1, \ldots, p_S}_{\vppara},
		\underbrace{p_{S+1}, \ldots, p_L}_{\vpperp}
		) \psep,
\end{eqnarray}
with the invariant manifolds in position space and phase space simply given by
\begin{eqnarray}
	\Sq = \bigl\{ \vq \in \mathbb{R}^L \bigm\vert \vqperp = \veknull \bigr\} \psep,\\
	{\SP} = \bigl\{ (\vq, \vp) \in \mathbb{R}^{2L} \bigm\vert \vqperp = \vpperp = \veknull \bigr\} \psep.
\end{eqnarray}
Since $\dot{\vq}_\perp = \veknull$ if $\vpperp = \veknull$ and $\vqperp = \veknull$, a trajectory that in position space stays within $\Sq$ for all times is characterized by the property that the perpendicular momenta $\vpperp$ vanish.
With the prerequisites given in \sref{sec:preliminaries} one can show that such a choice of phase-space coordinates is always possible (see \ref{app:symorientedqp}).
Moreover, in the presence of multiple symmetries, one can always find coordinates $\vq, \vp$ that simultaneously fulfil these requirements for each individual (sub-)symmetry.

Using these symmetry-oriented conjugate phase-space coordinates, the symmetric trajectory families $\gamma$ are exactly those for which setting the initial and final position of the representative trajectory to $\vq^\rmi, \vq^\rmf \in \mathcal{S}$, i.e., $\vq_\perp^\rmi = \vq_\perp^\rmf = \veknull$, implies $\vpigperp(\vq^\rmf, \vq^\rmi,t) = \vpfgperp(\vq^\rmf, \vq^\rmi,t) = \veknull$%
\footnote{%
	Note that we do not consider the possibility of symmetric trajectory families that are not connected to $\mathcal{S}$, i.e., which do not comprise trajectories with $\vq^\rmi, \vq^\rmf \in \mathcal{S}$.
	This may occur if the trajectories are separated from $\mathcal{S}$ by a caustic, outside of which the representatives could still be smoothly deformed into their symmetry partners, not leaving the family $\gamma$.
	Such families are, however, negligible in our case, where large separations of $\vq^\rmi$ and $\vq^\rmf$ from $\mathcal{S}$ are strongly suppressed due to the localization of the initial state and of the observable (or final state) on $\mathcal{S}$.
}.

\subsection{Local separation}
An essential feature of the dynamics close to the symmetric manifold ${\SP}$ is that in linear order the dynamics separate into the parallel and perpendicular degrees of freedom.
In particular, the unique time evolution of phase-space coordinates $\vq( \vRi, \vPi, t)$ and $\vp( \vRi, \vPi, t)$ obey
\begin{eqnarray} \label{eq:qpRPsiglam}
	\begin{aligned}
		&\left.\frac{\partial q_\lambda} {\partial R^\rmi_{\sigma} }\right\rvert_{\SP} = 0 \psep, \quad &
		&\left.\frac{\partial p_\lambda} {\partial R^\rmi_{\sigma} }\right\rvert_{\SP} = 0 \psep, \quad &
		&\left.\frac{\partial q_\lambda} {\partial P^\rmi_{\sigma} }\right\rvert_{\SP} = 0 \psep, \quad &
		&\left.\frac{\partial p_\lambda} {\partial P^\rmi_{\sigma} }\right\rvert_{\SP} = 0 \psep,
	\\
		&\left.\frac{\partial q_\sigma } {\partial R^\rmi_{\lambda}}\right\rvert_{\SP} = 0 \psep, &
		&\left.\frac{\partial p_\sigma } {\partial R^\rmi_{\lambda}}\right\rvert_{\SP} = 0 \psep, &
		&\left.\frac{\partial q_\sigma } {\partial P^\rmi_{\lambda}}\right\rvert_{\SP} = 0 \psep, &
		&\left.\frac{\partial p_\sigma } {\partial P^\rmi_{\lambda}}\right\rvert_{\SP} = 0 \psep.
	\end{aligned}
\end{eqnarray}
The validity of the first line of\eq\eref{eq:qpRPsiglam} is obvious, since a trajectory fully in ${\SP}$ will stay in ${\SP}$ as long as the initial phase-space point is also changed only inside ${\SP}$.
To show that also the other cross dependencies vanish [second line of\eq\eref{eq:qpRPsiglam}], one considers the initial and final momenta $\vp^\alpha_\gamma( \vRi, \vRf, t)$ for symmetric trajectories with $\vRi, \vRf \in \mathcal{S}$ and uses the relation between momenta and Hamilton's principal function~\eref{eq:palphaR} (see \ref{app:locsepdyn}).

\subsection{Extracting symmetric trajectory families}
\label{sec:extractsym}

We start from expression~\eref{eq:OmegatdiagWO} for the diagonal approximation.
The goal is to specifically address the contribution of only the symmetric trajectory families $\gamma$, which, at the level of the boundary value problem~\eref{eq:OmegatdiagWO}, we denote by replacing
\begin{equation}
	\sum_{\gamma} \rightarrow \sum_{\gamma \; \text{sym.}} \psep.
\end{equation}
The non-trivial part is to transform this contribution into an initial value problem in order to allow for efficient Monte Carlo simulations in practice.
To do so we devise a variant of \eref{eq:deltaRdeltaP} which selects only solutions in $\vPi$ that correspond to the symmetric trajectory families.
For this purpose, we introduce the approximate classical time evolution $\vqsap(\vRi, \vPi, t) \simeq \vq(\vRi, \vPi, t)$ close to ${\SP}$ given by
\begin{equation} \label{eq:qapproxlin}
	\eqalign{
		\vqsappara( \vRi, \vPi, t) = \vqpara( \vRin, \vPin, t) \psep,\\
		\vqsapperp( \vRi, \vPi, t) =
			\sum_{\lambda=S+1}^{L} \frac{\partial \vqperp}{\partial R^\rmi_\lambda} \biggr\rvert_{\SP} R^\rmi_\lambda
			+ \sum_{\lambda=S+1}^{L} \frac{\partial \vqperp}{\partial P^\rmi_\lambda} \biggr\rvert_{\SP} P^\rmi_\lambda
			\psep,
	}
\end{equation}
where the subscript $0$ is a notation introduced to indicate the projection onto ${\SP}$,
\begin{equation}
	\eqalign{
		\vR_0 &{}\equiv ( R_1, \ldots, R_S, 0, \ldots, 0 ) = ( \vR_\para, \veknull ) \psep,
	\\
		\vP_0 &{}\equiv ( P_1, \ldots, P_S, 0, \ldots, 0 ) = ( \vP_\para, \veknull ) \psep,
	}
\end{equation}
and we introduced the notation $(\cdot)\rvert_{\SP}$ for functions of the phase-space coordinates to be evaluated at $(\vRi, \vPi) = (\vRin, \vPin)$.
We will refer to the definition~\eref{eq:qapproxlin} simply as the \textit{linear approximation}, since it is the expansion of $ \vq(\vRi, \vPi, t) $ about $ {\SP} $ to linear order in $ \vRiperp $ and $ \vPiperp $, as is implied by the local separation of dynamics~\eref{eq:qpRPsiglam}.

The approximate evolution $\vqsap$ describes the dynamics of the system in the direct vicinity of the symmetric manifold ${\SP}$ and therefore reflects the behavior of only the symmetric trajectory families $\gamma$.
As particular representatives, the trajectories fully in ${\SP}$, which are themselves symmetric, are described exactly, while the other representatives of the symmetric families $\gamma$ in the vicinity of ${\SP}$ are only approximated.
The linear approximation~\eref{eq:qapproxlin} is based on infinitesimal variation of the initial condition out of ${\SP}$.
The corresponding infinitesimal change to the whole symmetric trajectory makes it nonsymmetric as an individual trajectory, but it cannot break its membership to the symmetric family.

One can formally derive (see \ref{app:symtrajDelta}) a variant of the Dirac-delta identity~\eref{eq:deltaRdeltaP} that selects only symmetric families $\gamma$, namely by using the approximate evolution $\vqsap( \vRi, \vPi, t )$ instead of the true time evolution.
Note that the explicit linear ap\-pro\-xi\-ma\-ti\-on~\eref{eq:qapproxlin} is not even necessary for that purpose and can be replaced by a less restrictive set of sufficient properties of $\vqsap$ as detailed in \ref{app:symtrajDelta}.
We find
\begin{equation} \label{eq:deltaRdeltaPsym}
	\fl
	\prod_l \delta\!\left[ R^{\rmf}_l - \sap{q}_l(\vRi, \vPi, t) \right] 
		=	\sum_{\gamma\ \text{sym.}}
			\left| \det \!\left(
				\frac{\partial \sap{p}_{\gamma, l}^{\rmi}}{\partial R_{l^\prime}^\rmf} \biggr\rvert_R
			\right)_{l,l^\prime} \right|
		\prod_l \delta\!\left[ P^{\rmi}_l - \sap{p}^\rmi_{\gamma,l}(\vRf, \vRi, t) \right] \psep,
\end{equation}
where $\vpsapig( \vRf, \vRi, t )$ are the momentum roots of the altered boundary value problem
\begin{equation} \label{eq:pgammasym}
	\vRf = \vqsap( \vRi, \vPi, t )
		\quad \Leftrightarrow \quad
		\vPi \in \left\{ \vpsapig( \vRf, \vRi, t) \right\}_{\gamma\ \text{sym.}}
\end{equation}
involving the linearized position evolution~\eref{eq:qapproxlin}.
They are enumerated by the symmetric trajectory families $\gamma$
and approximate the corresponding initial momenta $ \vpig( \vRf, \vRi, t ) $ close to $\vRi, \vRf \in \Sq$ with exact coincidence $\vpsapig \bigr\rvert_\Sq = \vpig \bigr\rvert_\Sq$ in the symmetric manifold.

\subsection{TWA of symmetric trajectory families}
\label{sec:TWAsym}
The identity~\eref{eq:deltaRdeltaPsym} lays the foundation for converting the boundary-value integral
\begin{equation} \label{eq:Omegatgsym}
	\fl
		\langle \hat{\Omega} \rangle_t^{\rm sym} = \int \!\rmd^L \vRi \int \!\rmd^L \vRf \;
		\sum_{\gamma\ \text{sym.}}
		\left| \det \!\left(
			\frac{\partial p_{\gamma, l}^{\rmi}}{\partial R_{l^\prime}^\rmf} \biggr\rvert_R
		\right)_{l,l^\prime} \right|
		\Wi\!\left( \vRi, \vp_{\gamma}^\rmi \big\rvert_R \right)
		\Ow\!\left( \vRf, \vp_{\gamma}^\rmf \big\rvert_R \right)
\end{equation}
of only the symmetric contributions into an integral over initial values similar to\eq\eref{eq:TWOmega}.
The selection of symmetric families in\eq\eref{eq:deltaRdeltaPsym} relies on the approximation of the dynamics close to ${\SP}$.
To assure that this does not lead to significant errors we need to introduce the requirement that both, the initial state $\Wi$ and the observable $\Ow$ be sufficiently localized around the symmetric manifold ${\SP}$ in phase space (as already mentioned in~\sref{sec:augTWAidea}).
In particular, we will consider the perpendicular coordinates
$ \vRiperp, \vRfperp, \vpigperp, \vpfgperp $ to be parametrically small in $\hbar$, due to a suppression of larger values by the weight terms $\Wi$ and $\Ow$ in the integral~\eref{eq:OmegatdiagWO}.
For instance, in the case of coherent states $W_\alpha(\vq^\alpha,\vp^\alpha)$ [see\eq\eref{eq:Wcoh}] located at $\vek{z} \in {\SP}$, the perpendicular coordinates are of order
\begin{equation} \label{eq:perpshbar}
	\eqalign{
		\vqperp^\alpha = \Ord(\sqrt{\hbar}) \psep, \\
		\vpperp^\alpha = \Ord(\sqrt{\hbar}) \psep.
	} \qquad \alpha \in \{ \rmi, \rmf \}
\end{equation}
The \qp-symmetric case~\eref{eq:perpshbar} is particularly interesting for the application to transition probabilities of Bose-Einstein condensates (see~\sref{sec:transprobcs}), where the inverse of the average filling factor $ L / N $ takes the role of $\hbar$ [see\eq\eref{eq:hbareff} of \sref{sec:BHsystems}].
However, it is not necessary to strictly impose \qp symmetry.
As shown in \ref{app:asym}, a less restrictive variant of the requirement~\eref{eq:perpshbar}, which is still in tune with the minimum uncertainty principle, can be formulated to allow for somewhat more asymmetric uncertainties between $ \vq $ and $ \vp $, which is especially interesting in view of possible applications to squeezed many-body states.
For simplicity, we will explicitly work here with the symmetric version~\eref{eq:perpshbar} while we note that a corresponding analysis with weaker assumptions leads to the same results (see \ref{app:ApproxSubMomenta} and \ref{app:swapandshift}).

Firstly, we relate the determinant in\eq\eref{eq:Omegatgsym} to its counterpart that involves the approximate time evolution, appearing in\eq\eref{eq:deltaRdeltaPsym}.
Because the determinant is a smoothly varying object, it is, for our purpose, sufficient to evaluate it on $\Sq$,
\begin{equation} \label{eq:detonS}
	\det \!\left(
			\frac{\partial p_{\gamma, l}^{\rmi}}{\partial R_{l^\prime}^\rmf}(\vRf, \vRi,t)
		\right)_{l,l^\prime} 
	\simeq \det \!\left(
			\frac{\partial p_{\gamma, l}^{\rmi}}{\partial R_{l^\prime}^\rmf}(\vRfn, \vRin,t)
		\right)_{l,l^\prime} \psep.
\end{equation}
The approximate equality, $\mathcal{A} \simeq \mathcal{B}$, appearing in\eq\eref{eq:OmegatgsymWOPi} and henceforth indicates equality to leading order in $\hbar$, i.e., $\mathcal{A} = \mathcal{B} \times (1 + \Ord(\hbar^\nu))$ for some $\nu > 0$, namely in particular $\nu = 1/2$ for\eq\eref{eq:perpshbar}.
Furthermore, as the full and approximate evolution, $\vq$ and $\vqsap$, coincide in linear order of $\vRiperp$ and $\vPiperp$, so do the corresponding momentum roots, $\vpig$ and $\vpsapig$, in linear order of $\vRiperp$  and $\vRfperp$ (see \ref{app:linequivp}), and we have
\begin{equation} \label{eq:dpdRfdsappdRf}
	\frac{\partial \sap{p}_{\gamma,l}^{\rmi}}{\partial 	R_{l^\prime}^\alpha} (\vRfn, \vRin,t) =
		\frac{\partial p_{\gamma,l}^{\rmi}}{\partial R_{l^\prime}^\alpha} (\vRfn, \vRin,t)
		\qquad  l,l^\prime \in \{ 1, \ldots, L \}, \alpha \in \{ \rmi, \rmf \} \psep.
\end{equation}
Together with the analogous expression of\eq\eref{eq:detonS} for the approximate solution $ \vpsapig $ this allows us to rewrite\eq\eref{eq:Omegatgsym} as
\begin{equation} \label{eq:OmegatgsymWOPi}
	\eqalign{
		\fl
		\langle \hat{\Omega} \rangle_t^{\rm sym} \simeq{}
		\int \!\rmd^L \vRi \int \!\rmd^L \vRf \int \!\rmd^L \vPi
		\sum_{\gamma\ \text{sym.}}
		\left| \det \!\left(
			\frac{\partial \sap{p}_{\gamma, l}^{\rmi}}{\partial R_{l^\prime}^\rmf} \biggr\rvert_R
		\right)_{l,l^\prime} \right|
	\\ \times
		\prod_l \delta \!\left[ P^\rmi_l - \sap{p}^\rmi_{\gamma, l}( \vRf, \vRi, t) \right]
		\Wi\!\left( \vRi, \vp_{\gamma}^\rmi \big\rvert_R \right)
		\Ow\!\left( \vRf, \vp_{\gamma}^\rmf \big\rvert_R \right) \psep,
	}
\end{equation}
where we have inserted the unity $ \int \!\rmd^L \vPi \prod_l \delta \left[ P^\rmi_l - \sap{p}^\rmi_{\gamma, l}( \vRf, \vRi, t) \right] $.

In order to liberate\eq\eref{eq:OmegatgsymWOPi} from the integral over $ \vRf $ by the use of\eq\eref{eq:deltaRdeltaPsym} we have to get rid of the explicit dependence on the family $\gamma$ that is inherent to $\Wi$ and $\Ow$ via the evaluation at $ \vp^{\rmi,\rmf}_\gamma \bigr\vert_R $.
Unlike the case of standard TWA, c.f.\eq\eref{eq:sumrule}, the discrepancy between the full and approximate dynamics close to ${\SP}$ inhibits the direct replacement of $ \vpig \bigr\vert_R $ by $ \vPi$ and $ \vpfg \bigr\rvert_R $ by the unique classical evolution of momenta $ \vp( \vRi, \vPi, t ) $%
\footnote{%
	This is a necessity rather than mere inaccuracy, since otherwise we would loose any imprint of the restriction to symmetric families and end up with the standard TWA~\eref{eq:TWOmega}.%
}%
.
Instead, we find that $\vpfg\bigr\rvert_R$ gets replaced by a version of $ \vp( \vRi, \vPi, t ) $ that is appropriately adapted to the approximate dynamics close to ${\SP}$, such that it, too, reflects the behavior of symmetric families only.
Note that the Dirac-delta function in\eq\eref{eq:OmegatgsymWOPi} uniquely determines the (symmetric) family $\gamma$ for given initial conditions $ ( \vRi, \vPi ) $.
This allows us to evaluate $\vpig$ and $\vpfg$ as functions of $ \vRi, \vPi $ and $t$ only. 
For the final momentum in\eq\eref{eq:OmegatgsymWOPi} we find the momentum analogue
\begin{eqnarray}
\label{eq:pfgparaNoshift}
	\fl
	\vpfgpara \!\left( \vqsap( \vRi, \vPi, t ), \vRi, t \right)
		= \vppara ( \vRin, \vPin, t ) + \Ord(\hbar) \psep, \\
\label{eq:pfgperpNoshift}
	\fl
	\vpfgperp \!\left( \vqsap( \vRi, \vPi, t ), \vRi, t \right)
		= \sum_{\lambda=S+1}^{L}
				\frac{\partial \vpperp}{\partial R^\rmi_\lambda}
			\biggr\rvert_{\SP} R^\rmi_\lambda
		+ \sum_{\lambda=S+1}^{L}
				\frac{\partial \vpperp}{\partial P^\rmi_\lambda}
			\biggr\rvert_{\SP} P^\rmi_\lambda
		+ \Ord(\hbar) \psep, 
\end{eqnarray}
of the linear approximation $\vqsap$ [see\eq\eref{eq:qapproxlin}]
up to corrections of order $\Ord (\hbar)$ for \qp-symmetric uncertainty~\eref{eq:perpshbar}, while the initial momentum becomes
\begin{equation}
\label{eq:pigNoshift}
	\vpig \!\left( \vqsap( \vRi, \vPi, t ), \vRi, t \right)
		= \vPi + \Ord(\hbar) \psep. \\
\end{equation}
What makes the statements~\eref{eq:pfgparaNoshift} and~\eref{eq:pfgperpNoshift} non-trivial is that they are significantly different from a mere expansion of the full $ \vp( \vRi,  \vPi, t ) $ in small $ \hbar $.
In systems involving chaotic dynamics, derivatives of $ \vp $ with respect to initial conditions \textit{cannot} be considered as $\Ord (1)$ because of exponential sensitivity to initial conditions.
This difference becomes conceivable at Ehrenfest time scales $ t = \tau \tE $, $\tau = \Ord (1) $.
For instance, generically one has $ \partial p_{\lambda} / \partial R^\rmi_{\lambda^\prime} \sim \rme^{ \lL t } \sim \hbar^{-\tau} \gg \Ord (1) $, where $\lL$ is the (classical) Lyapunov exponent.
For a detailed derivation of\eqs\eref{eq:pfgparaNoshift}--\eref{eq:pigNoshift}, see \ref{app:ApproxSubMomenta}.

In the following we will neglect the $ \Ord(\hbar) $-corrections to the function arguments of $ \Wi $ and $ \Ow $.
This is allowed if the latter are sufficiently well-behaved in the sense that they do not exhibit peaks that exceed certain upper bounds on their sharpness.
For instance, an overall \textit{sufficient} condition is the absence of peaks that are sharper than $\Ord(\hbar^{2/3})$ in any phase-space coordinate when $\hbar \to 0$ (see \ref{app:ApproxSubMomenta}).
This condition is not necessary and can be further relaxed, as shown in \ref{app:swapandshift}, thereby yielding the same result as the one that we will derive here by simply ignoring the small corrections of order $ \Ord(\hbar) $.

\subsection{Final result}

Before continuing we simplify the notation by subsuming position and momentum coordinates into single phase-space variables 
\begin{equation} \label{eq:defX}
	\vX \equiv ( \vR, \vP ) 
\end{equation}
of $ 2 L $ components and similarly
\begin{eqnarray}
	\vX_\para \equiv ( \vR_\para, \vP_\para ) \psep, \label{eq:defXpara}\\
	\vX_\perp \equiv ( \vR_\perp, \vP_\perp ) \label{eq:defXperp}
\end{eqnarray}
to solely address the $ 2 S $ parallel or $ 2 (L-S) $ perpendicular components, respectively.
We also adopt the notation $ \vX_0 = ( \vR_0, \vP_0 ) $ for a given point $ \vX = ( \vR, \vP ) $ in the full phase space to refer to its projected version in $ {\SP} $.

We separate $\Wi$ and $\Ow$ according to
\begin{eqnarray}
\label{eq:Wpara}
	W_\para(\vX_\para) = \int \!\rmd \, \vX_\perp \Wi(\vX) \psep,
		\qquad
		\Wi(\vX) = W_\para( \vX_\para ) \, W_\perp( \vX_\perp ; \vX_\para )
		\psep,\\
\label{eq:Opara}
	\Omega_\para(\vX_\para) = \int \!\rmd \, \vX_\perp \Ow(\vX) \psep,
		\qquad
		\Ow(\vX) = \Omega_\para( \vX_\para ) \, \Omega_\perp( \vX_\perp ; \vX_\para )
		\psep. 
\end{eqnarray}
into marginal distributions $W_\para$ and $\Omega_\para$ in ${\SP}$ and the distributions $W_\perp$ and $\Omega_\perp$ that encode the localization perpendicular to it.
Using the linearized dynamics in $ \vq $ and $ \vp $, according to\eqs\eref{eq:qapproxlin} and~\eref{eq:pfgparaNoshift}, and neglecting the $ \Ord(\hbar) $-corrections, gives
\begin{equation} \label{eq:Omegatsymparaperp}
	\eqalign{
		\fl
			\langle \hat{\Omega} \rangle_t^{\rm sym} \simeq
					\int \!\rmd \vXipara \;
						W_\para\!\left( \vXipara \right)
						\, \Omega_\para\!\left( \vx_\para( \vXin, t ) \right)
				\\ \times
					\int \!\rmd \vXiperp \;
						W_\perp\!\left( \vXiperp ; \vXipara \right)
						\, \Omega_\perp\!\left(  \Mstab \vXiperp ; \vx_\para( \vXin, t ) \right) \psep,
	}
\end{equation}
where $ \vx( \vXi, t ) $ is now the unique time evolution in phase space with initial condition $ \vx( \vXi, 0 ) = \vXi $, or in terms of position and momentum,
\begin{equation}
	\vx( \vXi, t ) = \vx\!\left( ( \vRi, \vPi ), t  \right)
		= \biggl( \begin{array}{c}
				\vq \!\left( \vRi, \vPi, t  \right) \\
				\vp \!\left( \vRi, \vPi, t  \right)
			\end{array} \biggr) \psep,
\end{equation}
and $ \Mstab = \Mstab( \vXin ) $ is the $2(L-S) \times 2(L-S)$ stability matrix of this evolution around $ {\SP} $,
\begin{equation} \label{eq:Mstab}
	\left( \Mstab \right)_{\xi, \xi^\prime}
		= \left.
				\frac{\partial x_{\perp,\xi}( \vXi, t )}{\partial X^\rmi_{\perp, \xi^\prime}}
			\right\rvert_{ \vXi = \vXin }
		\qquad \xi, \xi^\prime \in \{ 1, \ldots, 2(L-S) \} \psep,
\end{equation}
or, in terms of position and momentum coordinates,
\begin{equation} \label{eq:MStab}
	\Mstab = \left(\begin{array}{@{}c|c@{}}
			\biggl( \dfrac{\partial q_\lambda}{\partial R^\rmi_{\lambda^\prime}}
							\biggr\rvert_{\SP} \biggr)_{\lambda, \lambda^\prime}
			&
			\biggl( \dfrac{\partial q_\lambda}{\partial P^\rmi_{\lambda^\prime}}
										\biggr\rvert_{\SP} \biggr)_{\lambda, \lambda^\prime}
			\Bstrut{3ex}\\
			\hline
			\biggl( \dfrac{\partial p_\lambda}{\partial R^\rmi_{\lambda^\prime}}
										\biggr\rvert_{\SP} \biggr)_{\lambda, \lambda^\prime}
			&
			\biggl( \dfrac{\partial p_\lambda}{\partial P^\rmi_{\lambda^\prime}}
										\biggr\rvert_{\SP} \biggr)_{\lambda, \lambda^\prime}
			\Tstrut{4ex}
		\end{array}\right)
		\psep.
\end{equation}

A significant increase in formal simplicity and performance (in view of numerical implementations) is gained when $\Wi$ and $\Ow$ are approximated by normal distributions
\begin{equation} \label{eq:WOGauss0}
	\eqalign{
		\Wi(\vXi) \simeq \Wi(\vXin)
			\exp \bigl( - \case{1}{\hbar} (\vXiperp)^\rmT A^\rmi \vXiperp \bigr) \psep, \\
		\Ow(\vXf) \simeq \Ow(\vXfn)
			\exp \bigl( - \case{1}{\hbar} (\vXfperp)^\rmT A^\rmf \vXfperp \bigr)
	}
\end{equation}
in the perpendicular components, where $ A^\alpha = A^\alpha(\vX^\alpha_\para) = (A^\alpha)^\rmT $ are symmetric $ 2 (L-S) \times 2 (L-S) $ matrices encoding the local multivariate localization around $ {\SP} $.
They can be extracted by matching the covariances or second derivatives with respect to the perpendicular components of a given $\Wi$ and $\Ow$ (see \ref{app:GaussApprox}).
The integral over $ \vXiperp $ in \eref{eq:Omegatsymparaperp} can then be performed analytically, yielding the final result
\begin{equation} \label{eq:OmegatsymDet}
	\langle \hat{\Omega} \rangle_t^{\rm sym} \simeq
		( \pi \hbar )^{L-S}
		\int \!\rmd \vXipara \;
			\frac{\Wi\!\left( \vXin \right)
						\, \Ow\!\left( \vx( \vXin, t ) \right)}
				{\sqrt{\det\!\left( A^\rmi + \Mstab^\rmT A^\rmf \Mstab \right)}}
				\psep.
\end{equation}
Note that the obtained results~\eref{eq:Omegatsymparaperp} and~\eref{eq:OmegatsymDet} are invariant with respect to canonical transformations in the perpendicular phase-space coordinates (see \ref{app:GaussApprox}).

The significance of\eq\eref{eq:OmegatsymDet} is that we are left with a TWA-like \textit{sampling problem} within the symmetric subspace to determine the separate contribution of symmetric trajectory families, where only the unique classical time evolution of initial values inside (and infinitely close to) the symmetric subspace enters.
This enables a numerical implementation in form of a Monte-Carlo simulation, in an analogous manner to the standard TWA, by sampling $\Wi$ within the symmetric subspace ${\SP}$ and classically propagating each sample point $\vXi_0 = (\vXipara, \veknull)$ plus slightly displaced versions in all $2(L-S)$ perpendicular phase-space directions to numerically compute $\Mstab$.

\section{Application in many-body scenarios}
\label{sec:applicMB}
Here we focus on the application to transition probabilities between two (pure) coherent states---a situation that enjoys much attention especially in the vast field of many-body physics with ultracold bosonic atoms where coherent states are commonly used descriptions for Bose-Einstein condensates\cite{Lieb2007}.
In particular, we investigate return probabilities where the final and initial state are equal.
This is particularly interesting for the investigation of (dynamical) localization effects \cite{Anderson1958,Fishman1982,Shepelyansky1983,Bohigas1993,%
Altshuler1997,Gornyi2005,Basko2006,Oganesyan2007}, modal echo \cite{Altshuler1994,Weaver1994,Weaver2000MEreverb}, many-body coherent backscattering \cite{Engl2014,Schlagheck2017,Engl2018SpinEcho}, (dynamical) tunneling \cite{Davis1981,Bohigas1993,Tomsovic1994} in mean-field space \cite{Hensinger2001,Vanhaele2021}, or quantum many-body scarring \cite{Bernien2017,Turner2018,Zhao2020,Serbyn2021,Hummel2022unpub}, for which the associated characteristic enhancement signatures in the return probability can be discriminated from purely symmetry-related enhancement by our method.

\subsection{Transition probability between coherent states}
\label{sec:transprobcs}
We study the transition probability from a state $ \hat{\rho}_\rmi $ to a state $ \hat{\rho}_\rmf $ in time $ t $ by identifying
$ \hat{\Omega} = \hat{\rho}_\rmf $, giving $ \Ow(\vq, \vp ) = (2 \pi \hbar )^L \Wf( \vq, \vp ) $ with $ \Wf( \vq, \vp ) $ the final state Wigner distribution.
We furthermore consider coherent states $ \hat{\rho}_\alpha = \ket{\phi_{\vz^\alpha}} \bra{\phi_{\vz^\alpha}} $, centered around (quadrature) phase-space coordinates $ \vz^\rmi = ( \vq^\rmi, \vp^\rmi ) $ and $ \vz^\rmf = ( \vq^\rmf, \vp^\rmf ) $, respectively, both located on the symmetric subspace ${\SP}$. 
In symmetry-oriented phase-space coordinates [see \sref{sec:coordinates} and\eqs\eref{eq:defX}--\eref{eq:defXperp}] this means $ \vz^\rmi_\perp = \vz^\rmf_\perp = \veknull $.
The Wigner distributions are then given by Gaussian distributions~\eref{eq:WOGauss0} with symmetric shape $A^\rmi = A^\rmf = \mathbb{I}$:
\begin{eqnarray} \label{eq:Wcoh}
	\Wi(\vXi) =
		(\pi \hbar)^{-L} \exp \!\left[
				- \case{1}{\hbar} \left( \vXipara - \vz^\rmi_\para \right)^2
				- \case{1}{\hbar} \left( \vXiperp \right)^2
			\right] \psep,
	\\
	\Ow(\vXi) =
		2^{L} \exp \!\left[
				- \case{1}{\hbar} \left( \vXfpara - \vz^\rmf_\para \right)^2
				- \case{1}{\hbar} \left( \vXfperp \right)^2
			\right] \psep.
\end{eqnarray}
We choose here units such that the coherent states are those of a harmonic oscillator with unit mass and frequency parameters.
The contribution $P_{\mathrm{sym}}$ to the transition probability
\begin{equation}
	P(\vz^\rmf, \vz^\rmi; t) = \tr \!\left[ \hat{\rho}_\rmi(t) \hat{\rho}_\rmf \right]
\end{equation}
that arises solely due to symmetric trajectory families is then given by
\begin{equation} \label{eq:Psymcoh}
	\fl
	P_{\mathrm{sym}}(\vz^\rmf, \vz^\rmi; t)
		\simeq \frac{2^L}{(\pi \hbar)^S} \int \!\rmd^{2S} \vXipara \;
			\frac{\exp \!\left( - \frac{1}{\hbar} \Bigl[
						 ( \vXipara - \vz^\rmi_\para )^2
							+ \bigl( \vx_\para(\vXin, t) - \vz^\rmf_\para \bigr)^2
						\Bigr] \right)}
			{\sqrt{\det\!\left( 1 + \Mstab^\rmT  \Mstab \right)}} \psep.
\end{equation}
This is a TWA-like sampling problem within the symmetric manifold $ {\SP} $, where each classically evolved sample within $ {\SP} $ is weighted by a factor that reflects the stability in the directions pointing out of $ {\SP} $.

\subsection{Bose-Hubbard systems}
\label{sec:BHsystems}
The many-body models we specifically consider here for applying the augmented TWA are Bose-Hubbard systems whose degrees of freedom are represented by a set of discretized field operators $\hat\psi_l$ with $l=1,\ldots,L$, that fulfil bosonic commutation relations
\begin{eqnarray} \label{eq:psicommutators}
	\eqalign{
		\bigl[ \hat\psi_l , \hat\psi_{l^\prime} \bigr] = 0 \psep,\\
		\bigl[ \hat\psi_l , \hat\psi^\dagger_{l^\prime} \bigr] = \delta_{l l^\prime} \psep.
	}
\end{eqnarray}
The operators $\hat\psi_l$ ($\hat\psi^\dagger_l$) are understood as annihilators (creators) of bosonic quanta at the lattice site that is indexed by $l$, with local occupancy $ \hat n_l = \hat\psi^\dagger_l \hat\psi_l$ and the total number operator $ \hat N = \sum_l \hat n_l$.
We consider lattice systems governed by Hamiltonians of the form
\begin{equation} \label{eq:HBH}
	\hat{H} = \sum_{l} \epsilon_l \hat\psi^\dagger_l \hat\psi_l - \sum_{(l, l^\prime)} J_{l,l^\prime} \hat\psi^\dagger_{l^\prime} \hat\psi_l + \frac{1}{2} \sum_{l} U_l \hat\psi^\dagger_l \hat\psi^\dagger_l \hat\psi_l \hat\psi_l \,,
\end{equation}
where $\epsilon_l$ is the onsite energy of a single particle on site $l$, $J_{l,l^\prime} = J_{l^\prime,l}$ is the hopping strength along an available bond $(l,l^\prime)$, and $U_l$ is the coupling that controls the strength of interactions between particles on the same site $l$.

Without any interaction and hopping, the quadratic Hamiltonian $ \hat H_0 = \sum_l \epsilon_l \hat n_l $ is in full analogy with a set of uncoupled harmonic oscillators of frequencies $\omega_l = \epsilon_l / \hbar$ (apart from the constant zero point energy shift $\sum_l \hbar \omega_l / 2$).
One can identify Hermitian \textit{quadrature} operators $ \hat q_l, \hat p_l $, defined by
\begin{equation} \label{eq:psiqp}
	\hat\psi_l = \frac{1}{\sqrt{2}} \bigl( \hat q_l + \rmi \hat p_l \bigr) \psep,
\end{equation}
which would be related to the positions $\hat Q_l$ and momenta $\hat P_l$ of harmonic oscillators characterized by masses $m_l$ and frequencies $\omega_l$ according to
\begin{eqnarray} \label{eq:quadratures}
	\eqalign{
		\hat q_l = \sqrt{\frac{m_l \omega_l}{\hbar}} \hat Q_l\psep,\\
		\hat p_l = \sqrt{\frac{1}{\hbar m_l \omega_l}} \hat P_l \psep.
	}
\end{eqnarray}
The quadratures are dimensionless scaled versions of position and momentum, fulfilling the commutation relations
\begin{eqnarray} \label{eq:qpcommutators}
	\eqalign{
		\bigl[ \hat q_l, \hat q_{l^\prime} \bigr] = \bigl[ \hat p_l, \hat p_{l^\prime} \bigr] = 0 \psep,\\
		\bigl[ \hat q_l, \hat p_{l^\prime} \bigr] = \rmi \delta_{l l^\prime} \psep.
	}
\end{eqnarray}
The correspondence with the commutation relations of the actual position and momentum operators $\hat{Q}_l$ and $\hat{P}_l$ in\eq\eref{eq:quadratures} can be made exact by setting units to $\hbar = 1$ and $m_l = 1 / \omega_l$ for all $l$.

The role of the parameter $\hbar$ and its absence in the relation~\eref{eq:qpcommutators} might at first glance be contradictory to considering small $\hbar$, or formally $\hbar \to 0$, as the regime of quantum-to-classical correspondence, which is key to the TWA and its augmented version derived here.
However, the parameter to be small is not $\hbar$ itself (which would be a meaningless statement anyway, as it is not a dimensionless constant).
Instead, $\hbar$ \textit{compared} to the actions of the classical counterpart of the quantum system has to be small.
For a harmonic oscillator this means large excitation $n \gg 1$.
Likewise, instead of the harmonic oscillator's eigenstates one can consider the coherent states
\begin{equation}
	\ket{ \vek{z} } = \rme^{ ( \vek{z} \hat\psi^\dagger - \vek{z}^\ast \hat\psi ) /\sqrt{2} } \ket{ 0 } \psep,
\end{equation}
where $\ket{ 0 }$ is the vacuum state, i.e., the oscillator's ground state with $\hat\psi \ket{ 0 } = 0$.
They are the most classical states as they minimize the uncertainties of position and momentum simultaneously.
For this reason they are well-suited to describe coherent light and other macroscopic quantum states with well-defined phase and amplitude, realized, among others, with atomic Bose-Einstein condensates \cite{Lieb2007}.
Furthermore, their centroid $ \vek{z} $ follows the classical equations of motion without dispersing under action of the time evolution operator $\hat U (t) = \rme^{-\rmi \hat H t / \hbar}$.
For the harmonic oscillator this holds (anomalously) even down to arbitrarily small oscillations around the fix point.
But the quantum fluctuations in $ \ket{ \vek{z} } $ only become negligible in comparison to the motion of mean position and momentum (given by $\vek{z}$) when $ | \vek{z} | \gg 1 $, or equivalently, when the action $\int \rmd \vq \cdot \vp$ of the corresponding classical orbit becomes large compared to $\hbar$.
Translated into the many-body context via\eq\eref{eq:psiqp}, this means that quantum-to-classical correspondence is established for overall large occupancies $n_l$.
Correspondingly, one often considers the classical counterpart of a bosonic many-body quantum system to be reached in the limit $ N \to \infty $ (or more precisely $ N / L \to \infty $), such that the individual occupancies $n_l$ are large on average.
For this reason, the two notions of the mean-field limit and the formally classical limit of bosonic systems can be used synonymously.

To make this formally explicit, we introduce the rescaled version
\begin{equation} \label{eq:qptilde}
	\eqalign{
		\hat{\tilde{q}}_l = \sqrt{\tilde{\hbar}} \hat{q}_l \psep, \\
		\hat{\tilde{p}}_l = \sqrt{\tilde{\hbar}} \hat{p}_l
	}
\end{equation}
of the quadrature operators~\eref{eq:quadratures}, obeying commutation relations
\begin{equation} \label{eq:qptildecommutators}
	\eqalign{
		\bigl[ \hat{\tilde{q}}_l, \hat{\tilde{q}}_{l^\prime} \bigr]
			= \bigl[ \hat{\tilde{p}}_l, \hat{\tilde{p}}_{l^\prime} \bigr] = 0 \psep,\\
		\bigl[ \hat{\tilde{q}}_l, \hat{\tilde{p}}_{l^\prime} \bigr]
			= \rmi \tilde{\hbar} \delta_{l l^\prime} \psep,
	}
\end{equation}
where
\begin{equation} \label{eq:hbareff}
	\tilde{\hbar} = \frac{L}{N}
\end{equation}
serves as an effective Planck's quantum of action.
Here, $N$ is the average total number of particles participating in the process under consideration, such that $\hat{\tilde{q}}_l$ and $\hat{\tilde{p}}_l$ can be considered as objects of $\Ord(1)$.

The classical Hamiltonian that parallels the quantum one is then obtained by replacing the operators~\eref{eq:qptilde} by classical variables according to
\begin{equation}
	\eqalign{
		\hat{\tilde{q}}_l \mapsto \tilde{q}_l \psep,\\
		\hat{\tilde{p}}_l \mapsto \tilde{p}_l
	}
\end{equation}
with canonical Poisson brackets
\begin{equation}
	\eqalign{
		\{ \tilde{q}_l , \tilde{q}_{l^\prime} \} = \{ \tilde{p}_l , \tilde{p}_{l^\prime} \} = 0 \psep, \\
		\{ \tilde{q}_l , \tilde{p}_{l^\prime} \} = \frac{\tilde{\hbar}}{\hbar} \delta_{l,l^\prime} \psep.
	}
\end{equation}
This replacement is subject to ordering issues.
We follow the convention of using totally symmetric, i.e., Weyl ordering of operators as their classical counterparts.
After symmetric ordering of the Hamiltonian~\eref{eq:HBH} one obtains the ``classical'' mean-field limit
\begin{equation} \label{eq:Hcl}
	\fl
	H_{\mathrm{cl}} = \sum_l ( \epsilon_l - U_l ) \psi^\ast_l \psi_l - \sum_{(l, l^\prime)} J_{l,l^\prime} \psi^\ast_{l^\prime} \psi_l + \frac{1}{2} \sum_l U_l \psi^\ast_l \psi^\ast_l \psi_l \psi_l + \sum_l \Bigl( \frac{\epsilon_l}{2}  + \frac{U_l}{4}  \Bigr) 
\end{equation}
with Hamilton's equations of motion given by the discrete Gross-Pitaevskii equation
\begin{equation} \label{eq:GPE}
	\rmi \hbar \dot\psi_l = \epsilon_l \psi_l - \sum_{l^\prime} J_{l^\prime, l} \psi_{l^\prime} + U_l (\psi^\ast_l \psi_l - 1) \psi_l \psep,
\end{equation}
with the identification
\begin{equation} \label{eq:psiqptilde}
	\eqalign{
		\psi_l  = \frac{1}{\sqrt{2\tilde{\hbar}}} ( \tilde{q}_l + \rmi \tilde{p}_l ) \psep, \\
		\psi^\ast_l  = \frac{1}{\sqrt{2\tilde{\hbar}}} ( \tilde{q}_l - \rmi \tilde{p}_l ) \psep.
	}
\end{equation}

In contrast to semiclassical approximations in first quantized systems, $\hbar$ is here not considered as a small parameter.
As mentioned above, the semiclassical regime is instead indicated by overall large occupancies.
To make formal contact with the derivations---here presented in the form fitting to first quantized systems---of the TWA~\eref{eq:TWOmega} and its augmented version~\eref{eq:OmegatsymDet} for discrete symmetries, we set units to $\hbar=1$, meaning, $\epsilon_l$, $J_{l,l^\prime}$, and $U_l$ are given as frequencies instead of energies.
Additionally, we assume that the \textit{rescaled} interaction $U_l N/L$, as a physically more meaningful quantity than the bare $U_l$, has a magnitude roughly comparable with the hopping strengths $J_{l,l^\prime}$.
This way, the Hamiltonian~\eref{eq:Hcl}---and hence the corresponding Lagrangian and Hamilton's principal function~\eref{eq:Rgamma}---, can be viewed as a quantity that scales linearly with $N/L = 1/\tilde{\hbar}$.
Formally, this allows us to replace
\begin{equation}
	\frac{1}{\hbar} R_\gamma(\vq^\rmf, \vq^\rmi, t) = \frac{1}{\tilde{\hbar}} \tilde{R}_\gamma(\tilde{\vq}^\rmf, \tilde{\vq}^\rmi, t)
\end{equation}
in the expressions~\eref{eq:K} and~\eref{eq:Ag} for the van Vleck-Gutzwiller propagator that lays the foundation for the overall approach.
The small parameter that controls the semiclassical limit is then $\tilde \hbar$ while $\tilde{R}_\gamma$ is of order 1.
In analogy to first quantized systems, the principal function $\tilde{R}_\gamma$ can thus be viewed as a purely \textit{classical} object, given by
\begin{equation}
	\tilde{R}_\gamma( \tilde{\vq}^\rmf, \tilde{\vq}^\rmi, t )
		= \int \rmd t \left( \sum_l \frac{\rmd \tilde{q}_{\gamma,l}}{\rmd t} \tilde{p}_{\gamma, l}
		- \tilde{\hbar} H_{\mathrm{cl}}( \tilde{\vq}_\gamma, \tilde{\vp}_\gamma ) \right) \psep,
\end{equation}
evaluated along the classical solutions $\tilde{\vq}_\gamma(t), \tilde{\vp}_\gamma(t)$.
They are obtained as the solutions $\vek{\psi}(t)$ of\eq\eref{eq:GPE} that fulfil the boundary conditions $ \sqrt{2\tilde{\hbar}} \, \Re \{\psi_l(0)\} = \tilde{q}_l^\rmi$ and $\sqrt{2\tilde{\hbar}} \, \Re \{\psi_l(t)\} = \tilde{q}_l^\rmf$, where we use the identification~\eref{eq:psiqptilde}.
The generally multiple solutions are enumerated by the index $\gamma$.

Thus, formal equivalence with the treatment of first quantized systems, as presented in sections~\ref{sec:reviewTWA}--\ref{sec:augTWA} and~\ref{sec:transprobcs}, is established by understanding $\vq$ and $\vp$ as (eigenvalues of) the \textit{rescaled} quadratures~\eref{eq:qptilde} and $\hbar$ as the \textit{effective} quantity $\tilde{\hbar} = L/N$~\eref{eq:hbareff}.
Note that these replacements hold as well for the Wigner transforms~\eref{eq:Wi}, \eref{eq:Ow}, and~\eref{eq:Wcoh} and thus for the estimation of corrections due to finite separations from ${\SP}$ [see~\eref{eq:perpshbar}, \eref{eq:pfgparaNoshift}--\eref{eq:pigNoshift}, as well as \ref{app:asym}, \ref{app:ApproxSubMomenta}, and \ref{app:swapandshift}].

\section{Constructive many-body interference in Bose-Hubbard plaquettes}
\label{sec:BHapplication}
We consider spatially homogeneous Bose-Hubbard chains~\eref{eq:HBH} with constant $\epsilon_l = 0$, $U_l = U$, and $J_{l,l^\prime} = J$ along bonds that impose a one-dimensional lattice with nearest neighbour hopping, i.e., bonds $(l,l^\prime)$ are available only for $l^\prime = l \pm 1$.
We further establish periodic boundary conditions by understanding the indexes as $l,l^\prime \in \mathbb{Z}/L\mathbb{Z}$.
\Fref{fig:symmetriesL4}a illustrates the Bose-Hubbard ring as an idealization of atoms in an optical lattice.
The symmetries of the periodic Bose-Hubbard chain are given by translations, a parity/inversion operation, and combinations of translation and inversion (or equivalently inversions along different axes).

We choose an even number of sites, in particular $L=4$, that are alternately populated by a condensate, described by a coherent state~\eref{eq:Wcoh} centered about the classical (i.e., mean-field) state
\begin{equation} \label{eq:csalternating}
	( \psi^{(0)}_1, \psi^{(0)}_2, \psi^{(0)}_3, \psi^{(0)}_4 ) = ( \sqrt{n_A} \rme^{\rmi \theta_A}, \sqrt{n_B} \rme^{\rmi \theta_B}, \sqrt{n_A} \rme^{\rmi \theta_A}, \sqrt{n_B} \rme^{\rmi \theta_B} ) \in \mathbb{C}^4\psep,
\end{equation}
with average total number of particles $N = 2 n_A + 2 n_B$ for which we calculate the return or survival probability $P(t)$ after time $t$.
This state is symmetric under two distinct reflections that exchange sites $1\leftrightarrow3$ and $2\leftrightarrow4$, respectively, and the translation by two sites that is equivalent to the combination $(1,2) \leftrightarrow (3,4)$ of the two reflections.
The total number of respected symmetry operations is thus four, including the trivial ``symmetry'' under the identity.

\subsection{The augmented TWA in action}
When sampling initial conditions from the classical phase space in the vicinity of the centroid~\eref{eq:csalternating} one obtains points that belong to trajectory families of either one of the above-mentioned symmetry classes.
\begin{figure}
	\includegraphics[width=0.92\textwidth]{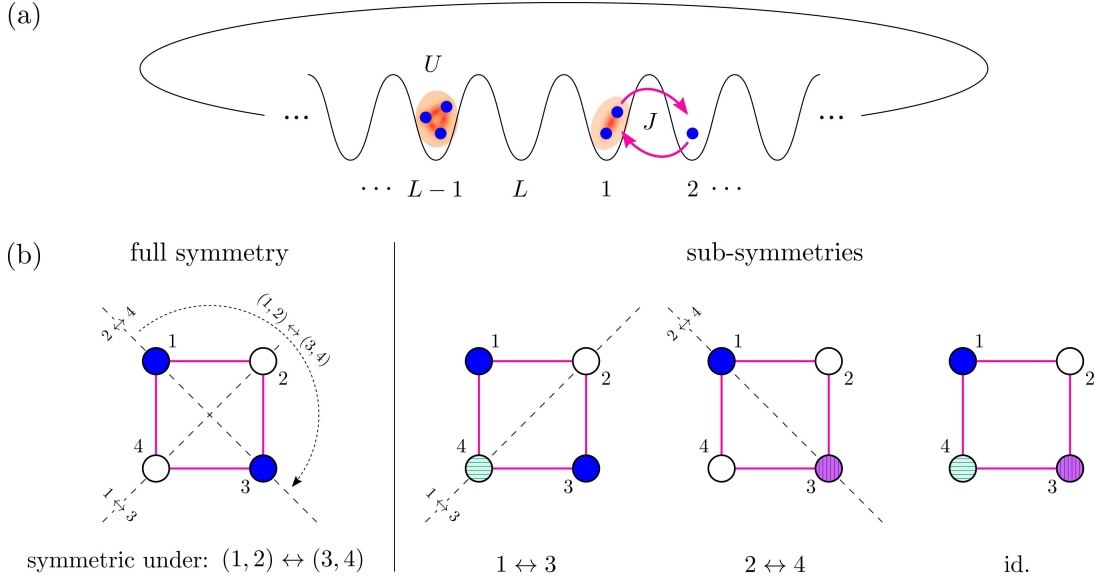}
	\caption{ \label{fig:symmetriesL4}
		(a)The homogeneous $L$-site Bose-Hubbard ring with onsite interaction (strength $U$) and nearest-neighbour hopping (strength $J$) as model of cold atoms in an optical lattice. 
		(b)The possible symmetries of a classical/mean-field state about the coherent-state center~\eref{eq:csalternating} in the case $L=4$.
		Different or equal colors/shadings indicate distinct or equal mean-field values $\psi_l \in \mathbb{C}$, respectively.
	}
\end{figure}
Figure \ref{fig:symmetriesL4}b illustrates these possible symmetries for the case $L=4$.
Our application of the augmented TWA thus involves a Monte-Carlo sampling~\eref{eq:Psymcoh} within each of the corresponding symmetric subspaces ${{\SP}}$.
We may denote these by $\SP^{1 \leftrightarrow 3}$, $\SP^{2 \leftrightarrow 4}$, $\SP^{(1,2) \leftrightarrow (3,4)}$, and $\SP^{\mathrm{id}}$.
For the trivial ``symmetry'', $\SP^{\mathrm{id}}$ is just the full phase space that yields the standard TWA, incorporating trajectory families of all possible symmetry classes, including the fully nonsymmetric ones.

The symmetries are related to each other by inclusion relations, meaning that certain ${\SP}$ are embedded in certain other ${\SP}$.
This hierarchical structure determines the multiplicities associated with the individual contributions.
In the case of $L=4$ and the alternating symmetry of the initial state~\eref{eq:csalternating}, the relevant inclusions are
\begin{eqnarray}
	\SP^{(1,2) \leftrightarrow (3,4)} \subset \SP^{1 \leftrightarrow 3} \psep, \qquad
		\SP^{1 \leftrightarrow 3} \subset \SP^{\mathrm{id}} \psep, \\
	\SP^{(1,2) \leftrightarrow (3,4)} \subset \SP^{2 \leftrightarrow 4} \psep, \qquad
		\SP^{2 \leftrightarrow 4} \subset \SP^{\mathrm{id}} \psep.
\end{eqnarray}
and the combined return probability is given by
\begin{equation} \label{eq:Pcombined}
	P(t) \simeq 4 P_{\mathrm{id}}(t)
		- 2 P_{1 \leftrightarrow 3}(t) - 2 P_{2 \leftrightarrow 4}(t)
		+ P_{(1,2) \leftrightarrow (3,4)}(t) \psep,
\end{equation}
where each $P_{s}(t)$ stands for an augmented TWA sampling~\eref{eq:Psymcoh} in the respective symmetric subspace $\SP^{s}$ and $P_{\mathrm{id}}$ gives the standard TWA as special case of~\eref{eq:Psymcoh} for $S=L$, $\det\!\left( 1 + \Mstab^\rmT  \Mstab \right) = 1$.
In the given case~\eref{eq:Pcombined} the prefactors can be easily understood:
Each fully nonsymmetric family $\gamma$ has four distinct families it can be paired with: $\gamma$ itself and the three families obtained by application of the symmetry operations $1 \leftrightarrow 3$, $2 \leftrightarrow 4$, and $(1,2) \leftrightarrow (3,4)$, hence a degeneracy factor of 4.
This overcounts the contribution from families $\gamma$ that are symmetric either under $1 \leftrightarrow 3$ or under $2 \leftrightarrow 4$.
These should contribute with a degeneracy of 2 corresponding to the pairing with $\gamma$ itself and one reflected counterpart, while in $4 P_{\mathrm{id}}$ they are counted with a factor of 4 instead.
The multiplicity $(-2)$ for $P_{1 \leftrightarrow 3}$ and $P_{2 \leftrightarrow 4}$ corrects this.
Finally, the families that are fully symmetric under $(1,2) \leftrightarrow (3,4)$ only have themselves as partner and should be counted exactly once, whereas they are so far counted with factors $4$, $(-2)$, and $(-2)$ as they participate in all of the three other contributions, resulting in a total of $4-2-2=0$.
They are thus completely gone and have to be added again with the proper multiplicity of $1$.
A corresponding treatment for arbitrary $L$ and arbitrary symmetry of the initial state is possible by analyzing the structure of available symmetry groups and sub-groups.

To implement the augmented TWA for either symmetry we use\eq\eref{eq:Psymcoh}, adapted to the many-body context by replacing $\hbar \mapsto \tilde{\hbar}$ and by using as phase-space variables the scaled quadratures~\eref{eq:psiqptilde}, $\vXipara \mapsto (\vek{\tilde{q}}_\para, \vek{\tilde{p}}_\para)$, transformed into symmetry-oriented coordinates (see below).
Expressed in the original (unscaled) field variables, denoted by $\vekpsi = (\psi_1, \ldots, \psi_L)$, this becomes
\begin{equation} \label{eq:Ps}
	P_{s}\bigl({\vekpsi}^{(0)}; t\bigr) \simeq 2^L \left\langle
		\frac{ \exp \Bigl[ -2 \sum_l \bigl\lvert \psi_l( \vekPsi, t) - \psi_l^{(0)} \bigr\rvert^2 \Bigr] }
			{\sqrt{\det\!\left( 1 + \Mstab^\rmT  \Mstab \right)}}
		\right\rangle_{\vekPsi, s} \,,
\end{equation}
where $\psi_l(\vekPsi,t)$ is the classical evolution of an initial field configuration $\vekPsi = (\Psi_1, \ldots, \Psi_L)$ under the (discrete) Gross-Pitaevskii equation~\eref{eq:GPE} and $\langle \cdot \rangle_{\vekPsi, s}$ means sampling of the initial field $\vekPsi$ in the symmetric subspace $\SP^s$ by means of a probability density function for the parallel components $\vekPhi_\para = ( \Phi_1, \ldots, \Phi_S )$, given by
\begin{equation}
	\fl
	\left\langle f(\vekPsi) \right\rangle_{\vekPsi, s} = \prod_{\sigma = 1}^{S} \int \rmd (\Re \Phi_\sigma) \int \rmd (\Im \Phi_\sigma) \; f(\vekPsi) \, \left( \frac{2}{\pi} \right)^S \!\exp \Bigl[ - 2 \sum_{\sigma=1}^{S} \bigl\lvert \Phi_\sigma - \varphi^{(0)}_\sigma \bigr\rvert^2 \Bigr] \psep.
\end{equation}
The symmetry-specific orthogonal transformations
\begin{equation}
	( \Psi_1 , \ldots, \Psi_L )^\rmT = \mathbb{O}_s \, ( \Phi_1, \ldots, \Phi_S, \Phi_{S+1}, \ldots, \Phi_L )^\rmT
\end{equation}
between original field variables $\vekPsi$ and symmetry-oriented coordinates $\vekPhi$, and anal\-o\-gous\-ly $\vekpsi = \mathbb{O}_s \vekphi$ and $\vekpsi^{(0)} = \mathbb{O}_s \vekphi^{(0)}$, can for example be chosen as
\begin{equation}
	\fl
	\mathbb{O}_s =
		\begin{pmatrix}
			0 & 0 & \frac{1}{\sqrt{2}} & \frac{1}{\sqrt{2}} \\
			1 & 0 & 0 & 0 \\
			0 & 0 & \frac{1}{\sqrt{2}} & - \frac{1}{\sqrt{2}} \\
			0 & 1 & 0 & 0
		\end{pmatrix}, \quad
		\begin{pmatrix}
			1 & 0 & 0 & 0 \\
			0 & 0 & \frac{1}{\sqrt{2}} & \frac{1}{\sqrt{2}} \\
			0 & 1 & 0 & 0 \\
			0 & 0 & \frac{1}{\sqrt{2}} & - \frac{1}{\sqrt{2}}
		\end{pmatrix}, \quad
		\frac{1}{\sqrt{2}}\begin{pmatrix}
			1 & 0 & 1 & 0 \\
			0 & 1 & 0 & 1 \\
			1 & 0 & -1 & 0 \\
			0 & 1 & 0 & -1
		\end{pmatrix}
\end{equation}
for the symmetries $1 \leftrightarrow 3\ (S=3)$, $2 \leftrightarrow 4\ (S=3)$, and $(1,2) \leftrightarrow (3,4)\ (S=2)$, respectively.
For the stability matrix $ \Mstab $ we simply replace $q_\lambda \mapsto \Re \varphi_\lambda$, $p_\lambda \mapsto \Im \varphi_\lambda$, $R^\rmi_\lambda \mapsto \Re \Phi_\lambda$, and $P^\rmi_\lambda \mapsto \Im \Phi_\lambda$ in\eq\eref{eq:MStab}, with $\vekphi = \mathbb{O}_s^\rmT \vekpsi( \mathbb{O}_s \vekPhi, t )$, implementing derivatives numerically by classically propagating slightly displaced initial fields $\vekPsi + \delta \vekPsi$, where we choose separately $\delta \vekPsi \propto \mathbb{O}_s \vek{e}_\lambda$ and $\delta \vekPsi \propto \rmi \mathbb{O}_s \vek{e}_\lambda$ for all $\lambda = S+1,\ldots,L$.

\begin{figure}
	\includegraphics[width=\linewidth]{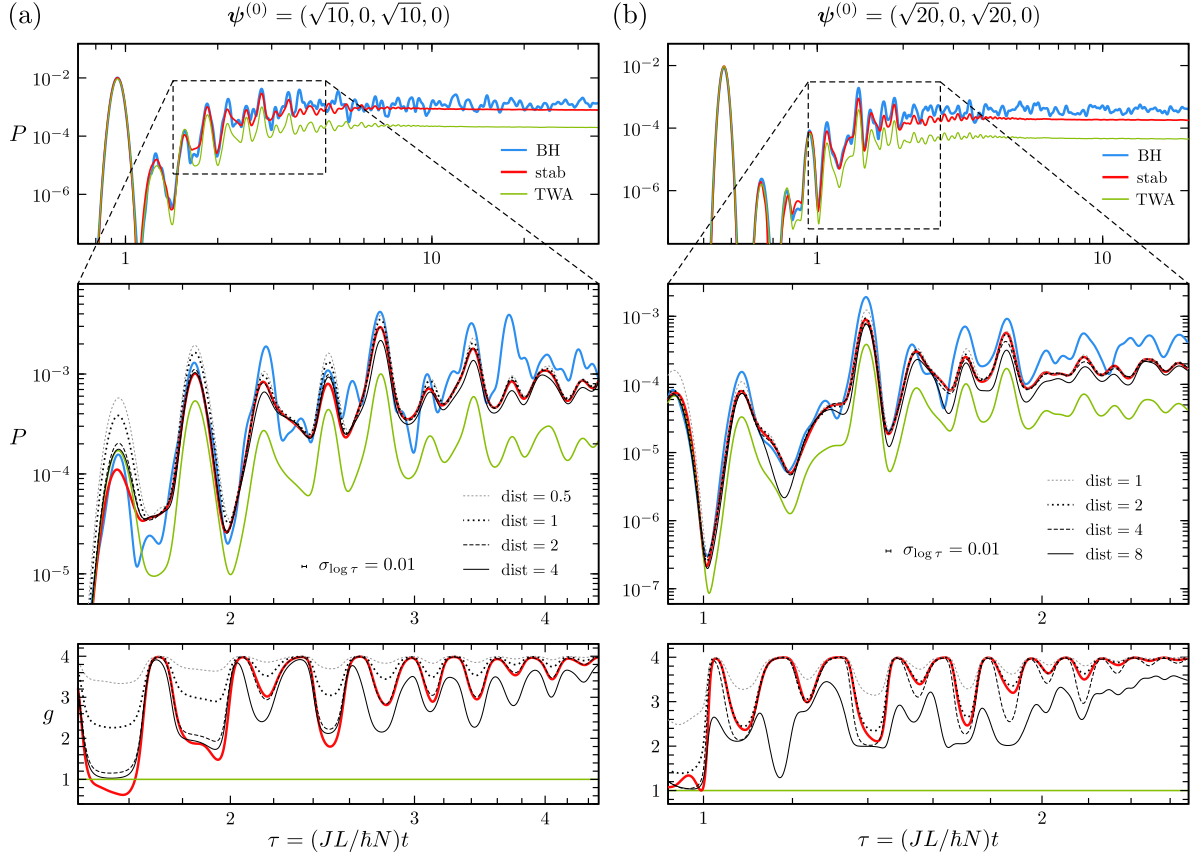}
	\caption{ \label{fig:plot_10_20}
		Return probability $P(t)$ of a coherent state centered at (a) $\vekpsi^{(0)} = ( \sqrt{10}, 0, \sqrt{10}, 0 )$ and (b) $\vekpsi^{(0)} = ( \sqrt{20}, 0, \sqrt{20}, 0 )$ in the homogeneous four-site Bose-Hubbard ring ($UN/JL = 0.5$ fixed) as a function of the scaled time $\tau=(JL/\hbar N)t$.
		Compared is the exact quantum evolution (blue) with the quasiclassical prediction obtained via the conventional TWA (green) and the semiclassical predictions of the augmented TWA (red) and its alternative implementation using the distance criterion \cite{Schlagheck2019}, for four different values of the distance threshold (grey dotted, black dotted, dashed and solid).
		The upper panels show the overall evolution, with the regimes of transient symmetry enhancement enlarged in the center panels.
		Evolution of the respective enhancement factors $g(t) \equiv P(t) / P_{\mathrm{TWA}}(t)$ are shown in the lower panels.
		All data for return probabilities are time-averaged with a width that linearly increases with $\tau$, by applying convolution with a Gaussian kernel in $\log \tau$ with standard deviation $\sigma_{\log\tau}=0.01$ (indicated by bars in the center panels).
	}
\end{figure}

\begin{figure}
	\includegraphics[width=\linewidth]{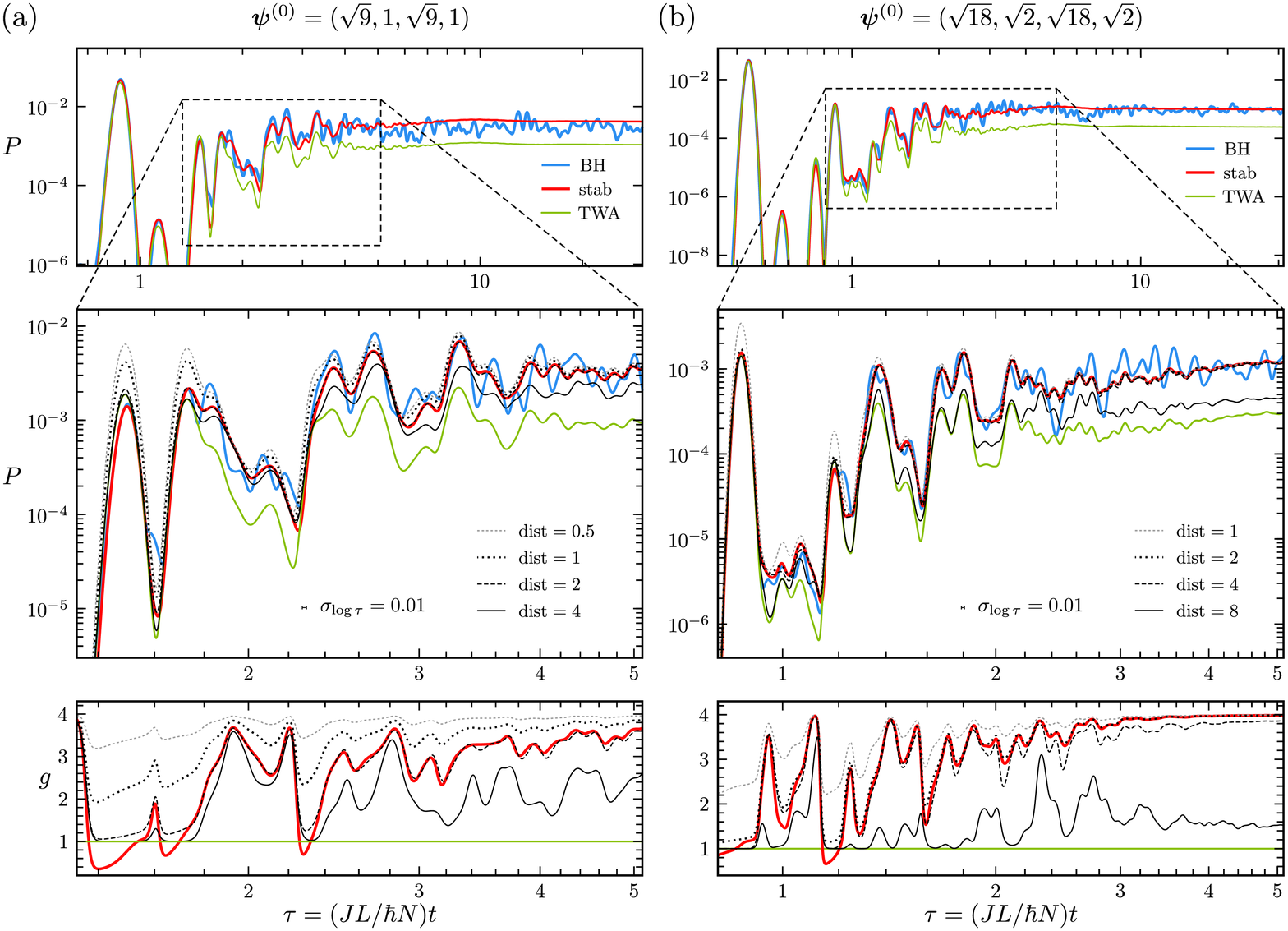}
	\caption{ \label{fig:plot_09_18}
		Return probability $P$ as a function of the scaled time $\tau=(JL/\hbar N)t$ of a coherent state centered at (a) $\vekpsi^{(0)} = ( \sqrt{9}, 1, \sqrt{9}, 1 )$ and (b) $\vekpsi^{(0)} = ( \sqrt{18}, \sqrt{2}, \sqrt{18}, \sqrt{2} )$ with otherwise the same parameters as in \fref{fig:plot_09_18}.
	}
\end{figure}

We calculate the return probability of the coherent state with centroid~\eref{eq:csalternating} after time $t$ for several choices of the average occupancies $n_A, n_B$.
The classical dynamics for a fixed number of particles $\sum_{l=1}^L \lvert \psi_l \rvert^2 \equiv N_\psi = \mathrm{const}$ is governed by the combined parameter $U N_\psi / J L$.
For the coherent state survival probabilities that we consider here, we thus determine the characteristic scale of this parameter by using the \textit{average} number of particles $N = 2 n_A + 2 n_B$ for $N_\psi$.
In particular, we fix $U N / J L = 0.5$ (that has also been used in \cite{Schlagheck2019}) for all shown simulations and plot the results against the scaled time $\tau = (JL/\hbar N) t$.
In \fref{fig:plot_10_20} and \fref{fig:plot_09_18} we show comparisons with converged quantum mechanical calculations confirming that the augmented TWA derived here, implemented as Monte-Carlo simulation, is able to capture the robust constructive interference between symmetry related mean-field solutions, whereas the standard TWA fails.
Especially in the transient regimes of intermediate time scales (center panels) we find very good agreement in all shown cases, whereas the long-time averages of the quantum evolution and the augmented TWA generally differ due to additional quantum effects not captured here.

\subsection{Discussion and comparison to the distance criterion}
Additionally, we compare our method with an earlier used variant of the augmented TWA that uses a heuristic distance criterion \cite{Schlagheck2019}:
Each simulated sample trajectory $\vekpsi(t)$ is thereby classified by computing the norm distances $\bigl(\sum_{l=1}^{L} \lvert \psi_l - (T_{s}\psi)_l \rvert^2 \bigr)^{1/2}$ from its symmetry transformed versions $ T_{s}\vekpsi(t) $ for each symmetry $s$.
The trajectory's family is then estimated to be symmetric with respect to $s$ if the corresponding distance never exceeds a specified threshold $d$ between initialization and measurement, and is assigned the corresponding degeneracy factor $g$.
This criterion was motivated by the generic behaviour of chaotic trajectories in the light of discrete symmetries (see \fref{fig:symnonsymgamma}).
Evidently, one would thereby erroneously count as nonsymmetric those members of a symmetric trajectory family whose starting or end points in phase space lie rather far away from the symmetric subspace.
However, as the Wigner distributions of the initial state and the final observable are both assumed to be tightly localized about the symmetric subspace, those particular trajectories do not significantly contribute to the expectation value under consideration in any case.

Intrinsically, this method is prone to the ambiguity of choosing $d$.
Nevertheless, in certain ranges of parameters and times $t$ the sensitivity of the result to $d$ is less pronounced than in others.
There, we find good agreement with the here-derived augmented TWA, which thereby confirms the validity of both approaches.
Moreover, in all ranges of the considered cases, the augmented TWA developed here coincides very well with the result obtained using the distance criterion when choosing the threshold $d$ appropriately.
In this sense, one can understand the augmented TWA as a means to determine the optimal distance threshold.
This is important because a wrong choice of $d$ can lead to misleading and spurious results.

To elaborate on this, we show in \fref{fig:plot_09_18}a a case where the distance criterion produces results that are very sensitive to the threshold $d$.
Comparison with the augmented TWA shows that choosing $d \simeq 2$ is optimal in reproducing the correct enhancement due to symmetry.
Without this validation, one might be tempted to determine the optimal choice for $d$ by maximizing the agreement with the full quantum mechanical evolution.
While an optimization on intermediate time scales $\tau \lesssim 5$ approximates the prediction of the unambiguous augmented TWA developed here, tuning $d$ towards a match of the long-time saturation (that is in discord with the augmented TWA) gives bad agreement in the transient regime.
Such a strategy effectively turns this method into a fitting procedure, loosing any predictive power.
The so-reached long-time agreement is then a mere coincidence, and the fact that it is not reproduced by the unambiguous augmented TWA strongly indicates that additional effects beyond symmetry enhancement for chaotic motion take place on longer time scales. 

To elaborate a strategy how to determine the distance $d$ in a reasonable manner, not relying on any fitting procedure, we have to account for the two constraints that the choice for $d$ has to respect.
On the one hand, $d$ should be chosen such that it practically encompasses all of the initial wave packet.
This would have to be a value that does not depend on $N$.
A reasonable minimal choice would, e.g., be $d = 2$, through which sample points located within two standard deviations of the Gaussian wave packet of the coherent state are counted as being ``close'' to the symmetry subspace.
Otherwise, choosing $d$ significantly below this value leads to a non-negligible fraction of contributing trajectories that would be falsely classified as nonsymmetric, resulting in an overestimation of the survival probability.
This discrepancy becomes especially evident on shorter time scales, since all families $\gamma$ are symmetric for $t \to 0$ (see \fref{fig:plot_10_20}a and \fref{fig:plot_09_18}a for $\tau \lesssim 2$ and $d \leq 1$).

On the other hand, the distance criterion is devised to discriminate symmetric from nonsymmetric families, which is a task that only involves the \textit{classical} dynamics.
Characteristic distances in the latter scale with $\sqrt{N}$, such that, in order to equally classify two equivalent trajectories of the same system that only differ by a scaling of $N$ (fixing $U N / J L$), one would have to scale $d \propto \sqrt{N}$ as well.
The reasoning that partially resolves this dilemma in fully chaotic systems is that, generically, the nonsymmetric families are exploring the available phase space on a \textit{global} scale, while the symmetric ones never depart further from $\SP$ than their initial or final point (see \fref{fig:symnonsymgamma}).
Under this assumption the discrimination is therefore rather insensitive to the actual value of the threshold $d$, as long as it is large enough to practically contain all points of the initial and final wave packets and smaller than the dimensions of the available phase space.

However, if the assumption of globally chaotic motion is not met, the discrimination can become more sensitive to $d$.
Consider, e.g., islands of regular motion and layers of locally chaotic motion that might be too small in extent to allow for the correct detection of nonsymmetric families for a certain value of $d$.
The corresponding underestimation of symmetry enhancement and corresponding breakdown of the distance criterion becomes evident in \fref{fig:plot_09_18}a,b for $d=4$ and $d=8$, respectively, where most-prominently even the saturation after long times is far off the value of the average long-time return probability to the initial state predicted by the unambiguous augmented TWA~\eref{eq:Pcombined}.
An analysis of the individual contributions for each symmetry to the augmented TWA further strongly suggests that indeed partially regular, stable motion plays a role here.
Since the corresponding analysis yields a similar picture for all relevant symmetries, we focus on the symmetry $1 \leftrightarrow 3$ in the following.

\begin{figure}
	\includegraphics[width=\linewidth]{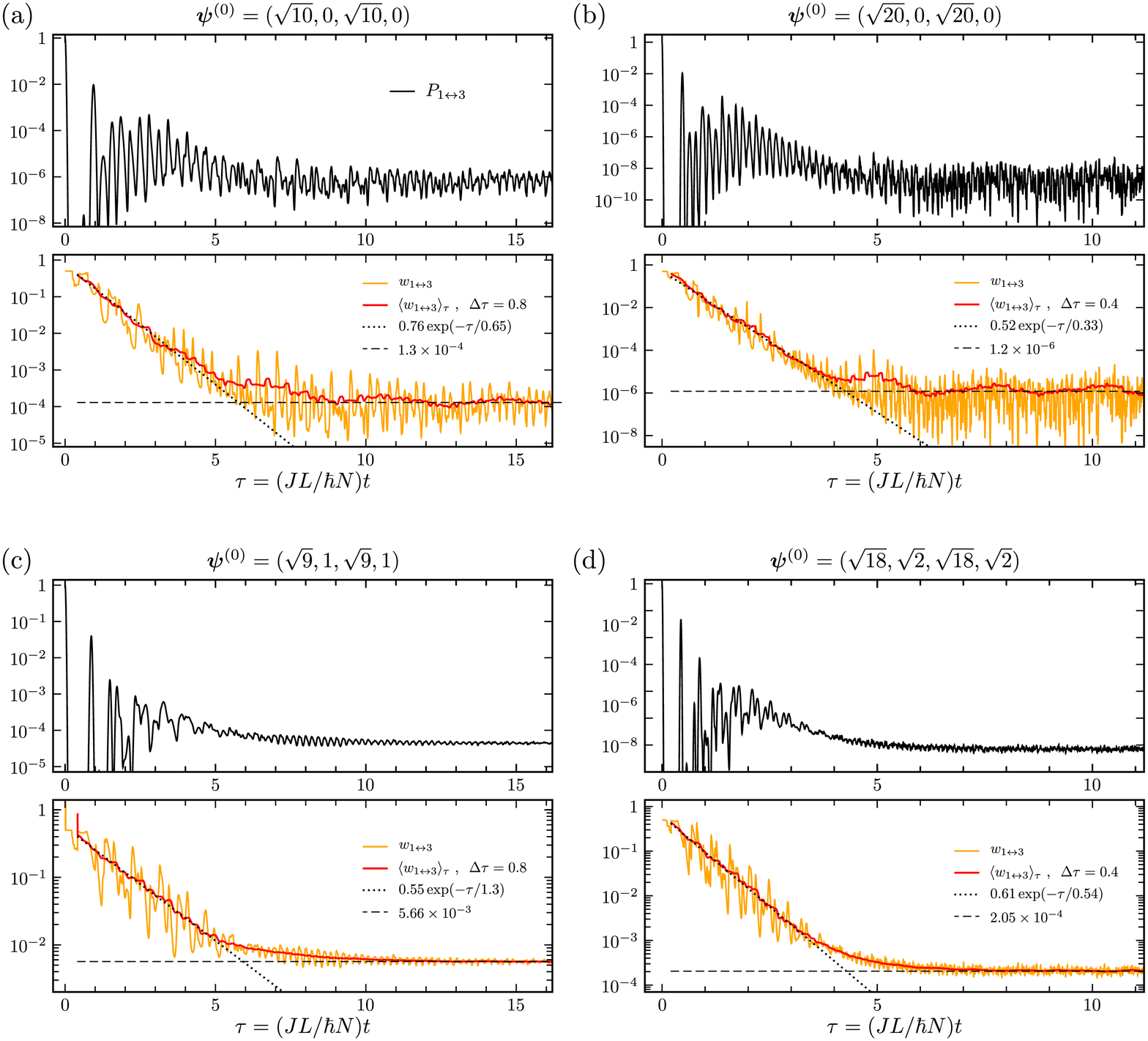}
	\caption{ \label{fig:S4_1}
		Individual contribution $P_{1 \leftrightarrow 3}$ of trajectory families symmetric under exchange of the opposite sites $l=1$ and $l=3$ to the augmented TWA as functions of the scaled time $\tau=(JL/\hbar N)t$ for the same parameters as in (a) \fref{fig:plot_10_20}a, (b) \fref{fig:plot_10_20}b, (c) \fref{fig:plot_09_18}a, and (d) \fref{fig:plot_09_18}b (upper panels, black).
		The corresponding average stability weights $w_{1 \leftrightarrow 3}$ are shown in the lower panels (orange), including a time average over a window of width $\Delta \tau$ (red).
		Fits for the exponential decay and long-time saturation are shown in dotted and dashed, respectively.
	}
\end{figure}

For the same parameters as in \fref{fig:plot_10_20} and \fref{fig:plot_09_18} the individual contribution $P_{1 \leftrightarrow 3}(t)$ associated with the symmetry $1 \leftrightarrow 3$, according to\eq\eref{eq:Ps}, is shown in \fref{fig:S4_1}.
In contrast to the overall symmetry-enhanced probability $P(t)$, equation~\eref{eq:Pcombined}, $P_{1 \leftrightarrow 3}(t)$ decays on intermediate and long time scales.
This is expected due to trajectories within $\SP^{1 \leftrightarrow 3}$ that are unstable, i.e. hyperbolic, in the direction(s) perpendicular to this particular symmetry subspace.
They will contribute with a stability weight $1 / \sqrt{\det\!\left( 1 + \Mstab^\rmT  \Mstab \right)}$ that decays exponentially since at least one eigenvalue of $\Mstab$ grows like $\sim \rme^{\lambda_\mathrm{L} t}$ with $\lambda_\mathrm{L}$ the corresponding Lyapunov exponent.
If only unstable trajectories were involved, $P_s(t)$ would thus decay to $0$ for $t\to\infty$ for all symmetries $s$ but the identity, resulting in a long-time saturation at the maximum degeneracy factor (here $g=4$).
Instead, we find that $P_s(t)$ decays to a finite value because a non-vanishing fraction of the sampled trajectories is stable, i.e., elliptic in all perpendicular directions, contributing with an oscillatory stability weight of constant $\mathcal{O}(1)$ magnitude.
In the four considered cases, this effect is most pronounced for $\vekpsi^{(0)}=(\sqrt{9},1,\sqrt{9},1)$ shown in \fref{fig:S4_1}c.

To address the phenomenon on a more quantitative basis, we additionally plot the average stability weight
\begin{equation} \label{eq:ws}
	w_s(t) \equiv P_s( t ) / P_s^{\mathrm{unweighted}}( t )
		= \Bigl\langle \sqrt{\det\!\left( 1 + \Mstab^\rmT  \Mstab \right)} \Bigr\rangle^{-1}_{P_s}
\end{equation}
by normalizing with the unweighted probability 
\begin{equation}
	P_s^{\mathrm{unweighted}}( t ) \equiv 2^L \Bigl\langle
				\exp \Bigl[ -2 \sum_l \bigl\lvert \psi_l( \vekPsi, t) - \psi_l^{(0)} \bigr\rvert^2 \Bigr]
				\Bigr\rangle_{\vekPsi, s}
\end{equation}
obtained by ignoring the stability weight, i.e., by artificially setting $\det\!\left( 1 + \Mstab^\rmT  \Mstab \right) \mapsto 1$ in \eref{eq:Ps}.
In the average $\langle \cdot \rangle_{P_s}$ of\eq\eref{eq:ws} each trajectory is weighted with the overall contribution to the augmented TWA~\eref{eq:Ps} for the symmetric subspace $\SP^s$, i.e., it accounts for the in-plane classical return probability as well as the sampling of initial conditions $\vekPsi$.
The average stability weights clearly show the exponential decay and the long-time saturation resulting from trajectories that are perpendicularly unstable or stable, respectively.
We find that the influence of (perpendicularly) stable dynamics on the case shown in \fref{fig:S4_1}c is at least one to two orders of magnitude stronger than in the other cases.
This observation is in agreement with the fact that in the scaled phase space, $\vekpsi_{\mathrm{sc}} = \vekpsi / \sqrt{N_\psi}$, the centroid of the two cases in \fref{fig:plot_09_18}, i.e.,  $\vekpsi_{\mathrm{sc}}^{(0)}=(\sqrt{9/20},\sqrt{1/20},\sqrt{9/20},\sqrt{1/20})$ is closer to the stable region located around the stable fix point $(\sqrt{1/4},\sqrt{1/4},\sqrt{1/4},\sqrt{1/4})$ than the centroid $\vekpsi_{\mathrm{sc}}^{(0)}=(\sqrt{1/2},0,\sqrt{1/2},0)$ of the other two cases (\fref{fig:plot_10_20}).
Additionally, the wave packets in the two cases \fref{fig:plot_10_20}b and \fref{fig:plot_09_18}b featuring a larger particle number $N = 40$ are localized more sharply (in scaled phase space) by a factor of $\sqrt{2}$ as compared to the corresponding $N = 20$ cases, such that the stable dynamics located in the tails has less weight.
The overall stronger influence of stable dynamics in the case \fref{fig:S4_1}c may also be indicated by the rather regular oscillations of $P_{1 \leftrightarrow 3}$, observable directly after the exponential decay at $5 \lesssim \tau \lesssim 10$.
Together, this results in a pronounced sensitivity to the choice for the distance threshold $d$ in \fref{fig:plot_09_18}a.

\subsection{Deviation in the long-time saturation}
For long time scales, we generally find a quantum survival probability that disagrees with the augmented TWA-based simulations, levelling off at values that differ significantly.
As we find otherwise full agreement up to intermediate times, where the average degeneracy factor $\langle g \rangle$ starts to saturate, this strongly indicates the existence of additional quantum effects beyond purely symmetry-related constructive interference, which are not captured by our method.
This is further substantiated by the fact that we also find enhancement that exceeds the theoretical maximum $\langle g \rangle = 4$ for constructive interference of symmetry partners (see \fref{fig:plot_10_20}a).
We do not attempt here to explain these discrepancies at long time scales.
But we see that the augmented TWA, which is a tool to unambiguously determine the effect of constructive symmetry-related interference, can further serve as a method to uncover the existence and the magnitude of additional quantum effects related to dynamical features of the system.

The individual contributions from sub-symmetries that are the ingredients to the augmented TWA can give additional information in order to reveal possible candidates for these discrepancies.
The influence of stable or mixed dynamics might lead to an overestimation of the long-time survival probability by the augmented TWA, due to quantum leakage out of classically stable regions, while the mean-field dynamics is trapped.
This tendency seems to be confirmed in the case shown in \fref{fig:plot_09_18}a, featuring an increased influence of stable dynamics, where correspondingly the augmented TWA slightly but nevertheless significantly overestimates the average quantum return probability on long time scales.
Opposite cases corresponding to an underestimation of the average long-time return probability by the augmented TWA, leading to a "super enhancement" beyond the maximal possible symmetry enhancement factor (see, e.g., \fref{fig:plot_10_20}a and \fref{fig:plot_10_20}b), can be indicative of an interference-induced inhibition of quantum transport due to dynamical features, related to (dynamical) localization \cite{Anderson1958,Fishman1982,Shepelyansky1983,Bohigas1993,%
Altshuler1997,Gornyi2005,Basko2006,Oganesyan2007} or quantum many-body scarring \cite{Bernien2017,Turner2018,Zhao2020,Serbyn2021,Hummel2022unpub}.
Coherent backscattering \cite{Engl2014,Schlagheck2017,Engl2018SpinEcho} and the manifestation of a modal echo \cite{Altshuler1994,Weaver1994,Weaver2000MEreverb} can be additional causes for such an enhancement of the return probability.

\section{Conclusion}
\label{sec:Conclusion}
In summary, we developed in this paper an augmented version of the Truncated Wigner method.
This augmented TWA is capable of accounting for constructive quantum interference effects due to the presence of discrete symmetries in the physical system at hand, and becomes practically relevant as soon as both the initial quantum state and the final observable to be evaluated are in phase space tightly localized about the corresponding symmetry subspaces.
The key idea of this method is to quantitatively separate the contributions of symmetric and nonsymmetric trajectory families to the TWA sampling, such that one can rescale the latter by the degeneracy factor that is appropriate for the symmetry under consideration.
To this end, separate TWA samplings within the corresponding symmetry subspaces can be performed, including weight factors that account for the transverse (i.e., nonsymmetric) stability of the involved symmetric trajectories.
The application to time evolution processes of mesoscopically populated Bose-Einstein condensates in optical square lattice plaquettes shows good agreement with the exact quantum results for the return probability of this quantum many-body system to its initial state, in stark contrast to the conventional TWA which grossly underestimates this return probability.
This application furthermore confirms the validity of an alternative implementation of the augmented TWA, which is based on the definition of a heuristic distance threshold with respect to the symmetry subspaces under consideration \cite{Schlagheck2019}.
The method can be readily applied to more complex quantum systems, such as Bose-Einstein condensates in hexagonal lattice plaquettes involving six sites, even though they may require a more sophisticated analysis of the involved symmetries and their respective treatment through subspace-restricted TWA simulations by appropriately choosing symmetry-oriented phase-space coordinates.

The comparison with the exact quantum return probabilities also revealed significant deviations for very long evolution times, which we attribute to additional quantum effects coming into play.
Possible causes could be (dynamical) quantum tunneling \cite{Davis1981,Bohigas1993,Tomsovic1994} in mean-field space \cite{Hensinger2001,Vanhaele2021}, many-body coherent backscattering \cite{Engl2014,Schlagheck2017,Engl2018SpinEcho}, a modal echo \cite{Altshuler1994,Weaver1994,Weaver2000MEreverb}, (dynamical) localization effects \cite{Anderson1958,Fishman1982,Shepelyansky1983,Bohigas1993,%
Altshuler1997,Gornyi2005,Basko2006,Oganesyan2007}, or many-body quantum scarring \cite{Bernien2017,Turner2018,Zhao2020,Serbyn2021,Hummel2022unpub}, depending on the underlying (regular or chaotic) classical phase-space structure.
Rather than considering the failure to reproduce those effects as an intrinsic shortcoming of the presented method, we believe, on the contrary, that the augmented TWA can serve as a valuable indicator for the relevance of one or several of those genuinely quantum phenomena.
Additional theoretical diagnostic tools, e.g., involving inverse participation ratios in phase space \cite{Hummel2022unpub}, can then be utilized in order to yield more insight into the nature of the quantum dynamical effect at work.

\ack
This research was supported by the University of Li{\`e}ge under Special Funds for Research, IPD-STEMA Programme.
We acknowledge funding of the Deutsche For\-schungs\-ge\-mein\-schaft through project  Ri681/15-1 within the Reinhart-Koselleck Programme.
We thank Klaus Richter, Steven Tomsovic, Denis Ullmo, and Juan Diego Urbina for
useful and inspiring discussions.

\appendix

\section{Symmetry-oriented phase-space coordinates}
\label{app:symorientedqp}

\subsection{Construction of symmetry-oriented variables}
Let us denote by $\vx$ and $\vk$ the original coordinates and conjugate momenta in which the degrees of freedom of the system are expressed, and by $\vek{f}: \mathbb{R}^L \rightarrow \mathbb{R}^L$, $\vx \mapsto \vek{f}(\vx)$ the action of a differentiable discrete symmetry transformation on the position coordinates $\vx$, which, after some finite number $n \geq 2$ of repetitions, gives the identity $\vek{f}^n \equiv \vek{f} \circ \vek{f} \circ \dots \circ \vek{f} = {\rm id}_{\mathbb{R}^L}$.
The latter property guarantees that $\vek{f}$ is invertible, with $\vek{f}^{n-1}$ being its inverse.
We further demand that the linear approximation of $\vek{f}$ is invertible, i.e., that the matrix
\begin{equation} \label{eq:Ddef}
	D(\vx) = \left( \frac{\partial f_{l}}{\partial x_{l^\prime}} \right)_{l, l^\prime} \psep,
\end{equation}
is regular for any $\vx$.
The point transformation $\vek{f}$ of position coordinates induces a canonical transformation $(\vx,\vk) \mapsto (\vx^\prime, \vk^\prime)$ with $\vx^\prime = \vek{f}(\vx)$ and $\vk^\prime = [\mat{D}(\vx)]^{-\rmT} \cdot \vk$, where $(\cdot)^{-\rmT}$ denotes the transpose of the inverse matrix. [see\eq\eref{eq:qptrafo} in the main text and\eqs\eref{eq:kp1} and \eref{eq:kp2} below].
By assumption, the Hamiltonian is symmetric under this transformation, $H(\vx^\prime, \vk^\prime,t) = H(\vx, \vk, t)$%
\footnote{%
	This ensures that the equations of motion for $(\vx^\prime, \vk^\prime)$ are identical to the ones of $(\vx,\vk)$, such that the transformed version $(\vx^\prime, \vk^\prime)(t)$ of a solution $(\vx, \vk)(t)$ gives another solution to the equations of motion.%
}.
The symmetric manifold in configuration space is then defined as the set $\Sq \equiv \{ \vx \in \mathbb{R}^L \vert \vek{f}(\vx) = \vx \}$ of fix points and we denote its dimension by $S < L$.
The tangent space $\mathcal{T}_{\vx}\Sq$ of $\Sq$ in $\vx \in \Sq$ is given by the nullspace $\ker( \mat{D} - 1 )$, which can be seen from linearly expanding the defining fix-point equation $\vek{f}( \vx + \delta \vx_0 ) - \vx - \delta \vx_0 = \veknull$ in $\delta \vx_0$.
Its dimension is $S \equiv \dim \ker ( D - 1 )$ by definition, while the image $\mathrm{im}( D^\rmT - 1 )$, i.e., the orthogonal complement to the tangent space, has dimension $L - S$, as implied by the rank-nullity theorem (we assume that the space of position coordinates is equipped with a scalar product).

We can now use the fact that it is possible to find a canonical point transformation to new phase-space variables fulfilling $\vq = \vq(\vx)$ for an arbitrarily defined, locally differentiable and invertible [also in linearized form, c.f.\eq\eref{eq:Ddef}] position coordinate transformation $\vx \mapsto \vq(\vx)$.
This canonical transformation can be associated with the type-two generating function of $F(\vx,\vp) = \vp \cdot \vq(\vx)$ yielding
\begin{equation} \label{eq:kp1}
	k_l = \frac{\partial F}{\partial x_l}(\vx,\vp) = \sum_{j=1}^L p_j \frac{\partial q_j(\vx)}{\partial x_l} \qquad  l=1,\ldots,L \psep.
\end{equation}
We infer from this latter relation
\begin{equation} \label{eq:kp2}
		\vp = \left[\mat{T}(\vq)\right]^{-\rmT} \cdot \vk
		\qquad \text{with} \qquad \mat{T}_{l,l^\prime}(\vq) = \frac{\partial q_l(\vx)}{\partial x_{l^\prime}} \biggr\rvert_{\vx=\vx(\vq)} \psep.
\end{equation}
Specifically, $\vq(\vx)$ shall be defined such that the first $S$ coordinates of the position and momentum vectors (denoted by the index $\sigma$ and referred to as ``parallel'' components in the following) represent the degrees of freedom that lie within the symmetric manifold, while the remaining $L-S$ coordinates (denoted by the index $\lambda$ and referred to as ``perpendicular'' components in the following) comprise the degrees of freedom that describe the motion out of $\Sq$.
To be more specific, we may write $\vq = ( \vqpara, \vqperp )$ with
\begin{equation}
	\eqalign{
		\vqpara = ( q_{\para,1}, \ldots, q_{\para,S} ) = ( q_1, \ldots, q_S ) \psep,\\
		\vqperp = ( q_{\perp,1}, \ldots, q_{\perp,L-S} ) = ( q_{S+1}, \ldots, q_L )
	}
\end{equation}
and correspondingly $\vp = ( \vppara, \vpperp )$ with
\begin{equation}
	\eqalign{
		\vppara = ( p_{\para,1}, \ldots, p_{\para,S} ) = ( p_1, \ldots, p_S ) \psep,\\
		\vpperp = ( p_{\perp,1}, \ldots, p_{\perp,L-S} ) = ( p_{S+1}, \ldots, p_L ) \psep.
	}
\end{equation}
As outlined in the following, one can construct $\vq(\vx)$ such that i) $\vqperp(\vx) = \veknull$ if and only if $\vx \in \mathcal{S}$ as well as ii) $\vqpara( \vek{f}( \vx_0 + \epsilon \vek{y} ) ) = \vqpara( \vx_0 + \epsilon \vek{y} ) + \Ord(\epsilon^2)$ for any $\vx_0 \in \Sq$, i.e., such that in the vicinity of $\Sq$ the parallel components $\vqpara$ are invariant under $\vek{f}$ (at least up to second-order corrections).
Moreover, one can show that within this construction Hamilton's equations of motion take a form such that iii) for $\vq \in \Sq$, $\dot{\vq}_\perp = \veknull$ is equivalent to $\vpperp = \veknull$.
Hence, motion $\vq(t)$ that stays in $\mathcal{S}$ for all times is characterized by vanishing perpendicular momenta $\vpperp = \veknull$.
The special properties i)--iii) are all requirements on only the the \emph{local} behaviour of $\vq(\vx)$ in the immediate vicinity of $\Sq$.
We assume that a corresponding global transformation $\vq(\vx)$ is possible whenever the local conditions are fulfilled%
\footnote{%
In case that $\Sq$ is not contractible this may involve resolving discontinuities in $\vq(\vx)$ by introducing multiple charts.%
}.
By convention~\eref{eq:siglam}, we denote parallel components by indexes $\sigma = 1, \ldots, S$ and perpendicular components by $\lambda = S + 1, \ldots, L$, while Roman letters index all components, e.g., $l= 1, \ldots, L$.

Property i) can be fulfilled by choosing $q_\lambda(\vx) = Q_\lambda( \vek{f}(\vx) - \vx )$ for a set of $L-S$ (differentiable) functions $Q_\lambda: \mathbb{R}^L \to \mathbb{R}$ with $Q_\lambda( \veknull ) = 0$.
For $\vx \in \Sq$, the $L-S$ vectors $\vnabla q_\lambda = (\mat{D}^\rmT -1) \cdot \vnabla Q_\lambda$, where $\vnabla$ denotes the $L$ derivatives with respect to the respective function arguments, can be chosen linearly independent by an appropriate choice of the functions $Q_\lambda$, e.g., using linear combinations $Q_\lambda(\vek{z})=\sum_l c_{l}^{(\lambda)} z_l$ with properly defined coefficients $c_l^{(\lambda)}$.
This is guaranteed by the fact that the rank $\mathrm{rk}(\mat{D}^\rmT-1) = \mathrm{rk}(\mat{D}-1) = L - S$ equals the number of perpendicular components $\lambda$, such that $\{ \vnabla q_\lambda \}_\lambda$ form a basis of the orthogonal space $\mathrm{im}(\mat{D}^\rmT-1)$.

Similarly, in order to fulfil property ii) one can choose $\vqpara$ such that the linearly independent vectors $\{ \vnabla q_\sigma \}_\sigma$ form a basis of $\ker ( \mat{D}^\rmT - 1 )$ at any $\vx \in \Sq$%
\footnote{%
	Note that the nullspace of $D^\rmT - 1$ does not necessarily coincide with the tangent space $\ker ( \mat{D} - 1 )$, unless $\vek{f}$ is an orthogonal transformation, i.e., $\mat{D}^\rmT \mat{D} = 1$, since then $\ker( D^\rmT - 1 ) = \ker( \mat{D} ( \mat{D}^\rmT - 1 ) ) = \ker( 1 - \mat{D} )$.%
}.
Again, the dimension $\dim \ker ( \mat{D}^\rmT - 1) = S$ equals the number of parallel components $\sigma$, implying that such a choice for the $\vnabla q_\sigma$ is possible.
A point $\vx = \vx_0 + \epsilon \vek{y}$ in the vicinity of $\Sq$, with $\vx_0 \in \Sq$, transforms as $\vek{f}(\vx) = \vek{f}(\vx_0) + \epsilon \sum_j y_j (\partial \vek{f} / \partial x_j)\vert_{\vx_0} + \Ord(\epsilon^2) = \vx_0 + \epsilon D \cdot \vek{y} + \Ord(\epsilon^2)$, such that the parallel coordinates transform as $q_\sigma( \vek{f}(\vx) ) = q_\sigma + \epsilon \vek{y}^\rmT \cdot ( \mat{D}^\rmT - 1) \cdot \vnabla q_\sigma + \Ord(\epsilon^2)$.
Property ii) follows directly with $\vnabla q_\sigma$ in the nullspace of $\mat{D}^\rmT - 1$.

This construction yields an overall locally invertible coordinate transformation with regular matrix $( \partial q_l / \partial x_{l^\prime} )_{l,l^\prime}$ if the tangent space $\mathrm{im}( \mat{D}^\rmT - 1 )$ is linearly independent from $\ker( \mat{D}^\rmT - 1 )$.
Indeed, the latter can be shown to be fulfilled in our case as follows.
Since $n$-fold composition $\vek{f}^n$ of $\vek{f}$ gives the identity, the same holds for the linearization represented by the matrix $\mat{D}$, i.e., $\mat{D}^n =  1$, through differentiation.
As a consequence, it can be easily shown that $ \mat{A} \cdot (\mat{D}^\rmT - 1)^2 = (\mat{D}^\rmT - 1) $ with $\mat{A} = -\sum_{k=1}^{n-1} \frac{k}{n} (\mat{D}^\rmT)^{n-1-k}$.
Let now $\vx \in \ker(\mat{D}^\rmT - 1)$ and at the same time $\vx \in \mathrm{im}(\mat{D}^\rmT-1)$, i.e., $(\mat{D}^\rmT-1)\cdot \vx = \veknull$ and $\exists \vek{y} \in \mathbb{R}^L : (\mat{D}^\rmT-1)\cdot \vek{y} = \vx$.
It follows that $(\mat{D}^\rmT-1)^2 \cdot \vek{y} = \veknull$.
After multiplication with $\mat{A}$ from the left we get $(\mat{D}^\rmT-1) \cdot \vek{y} = \veknull$, i.e., $\vx = \veknull$, and thus $\mathrm{im}(\mat{D}^\rmT - 1) \cap \ker (\mat{D}^\rmT-1) = \{\veknull\}$.

It remains to show property iii), i.e., that motion inside $\Sq$ is equivalent to $\vx \in \Sq$ and $p_\lambda = 0$ for all $\lambda=S+1,\ldots,L$.
Henceforth we consider a point $\vx$ in position space that sits on the symmetric manifold $\Sq$ at time $t$.
For $\vx(t)$ to stay inside of $\Sq$, $\dot{\vx}$ needs to lie in the tangent space of $\Sq$ at $\vx$, i.e., $\dot{\vx} \in \ker( D - 1 )$, or equivalently $\rmd / \rmd t [ \vx - \vek{f}(\vx) ] = \veknull$.
The latter is, due to the symmetry of $H$ and recalling the transformation $\vk^\prime = [\mat{D}(\vx)]^{-\rmT}\cdot \vk$ of momenta, equivalent to the invariance $\vek{v}( \vx, \vk, t ) = \vek{v}( \vx, \vk^\prime, t )$ of the velocities under the symmetry transformation, where
\begin{equation}
	v_l( \vx, \vk, t ) \equiv \frac{\partial}{\partial k_l} H( \vx, \vk, t )
\end{equation}
is the function that (for given $\vx$) uniquely maps the velocity vector $\dot\vx$ to the momentum vector $\vk$ and vice versa.
Motion $\vx(t)$ that stays inside of $\Sq$ (given that $\vx^\prime = \vx$ at time $t$) is thus characterized by
$\vk = \vk^\prime$.
Expressing $\vk$ and $\vk^\prime$ for a given $\vx = \vx^\prime$ in terms of the symmetry-oriented momentum $\vp$ via\eq\eref{eq:kp2}, gives
\begin{equation}
	\fl
	\vk - \vk^\prime = \mat{D}^{-\rmT} \cdot ( \mat{D}^\rmT - 1 ) \cdot \sum_l (\vnabla q_l) p_l
	= \mat{D}^{-\rmT} \cdot ( \mat{D}^\rmT - 1 ) \cdot \sum_{\lambda} (\vnabla q_{\lambda}) p_{\lambda} \psep,
\end{equation} 
where in the last step we used the fact that $( \mat{D}^\rmT - 1 ) \cdot \vnabla q_\sigma = \veknull$ for all $\sigma = 1, \ldots, S$ by construction.
On the one hand, if all \emph{perpendicular} momenta vanish, i.e., $p_\lambda = 0$ for all $\lambda = S + 1, \ldots, L$, then $\vk = \vk^\prime$ and we get motion restricted to $\Sq$.
On the other hand, if motion is inside $\Sq$, then we have $\vk = \vk^\prime$, implying that $\sum_\lambda (\vnabla q_\lambda) p_\lambda \in \ker( \mat{D}^\rmT - 1 )$.
At the same time we know by construction that $\sum_\lambda (\vnabla q_\lambda) p_\lambda \in \mathrm{im}( \mat{D}^\rmT - 1 )$ and that $\mathrm{im}( \mat{D}^\rmT - 1 ) \cap \ker( \mat{D}^\rmT - 1 ) = \{ \veknull \}$, such that $\vk = \vk^\prime$ implies $\sum_\lambda (\vnabla q_\lambda) p_\lambda = \veknull$ and thus all $p_\lambda = 0$ due to linear independence of the $\vnabla q_\lambda$.
In total, $\vk = \vk^\prime$ is equivalent to $\vpperp = \veknull$ for $\vx \in \Sq$, i.e., property iii) is fulfilled by our construction of symmetry-oriented variables $\vq$ and $\vp$.
It is worth to note that, for $\vx \in \Sq$, $\mathrm{span} ( \{\vnabla q_\sigma\}_\sigma ) = \ker( \mat{D}^\rmT - 1 )$ is not only a sufficient condition but also necessary to achieve property iii) together with linear independence of all $\{\vnabla q_l\}_l$ and assuming $\mathrm{span} ( \{\vnabla q_\lambda\}_\lambda ) = \mathrm{im}( \mat{D}^\rmT - 1 )$.

If multiple symmetry transformations $\vek{f}^{(1)}, \ldots, \vek{f}^{(m)}$ are at work simultaneously---which is a typical situation in the context of the augmented TWA---, then $\Sq$ is the manifold that is simultaneously invariant with respect to all transformations $\vek{f}^{(j)}$, $j=1,\ldots,m$.
In analogy with the above analysis, the construction of symmetry-oriented variables would then be given by a canonical point transformation $\vx \mapsto \vq(\vx)$ together with\eq\eref{eq:kp2} such that $\{ \vnabla q_\lambda \}_\lambda$ are a basis of the \emph{common} image $\mat{I} \equiv  \mathrm{span}( \cup_{j=1}^m \mathrm{im}( \mat{D}_j^\rmT - 1 ) )$ and such that $\{ \vnabla q_\sigma \}_\sigma$ are a basis of the \emph{common} nullspace $\mat{K} \equiv \cap_{j=1}^m \ker( \mat{D}_j^\rmT - 1 )$, where $\mat{D}_j$ is comprised of the derivatives of $\vek{f}^{(j)}$ analogously to\eq\eref{eq:Ddef}.
In order to yield a valid transformation one needs $\dim(\mat{I}) + \dim(\mat{K}) = L$ and $\mat{I} \cap \mat{K} = \{ \veknull \}$.
As shown in the following, the case $m > 1$ seems to be more restrictive regarding the functions $\vek{f}^{(j)}$ than $m=1$.
The tangent space of $\Sq$ is given by the common nullspace of all matrices $(\mat{D}_j - 1)$,
\begin{equation}
	\mathcal{T}_\vx \Sq = \bigcap_{j=1}^m \ker ( \mat{D}_j - 1 ) = \ker\underbrace{\left(
		\begin{array}{c}
			\mat{D}_1 - 1 \\
			\vdots \\
			\mat{D}_m - 1
		\end{array}
		\right)}_{\equiv \mat{M}} \psep,
\end{equation}
where $\mat{M}$ is a $mL \times L$ matrix.
By the rank-nullity theorem, the dimension of $\mathcal{T}_{\vx} \Sq$, denoted by $S \equiv \dim \ker(\mat{M})$, is then given by
\begin{equation}
	S = L - \dim \mathrm{im}(\mat{M}) = L - \dim \mathrm{im}(\mat{M}^\rmT) = L - \dim(\mat{I}) \psep,
\end{equation}
where in the last step we used that $\mat{I} = \mathrm{im}\bigl( ( \mat{D}_1^\rmT - 1 ) \ldots ( \mat{D}_m^\rmT - 1 ) \bigr) = \mathrm{im}(\mat{M}^\rmT)$.
The number of perpendicular coordinates $q_\lambda$ would thus match the dimension $L-S$ of the subspace $\mat{I}$ orthogonal to $\mathcal{T}_{\vx} \Sq$.
In contrast, the dimension of $\mat{K}$ is in general not identical to the dimension $S$ of the tangent space, which inhibits the above construction of symmetry-oriented variables.
However, if we restrict ourselves to symmetry transformations that are all locally orthogonal, i.e., $\mat{D}_j^\rmT \mat{D}_j = 1$ for all $j=1,\ldots,m$, then $\ker( \mat{D}_j - 1 ) = \ker\bigl( \mat{D}_j^\rmT \cdot ( \mat{D}_j - 1 ) \bigr) = \ker( 1 - \mat{D}_j^\rmT ) = \ker( \mat{D}_j^\rmT - 1 )$ and thus $\mat{K} = \cap_{j=1}^m \ker( \mat{D}_j - 1 ) = \ker(\mat{M}) = \mathcal{T}_{\vx} \Sq$ with correct dimension $S$ and automatically fulfilling $\mat{I} \cap \mat{K} = \{ \veknull \}$ since $\ker(\mat{M})$ is the orthogonal complement of $\mathrm{im}(\mat{M}^\rmT)$ for any matrix $\mat{M}$.
Note that all discrete symmetries that are composed of reflections, rotations and/or permutations of coordinates correspond to orthogonal transformations and thus comply with the mentioned requirements.
This is the case for all our presented applications within the augmented TWA.

\subsection{Symmetry-oriented variables in the TWA}
Let us adapt the expression~\eref{eq:OmegatdiagWO} to the symmetry-oriented coordinates $\vq,\vp$.
We assume that originally the expression is formulated in the canonical coordinates $\vx, \vk$ with the initial Wigner function and the observable given as $\Wi(\vx,\vk)$ and $\Ow(\vx,\vk)$, respectively:
\begin{equation} \label{eq:OmegatdiagWOxk}
	\fl
	\langle \hat{\Omega} \rangle_t^{\rm diag} \simeq{} \int \!\rmd^L \vx^\rmi \int \!\rmd^L \vx^\rmf \;
	\sum_{\gamma}
	\left| \det \!\left(
	\frac{\partial^2 R_\gamma}{\partial x^\rmi_l \partial x^\rmf_{l^\prime} }
	\right)_{l,l^\prime} \right|
	\Wi\!\left( \vx^\rmi, \vk_{\gamma}^\rmi \big\rvert_{\vx} \right)
	\Ow\!\left( \vx^\rmf, \vk_{\gamma}^\rmf \big\rvert_{\vx} \right) \psep.
\end{equation}
In the following we simplify the notation by identifying $\Ralt_\gamma(\vq^\rmf, \vq^\rmi,t) = R_\gamma\!\left(\vx^\rmf,\vx^\rmi,t\right)$ with $\vx^{\rmi,\rmf}=\vx(\vq^{\rmi,\rmf})$ and $\vx(\vq)$ denoting the inversion of the coordinate transformation $\vq(\vx)$.
Deriving this principal function with respect to $\vq^{\rmi, \rmf}$ gives the momenta conjugate to $\vq$ that result from the above canonical transformation applied to $\vk^{\rmi,\rmf}$.
Denoting these initial ($\alpha = \rmi$) and final ($\alpha = \rmf$) momenta of trajectory $\gamma$ by
\begin{equation}
	\eqalign{
		k^{\alpha}_{\gamma,l} \big\rvert_\vx = \sigma_\alpha \frac{\partial R_\gamma}{\partial x^\alpha_l}(\vx^\rmf,\vx^\rmi,t) \psep,
	\\
		p^{\alpha}_{\gamma,l} \big\rvert_\vq = \sigma_\alpha \frac{\partial \Ralt_\gamma}{\partial q^\alpha_l}(\vq^\rmf,\vq^\rmi,t) \psep,
	}
\end{equation}
one finds easily that they result from each other through the canonical transformation~\eref{eq:kp1}, \eref{eq:kp2} from $(\vx,\vk) \rightarrow (\vq, \vp)$, i.e.,
\begin{equation}
	\vk^\alpha_\gamma \big\rvert_{\vx(\vq)} = \mat{T}(\vq^\alpha)^\rmT \cdot \vp^\alpha_\gamma \big\rvert_\vq \psep.
\end{equation}
The determinant in \eref{eq:OmegatdiagWO} is then transformed to the symmetry-oriented variables as
\begin{equation}
	\fl
	\left| \det\!\left( \frac{\partial^2 R_\gamma}{\partial x^\rmi_l \partial x^\rmf_{l^\prime} } \right)_{l,l^\prime} \right|
		= \left| \det\!\left( \frac{\partial^2 \Ralt_\gamma}{\partial q^\rmi_l \partial q^\rmf_{l^\prime} } \right)_{l,l^\prime} \;
			\det\!\left( \frac{\partial q_l}{\partial x_{l^\prime} }\biggr\rvert_{\vx(\vq^\rmf)} \right)_{l,l^\prime} \;
			\det\!\left( \frac{\partial q_l}{\partial x_{l^\prime} }\biggr\rvert_{\vx(\vq^\rmi)} \right)_{l,l^\prime} \right| \psep.
\end{equation}
The last two terms serve as Jacobians for the transformation of the integration variables from $\vx^{\rmi,\rmf}$ to $\vq^{\rmi,\rmf}$, giving
\begin{equation} \label{eq:Omegatqp}
	\fl
	\langle \hat{\Omega} \rangle_t^{\rm diag} \simeq \int \!\rmd^L \vq^\rmi \int \!\rmd^L \vq^\rmf \;
		\sum_{\gamma}
		\left| \det \!\left( \frac{\partial p^\rmi_{\gamma,l}}{\partial q^\rmf_{l^\prime} }(\vq^\rmf,\vq^\rmi,t) \right)_{l,l^\prime} \right|
		\Wi^\mathcal{S}\!\left( \vq^\rmi, \vp_{\gamma}^\rmi \big\rvert_{\vq} \right)
		\Ow^\mathcal{S}\!\left( \vq^\rmf, \vp_{\gamma}^\rmf \big\rvert_{\vq} \right) \psep.
\end{equation}
Here the new Wigner representations of the initial state and the observable are the old ones expressed as functions of the new variables, i.e.,
\begin{equation} \label{eq:WiSOmegaiS}
	\eqalign{ 
		\Wi^\mathcal{S}\!\left( \vq^\rmi, \vp_{\gamma}^\rmi \big\rvert_{\vq} \right)
			= \Wi\!\left( \vx(\vq^\rmi), \mat{T}(\vq^\rmi)^\rmT \cdot \vp_{\gamma}^\rmi \big\rvert_{\vq} \right) \psep,
	\\
		\Ow^\mathcal{S}\!\left( \vq^\rmf, \vp_{\gamma}^\rmf \big\rvert_{\vq} \right) 
			= \Ow\!\left( \vx(\vq^\rmf), \mat{T}(\vq^\rmf)^\rmT \cdot  \vp_{\gamma}^\rmf \big\rvert_{\vq} \right) \psep.
	}
\end{equation}
It confirms that we can rely on the notation used throughout the manuscript, where $\vR^\alpha$ and $\vp^\alpha_\gamma$ are the symmetry-oriented phase-space coordinates.
We have to keep in mind that $\Wi$ and $\Ow$ have to be expressed in those, possibly using \eref{eq:WiSOmegaiS} when they are originally given in different phase-space variables $\vx,\vk$.
Furthermore, instead of writing $\vx \in \mathcal{S}$, which is equivalent to $\vq(\vx) \in \vq(\mathcal{S})$ we may introduce the simpler notation $\vq \in \mathcal{S}$.

\section{Local separation of dynamics}
\label{app:locsepdyn}
To show that cross dependencies between parallel and perpendicular phase-space coordinates vanish on ${\SP}$~\eref{eq:qpRPsiglam}, we consider a symmetric trajectory family $\gamma$ and in particular its representative trajectories that are themselves symmetric and thus fully in ${\SP}$.
Consequently, the perpendicular initial and final momenta are identically zero, as long as we choose $\vRi$ and $\vRf$ to be in the symmetric manifold $\mathcal{S}$, i.e., $\vp^\alpha_{\gamma, \perp} \bigr\rvert_\Sq = \veknull$ for $\alpha=\rmi,\rmf$.
Put in differential form, this identity implies
\begin{equation} \label{eq:pgRlamsig}
	\frac{\partial p^{\alpha}_{\gamma,\lambda}} {\partial R^{\alpha^\prime}_{\sigma} }\biggr\rvert_\Sq = 0
		\qquad  \alpha, \alpha^\prime \in \{ \rmi, \rmf \} \psep.
\end{equation}
The relation between initial (final) momenta and Hamilton's principal function~\eref{eq:palphaR} further implies
\begin{equation} \label{eq:dpdRswitch}
	\frac{\partial p^{\alpha}_{\gamma,l}} {\partial R^{\alpha^\prime}_{l^\prime} }
		= \sigma_\alpha \frac{\partial^2 R_{\gamma}} { \partial R^\alpha_{l} \partial R^{\alpha^\prime}_{l^\prime} }
		= \sigma_{\alpha^\prime} \sigma_\alpha \frac{\partial p^{\alpha^\prime}_{\gamma,l^\prime}} {\partial R^\alpha_{l} } \psep,
\end{equation}
which, setting $l= \sigma, l^\prime = \lambda$ and using \eref{eq:pgRlamsig} immediately leads to
\begin{equation} \label{eq:pgRsiglam}
	\frac{\partial p^{\alpha}_{\gamma,\sigma}} {\partial R^{\alpha^\prime}_{\lambda} }\biggr\rvert_\Sq = 0
		\qquad \alpha, \alpha^\prime \in \{ \rmi, \rmf \} \psep.
\end{equation}
In other words, if one writes the total differentials as
\begin{equation} \label{eq:dpgM}
	\eqalign{
		\rmd \vpig \equiv \mat{M}^{(\rmi \rmi)} \rmd \vRi
			+ \mat{M}^{(\rmi \rmf)} \rmd \vRf \psep, \\
		\rmd \vpfg \equiv \mat{M}^{(\rmf \rmi)} \rmd \vRi
					+ \mat{M}^{(\rmf \rmf)} \rmd \vRf \psep,
	}
\end{equation}
the matrices $\mat{M}^{(\alpha \alpha^\prime)}$ assume a block diagonal form
\begin{equation} \label{eq:blockM}
	\mat{M}^{(\alpha \alpha^\prime)} \equiv
		\left(\begin{array}{@{}c|c@{}}
			\mat{M}^{(\alpha \alpha^\prime)}_\para & \bigzero \Bstrut{1.3ex} \\
			\hline
			\bigzero & \mat{M}^{(\alpha \alpha^\prime)}_\perp \Tstrut{2.6ex}
		\end{array}\right)
		\begin{array}{@{\hspace*{-1em}}l}
			\left.\phantom{\begin{matrix} M \end{matrix}} \right\}S\\
			\left.\phantom{\begin{matrix} M \end{matrix}} \right\}L-S
		\end{array} \psep.
\end{equation}
We note that the $(l, l^\prime)$-element of $\mat{M}^{(\alpha \alpha^\prime)}$ is given by \eref{eq:dpdRswitch}, evaluated on $\Sq$.
These matrices are functions of the trajectory, i.e., $\gamma$, $\vRi$, and $\vRf$, which we do not denote explicitly.
The block diagonal structure holds as well for the inverse matrices (assuming their existence), namely
\begin{equation} \label{eq:blockMinv}
	\bigl( \mat{M}^{(\alpha \alpha^\prime)} \bigr)^{-1} =
		\left(\begin{array}{@{}c|c@{}}
			\bigl( \mat{M}^{(\alpha \alpha^\prime)}_\para \bigr)^{-1} & \bigzero \Bstrut{1.3ex}\\
			\hline
			\bigzero & \bigl( \mat{M}^{(\alpha \alpha^\prime)}_\perp \bigr)^{-1} \Tstrut{2.6ex}
		\end{array}\right)
		\begin{array}{@{\hspace*{-1em}}l}
			\left.\phantom{\begin{matrix} M \end{matrix}} \right\}S\\
			\left.\phantom{\begin{matrix} M \end{matrix}} \right\}L-S
		\end{array} \psep.
\end{equation}

It remains to write the property~\eref{eq:pgRsiglam} of the \textit{boundary value} problem as a corresponding property of the \textit{initial value} problem.
This is made explicit by inversion of the momentum root functions $\vpig( \vRf, \vRi, t )$ and $\vpfg( \vRf, \vRi, t )$ such that $ \vRf $ and $ \vpfg $ are expressed as functions of $ \vRi $ and $ \vpig $ (and $t$):
\begin{equation} \label{eq:inversionqp}
	\begin{array}{l}
		\vRf = \vq( \vRi, \vPi, t ) \\
		\vpfg ( \vRf, \vRi, t ) = \vp( \vRi, \vPi, t )
	\end{array}
	\qquad \Leftrightarrow \qquad
	\vPi = \vpig ( \vRf, \vRi, t ) \,,
\end{equation}
where the dependence on the family $\gamma$ vanishes for the inverted functions $ \vq $ and $ \vp $, since they denote the \textit{unique} time evolution of an initial point $ ( \vRi, \vPi )$ in phase space.
By identifying the total differentials of the involved functions~\eref{eq:inversionqp} correspondingly, the differential form~\eref{eq:dpgM} becomes:
\begin{equation} \label{eq:dPidpM}
	\eqalign{
		\rmd \vPi \equiv \mat{M}^{(\rmi \rmi)} \rmd \vRi
			+ \mat{M}^{(\rmi \rmf)} \rmd \vq \psep, \\
		\rmd \vp \equiv \mat{M}^{(\rmf \rmi)} \rmd \vRi
					+ \mat{M}^{(\rmf \rmf)} \rmd \vq \psep.
	}
\end{equation}
Solving the linear relations~\eref{eq:dPidpM} for the final position $\rmd \vq$ and momentum $\rmd \vp$ gives then the differential form of the unique time evolution $\vq( \vRi, \vPi, t)$, $\vp( \vRi, \vPi, t)$ of an initial point $(\vRi, \vPi)$ in phase space about ${\SP}$,
\begin{eqnarray}
	\rmd \vq = - \left( \mat{M}^{(\rmi \rmf)} \right)^{-1} \mat{M}^{(\rmi \rmi)} \rmd \vRi + \left( \mat{M}^{(\rmi \rmf)} \right)^{-1} \rmd \vPi \psep,\\
	\rmd \vp = \left(  \mat{M}^{(\rmf \rmi)} - \mat{M}^{(\rmf \rmf)} \left( \mat{M}^{(\rmi \rmf)} \right)^{-1} \mat{M}^{(\rmi \rmi)} \right) \rmd \vRi + \mat{M}^{(\rmf \rmf)} \left( \mat{M}^{(\rmi \rmf)} \right)^{-1} \rmd \vPi \psep.
\end{eqnarray}
This immediately results, through the block structure~\eref{eq:blockM} and~\eref{eq:blockMinv}, in the local separation of dynamics as formulated in\eq\eref{eq:qpRPsiglam}.

\section{A sum rule for symmetric trajectory families}
\label{app:symtrajDelta}
In the following we show that replacing the full dynamics of position coordinates by an appropriate approximate version close to ${\SP}$ in~\eref{eq:deltaRdeltaP} of the main text provides a variant of the sum rule that selects precisely the symmetric trajectory families.
In particular, the linearized position evolution given by\eq\eref{eq:qapproxlin}, fulfils all requirements.
Here, we demand a set of less restrictive properties, namely
\newlength{\lendmy}\settowidth{\lendmy}{$\text{(iii)} \quad$}
\begin{eqnarray}
\label{eq:qapprox1}
	&\hspace*{-\lendmy}\mathrlap{\text{(i)}}\hphantom{\text{(iii)}} \quad
			\vqsap( \vRin, \vPin, t ) \equiv
			\vq( \vRin, \vPin, t ) \psep,
	\\
\label{eq:qapprox2}
	&\hspace*{-\lendmy}\eqalign{
			\mathrlap{\text{(ii)}}\hphantom{\text{(iii)}} \quad
			&\frac{\partial \vqsap}{\partial R^\rmi_\lambda}(\vRin, \vPin, t) = \frac{\partial \vq}{\partial R^\rmi_\lambda}(\vRin, \vPin, t) \psep,
		\\
			&\frac{\partial \vqsap}{\partial P^\rmi_\lambda}(\vRin, \vPin, t) = \frac{\partial \vq}{\partial P^\rmi_\lambda}(\vRin, \vPin, t) \psep,
	}
	\\
\label{eq:qapprox3}
	&\hspace*{-\lendmy}\text{(iii)} \quad
		\vqsapperp(\vRi, (\vPipara,\vPiperp), t) =
		\vqsapperp\!\big( \vRi, ( \vPipara, {\vPiperp}^\prime), t \big)
		\; \Rightarrow \;
		\vPiperp = {\vP^{\rmi}_\perp}^\prime \psep.
\end{eqnarray}
Besides the coincidence on ${\SP}$ of the function values~\eref{eq:qapprox1} and also their first derivatives~\eref{eq:qapprox2} the perpendicular components need to be injective in $\vPiperp$~\eref{eq:qapprox3}.
The linear approximation~\eref{eq:qapproxlin} used in the main text is one particular choice, where to guarantee injectivity we exclude the exceptional case of hitting a focal point or caustic where $\left(\partial q_\lambda / \partial P^\rmi_{\lambda^\prime}\right)_{\lambda, \lambda^\prime}\bigr\rvert_{\SP}$, and consequently $\left(\partial q_l / \partial P^\rmi_{l^\prime}\right)_{l, l^\prime}\bigr\rvert_{\SP}$, would not have full rank.

As explained in \sref{sec:extractsym}, the linear approximation implements a finite version of the just infinitesimal change of symmetric trajectories out of the subspace ${\SP}$ and thereby prevents them from leaving the symmetric family.
Likewise, the more general criterion of injectivity~\eref{eq:qapprox3} guarantees that increasing $\vPiperp$ from $\veknull$ to finite values cannot change the (symmetric) family but artificially extends its support to arbitrary perpendicular initial conditions, gained at the cost of loosing the exact description of the dynamics at larger distance from ${\SP}$.

We consider transforming the multi-dimensional Dirac delta distribution
\begin{equation} \label{eq:DiracMod}
	\delta^{(L)} \!\left[ \vRf - \vqsap( \vRi, \vPi, t ) \right] = \prod_{l=1}^L \delta \!\left[ R^\rmf_l - \sap{q}_l( \vRi, \vPi, t ) \right]
\end{equation}
into a sum of the corresponding roots of its argument in $\vPi$.
We denote these roots by $\sap{\vp}^\rmi_\beta$, enumerated by an index $\beta$, the precise definition being
\begin{equation} \label{eq:pbeta}
	\vRf = \vqsap( \vRi, \vPi, t )
		\quad \Leftrightarrow \quad
		\vPi \in \left\{ \sap{\vp}^\rmi_\beta( \vRf, \vRi, t) \right\}_\beta \psep,
\end{equation}
where $\beta$ enumerates all possible \textit{families} of solutions, with representatives fixed by a specific choice of $\vRi$, $\vRf$, and $t$, which smoothly transform into one another under smooth variations of initial and final position.
If a family $\beta$ of solutions does not contain a representative with a given $\vRi$ and $\vRf$, we simply exclude it from the set in \eref{eq:pbeta}.
We demand that each family of solutions is counted only once, 
\begin{equation}
	\sap{\vp}^\rmi_\beta \neq \sap{\vp}^\rmi_{\beta^\prime} \qquad \forall \beta \neq \beta^\prime \psep,
\end{equation}
and we further restrict ourselves to the generic case of point-wise distinction
\begin{equation} \label{eq:bneqbprime}
	\sap{\vp}^\rmi_\beta(\vRf, \vRi, t) \neq \sap{\vp}^\rmi_{\beta^\prime}(\vRf, \vRi, t) \qquad \forall \beta \neq \beta^\prime \psep,
\end{equation}
which is true for almost all $\vRi, \vRf \in \mathbb{R}^L$, excluding only exceptional points, like, e.g., bifurcations within the symmetric subspace ${\SP}$.
Because of the injectivity~\eref{eq:qapprox3} of $\vqsapperp$ in $\vPiperp$, we can infer that two solutions belonging to different families $\beta \neq \beta^\prime$ must be distinct in their tangent components, which is a stronger statement than the general point-wise distinction~\eref{eq:bneqbprime}:
\begin{equation} \label{eq:bneqbprimepara}
		\sap{\vp}^\rmi_{\beta, \para}(\vRf, \vRi, t) \neq \sap{\vp}^\rmi_{\beta^\prime, \para}(\vRf, \vRi, t) \qquad \forall \beta \neq \beta^\prime \psep;
\end{equation}
otherwise, due to\eq\eqref{eq:qapprox3} they would also be identical in the perpendicular components%
\footnote{%
		Note that the final positions involving both solutions are identical by definition~\eref{eq:pbeta}, $\vqsap( \vRi, \sap{\vp}^\rmi_\beta, t ) = \vRf = \vqsap( \vRi, \sap{\vp}^\rmi_{\beta^\prime}, t )$.
}
and therefore be equal in all their components, contradicting\eq\eref{eq:bneqbprime}.

To ease notation we will henceforth drop the explicit dependence on time $t$ in the unique time evolutions of $\vq$, $\vp$, the approximate time evolution $\vqsap$ and also in the solutions $\vp^\alpha_\gamma$ and $\sap{\vp}^\alpha_\beta$.
To show that the solutions $\beta$ correspond to symmetric trajectory families, one can take the initial and final position onto $\Sq$, i.e., we take $\vR^{\rmi,\rmf} \rightarrow \vR^{\rmi,\rmf}_0$, and analyse the corresponding solution, fulfilling
\begin{equation}
	\vRfn = \vqsap\big( \vRin, \sap{\vp}^\rmi_\beta( \vRfn, \vRin ) \big) \psep,
\end{equation}
or, written separately in parallel and perpendicular components,
\begin{eqnarray}
	\vRfpara = \vqsappara\big( \vRin, \sap{\vp}^\rmi_\beta( \vRfn, \vRin ) \big) \psep, \\
	\veknull = \vqsapperp\big( \vRin, \sap{\vp}^\rmi_\beta( \vRfn, \vRin ) \big) \psep. \label{eq:qsapperpnull}
\end{eqnarray}
Since the approximate time evolution of the perpendicular position~\eref{eq:qapprox1}--\eref{eq:qapprox3} is unique in $\vPiperp$, the solution of \eref{eq:qsapperpnull} becomes trivial in the perpendicular momentum components,
\begin{equation} \label{eq:pbetaperpnull}
	\sap{\vp}^\rmi_{\beta,\perp}( \vRfn, \vRin ) = \veknull \psep.
\end{equation}
Indeed, whatever is $\sap{\vp}^\rmi_{\beta, \para}$, we know from \eref{eq:qapprox1} that
\begin{equation}
	\vqsapperp \big( \vRin , ( \sap{\vp}^\rmi_{\beta, \para}, \veknull ) \big) = \vqperp \big( \vRin , ( \sap{\vp}^\rmi_{\beta, \para}, \veknull ) \big) = \veknull	
\end{equation}
and thus $ \vqsapperp \big( \vRin , ( \sap{\vp}^\rmi_{\beta, \para}, \veknull ) \big) = \vqsapperp ( \vRin , \sap{\vp}^\rmi_{\beta} ) $, from which \eref{eq:pbetaperpnull} directly follows via \eref{eq:qapprox3}.
Moreover, \eref{eq:pbetaperpnull} implies [via \eref{eq:qapprox1}] that
\begin{equation}
	\vRfn = \vqsap\big( \vRin, \sap{\vp}^\rmi_\beta( \vRfn, \vRin ) \big) = \vq \big( \vRin, \sap{\vp}^\rmi_\beta( \vRfn, \vRin ) \big) \psep,
\end{equation}
meaning that on $\Sq$ the roots of the Dirac-delta argument in\eq\eref{eq:DiracMod}, using the approximate time evolution, are also roots of the unmodified Dirac-delta argument in\eq\eref{eq:deltaRdeltaP}, using the full time evolution $ \vq ( \vRi, \vPi ) $.
Thus for every approximate solution $\beta$ there is exactly one trajectory family $\gamma$ such that $ \sap{\vp}^\rmi_\beta \equiv \vpig $ for  $ \vRi, \vRf \in \Sq $.
This $ \gamma $ is further a symmetric trajectory family, since \eref{eq:pbetaperpnull} implies $ \vpigperp( \vRfn, \vRin ) = \veknull $.
Also, for every symmetric trajectory family $ \gamma $ there is exactly one such family of approximate solutions $\beta$.
This mutual uniqueness is inferred from the point-wise distinction~\eref{eq:bneqbprime} for both the $\beta$ and $\gamma$ families.

From the exact equivalence $ \sap{\vp}^\rmi_\beta \equiv \vpig $, which is restricted to $ \Sq $, we infer
the general one-to-one correspondence of the families $\beta$ of approximate solutions with the symmetric trajectory families $\gamma$, which is
guaranteed to hold in a vicinity%
\footnote{%
	It might happen that some of the symmetric families $\gamma$ don't support trajectories when moving $\vRi$ and $\vRf$ too far away from $\Sq$, whereas the corresponding families $\beta$ still do.
	This potential discrepancy becomes negligible when we finally consider separations of $\vRi$ and $\vRf$ from $\Sq$ that are	parametrically small in $ \hbar $%
	, see\eq\eref{eq:perpshbar}.
}
around $\Sq$.
Due to the identification of $\beta$'s and symmetric $\gamma$'s we may adapt the indexation of $\vpsap^\rmi$, writing $\sap{\vp}^\rmi_\gamma \equiv \sap{\vp}^\rmi_{\beta(\gamma)}$ for all symmetric families $\gamma$, while for nonsymmetric families $\gamma$ there is no corresponding $\beta$ and thus $\vpsapig$ is not defined.
We thus simply write
\begin{equation}
	\vRf = \vqsap( \vRi, \vPi, t )
		\quad \Leftrightarrow \quad
		\vPi \in \left\{ \vpsapig( \vRf, \vRi, t) \right\}_{\gamma\ \text{sym.}}
\end{equation}
for the roots of the Dirac-delta argument~\eref{eq:DiracMod}, each being uniquely assigned to a symmetric trajectory family by the equivalence on $\Sq$,
\begin{equation} \label{eq:pgammasymS}
	\vpsapig( \vRfn, \vRin, t) = \vpig( \vRfn, \vRin, t) \qquad \text{for } \gamma \text{ sym} \psep.
\end{equation}
Using this correspondence we arrive at the counterpart~\eref{eq:deltaRdeltaPsym} of the Dirac-delta identity~\eref{eq:deltaRdeltaP} that selects only symmetric trajectory families by implementing the approximate near-${\SP}$ classical time evolution~\eref{eq:qapprox1}--\eref{eq:qapprox3}, i.e.,
\begin{equation}
	\fl
	\prod_l \delta\!\left[ R^{\rmf}_l - \sap{q}_l(\vRi, \vPi, t) \right] 
		=	\sum_{\gamma\ \text{sym.}}
			\left| \det \!\left(
				\frac{\partial \sap{p}_{\gamma, l}^{\rmi}}{\partial R_{l^\prime}^\rmf} \biggr\rvert_R
			\right)_{l,l^\prime} \right|
		\prod_l \delta\!\left[ P^{\rmi}_l - \sap{p}^\rmi_{\gamma,l}(\vRf, \vRi, t) \right] \psep.
\end{equation}

\section{Linear equivalence of $\vpig$ and $\vpsapig$ around $ \Sq $}
\label{app:linequivp}

To derive\eq\eref{eq:dpdRfdsappdRf}, one can expand the defining equations for both, $ \vpig $ and $ \vpsapig $,
\begin{eqnarray}
	\vq( \vRi, \vpig, t ) = \vRf \psep, \label{eq:defpgamma} \\
	\vqsap( \vRi, \vpsapig, t ) = \vRf \psep, \label{eq:defsappgamma}
\end{eqnarray}
around $ \vR^{\rmi, \rmf} = \vR^{\rmi, \rmf}_0 $ and equate the left-hand and right-hand sides of the corresponding equations in linear order in $ \vR^{\rmi,\rmf}_\perp $, or, in other words, one evaluates\eqs\eref{eq:defpgamma} and~\eref{eq:defsappgamma} for infinitesimal $ \vR^{\rmi,\rmf}_\perp $.
Using the local equivalence of time evolution~\eref{eq:qapprox1}, of the solutions $ \vpig \big\rvert_\Sq = \vpsapig \big\rvert_\Sq $, of the first derivatives~\eref{eq:qapprox2}, as well as the local separation of dynamics~\eref{eq:qpRPsiglam}, one finds
\begin{eqnarray} \label{eq:linearsapp}
	\fl
		\frac{\partial \sap{p}_{\gamma, \sigma}^{\rmi}}{\partial R_{\lambda}^\alpha}(\vRfn, \vRin,t) = 0 \psep,
	\\ \fl
		\left(
			\frac{\partial \sap{p}_{\gamma, \lambda}^{\rmi}}{\partial R_{\lambda^\prime}^\rmf}(\vRfn, \vRin,t)
		\right)_{\lambda,\lambda^\prime}
		= 
		\left[ \left(
			\frac{\partial q_{\lambda}}{\partial P_{\lambda^\prime}^\rmi}(\vRin, \vPin,t)
		\right)_{\lambda,\lambda^\prime} \right]^{-1} \psep,
	\\ \fl		
		\left(
			\frac{\partial \sap{p}_{\gamma, \lambda}^{\rmi}}{\partial R_{\lambda^\prime}^\rmi}(\vRfn, \vRin,t)
		\right)_{\lambda,\lambda^\prime}
		=
		-\left[ \left(
			\frac{\partial q_{\lambda}}{\partial P_{\lambda^\prime}^\rmi}(\vRin, \vPin,t)
		\right)_{\lambda,\lambda^\prime} \right]^{-1}
		\left(
			\frac{\partial q_{\lambda}}{\partial R_{\lambda^\prime}^\rmi}(\vRin, \vPin,t)
		\right)_{\lambda,\lambda^\prime}
\end{eqnarray}
from~\eref{eq:defsappgamma} and the same for the perpendicular derivatives of $ \vpig $
from~\eref{eq:defpgamma}, where $ \vPin = \vpig ( \vRfn, \vRin, t ) $.
In addition, $ \vpig \big\rvert_\Sq = \vpsapig \big\rvert_\Sq $ directly implies the equivalence of parallel derivatives
\begin{equation} \label{eq:linearsappppara}
	\frac{\partial \sap{p}_{\gamma,l}^{\rmi}}{\partial R_{\sigma}^\alpha}(\vRfn, \vRin,t)
		= \frac{\partial p_{\gamma,l}^{\rmi}}{\partial R_{\sigma}^\alpha}(\vRfn, \vRin,t)
		\psep.
\end{equation}
Together, \eref{eq:linearsapp}--\eref{eq:linearsappppara} give the statement~\eref{eq:dpdRfdsappdRf}, i.e.,
\begin{equation}
	\frac{\partial \sap{p}_{\gamma,l}^{\rmi}}{\partial 	R_{l^\prime}^\alpha} (\vRfn, \vRin,t) =
		\frac{\partial p_{\gamma,l}^{\rmi}}{\partial R_{l^\prime}^\alpha} (\vRfn, \vRin,t) \psep.
\end{equation}

\section{Asymmetric uncertainties in $q$ and $p$}
\label{app:asym}
A more general version of\eq\eref{eq:perpshbar}, which also complies with minimum uncertainty, would be
\begin{equation} \label{eq:perpshbar2}
	\eqalign{
		\vqperp^\alpha &{}= \Ord(\hbar^{\nu_\alpha}) \psep, \\
		\vpperp^\alpha &{}= \Ord(\hbar^{1 - \nu_\alpha}) \psep,
	}
\end{equation}
which admits a somewhat more asymmetric distribution of uncertainty between $ \vqperp $ and $ \vpperp $.
With the weaker assumption~\eref{eq:perpshbar2} we will leave the door open for the application to squeezed states in hindsight of calculating transition probabilities of Bose-Einstein condensates.
In the general case that the uncertainty asymmetry parameters $ \nu_\rmi $ and $ \nu_\rmf $ differ for the different components, one should consider them as multi-indexes.
Also, equation \eref{eq:perpshbar2} requires that the perpendicular phase-space coordinates are chosen along the main axes of covariance in $ \Wi $ and $ \Ow $.
We assume this as a prerequisite enabled by the freedom of canonical transformations among the perpendicular (and, separately, also among the parallel) components.
This can in fact always be achieved simultaneously with a sole transformation as long as the perpendicular localization for each of the two distributions saturates (up to a free global constant) the minimum uncertainty principle in \textit{some} \textit{individual} basis of conjugate phase-space coordinates.

To be precise, we write the saturation of uncertainty as a condition on the covariance matrices:
\begin{equation}  \label{eq:uncertsat}
	\Sigma^\alpha = c_\alpha \frac{\hbar}{2} S_\alpha S_\alpha^\rmT  \qquad \alpha \in \{\rmi, \rmf\} \psep,
\end{equation}
with two individual locally defined linear canonical transformations $ \vx_\perp \mapsto \vx_\perp^\prime = S_\alpha \vx_\perp $, where $\vx_\perp = (\vqperp, \vpperp)$, and two arbitrary positive global dimensionless factors $ c_\alpha \geq 1 $.
Equation~\eref{eq:uncertsat} states that there exists a canonical transformation $ S_\rmi $ to new perpendicular phase-space variables $ (\vqperp^\prime, \vpperp^\prime) $ whose variances
$ \langle {q_\lambda^\prime}^2 \rangle $ and $ \langle {p_\lambda^\prime}^2 \rangle $ in the initial state Wigner distribution $ \Wi $ fulfil
\begin{equation} \label{eq:qpvar1}
	\sqrt{ \langle {q_\lambda^\prime}^2 \rangle \langle {p_\lambda^\prime}^2 \rangle } = c_\rmi \frac{\hbar}{2}
\end{equation}
and moreover whose individual variances are
\begin{equation} \label{eq:qpvar2}
	\eqalign{
		\langle {q_\lambda^\prime}^2 \rangle &{}= c_\rmi \frac{\hbar}{2} \psep, \\
		\langle {p_\lambda^\prime}^2 \rangle &{}= c_\rmi \frac{\hbar}{2} \psep,
	}
\end{equation}
while the covariances of different variables vanish, i.e., $ \langle q_\lambda^\prime q_{\lambda^\prime}^\prime \rangle = \langle p_\lambda^\prime p_{\lambda^\prime}^\prime \rangle = 0 $ for $ \lambda \neq \lambda^\prime $ and $ \langle q_\lambda^\prime p_{\lambda^\prime}^\prime \rangle = 0 $ for all $ \lambda, \lambda^\prime $.
The potential asymmetry of the uncertainties between $ q $ and $ p $ are encoded in a scaling transformation that is part of the canonical transformation $ S_\rmi $.
Analogously there exists a \textit{different} canonical transformation $ S_\rmf $ to perpendicular phase-space variables in which the Weyl symbol $\Ow$ of the final observable assumes the ``diagonal'' and symmetric variances~\eref{eq:qpvar1}, \eref{eq:qpvar2}, just possibly with a different global factor $ c_\rmf $ that describes an overall broadening of perpendicular uncertainty.
It is then always possible to find yet another canonical transformation $ \tilde{S} $ under which both covariance matrices simultaneously become diagonal, i.e.,
\begin{equation} \label{eq:SigmaDiag}
	\tilde{S}^{-1} (\Sigma^\alpha) \tilde{S}^{-\rmT} = \frac{c_\alpha \hbar}{2}
		\left( \begin{array}{c|c}
			D_\alpha & 0 \\
			\hline
			0 & D_\alpha^{-1}
		\end{array}\right)
\end{equation}
with individual \textit{diagonal} matrices $ D_\alpha $ for $ \alpha = \rmi $ and $ \alpha = \rmf $ but a common transformation $ \tilde{S} $.
This shows the validity of writing the estimates for perpendicular phase-space coordinates in the simultaneously diagonal form~\eref{eq:perpshbar2}.

The uncertainty asymmetry parameters $ \nu_\alpha $ are encoded in the $\hbar$-dependence of the diagonal entries of the matrices $ D_\alpha $,
\begin{equation}
	( D_\alpha )_{\lambda \lambda} = \Ord\!\left( \hbar^{-1 + 2 ( \nu_\alpha )_\lambda } \right) \psep,
\end{equation}
where $ \nu_\alpha $ is understood as multi-index.
The symmetric case $ (\nu_\alpha )_\lambda = 1/2 $ corresponds to $ D_\alpha $ featuring no dependence on $\hbar$.
Indeed, one could choose $ \tilde{S} $ such that the covariance matrix of the initial Wigner distribution becomes proportional to the identity matrix with $ D_\rmi = \mathbb{I}_{(L-S) \times (L-S)} $.
For this choice one assumes a frame in which the initial state always has symmetrically scaling perpendicular variances~\eref{eq:perpshbar}, i.e., all $ \nu_\rmi = 1 / 2 $, while the final observable in general shows asymmetries in this frame that are still diagonal~\eref{eq:perpshbar2}.
To see this, first apply the symplectic transformation $ S_\rmi $ from\eq\eref{eq:uncertsat}.
The initial covariance matrix trivially becomes $ S_\rmi^{-1} \Sigma^\rmi S_\rmi^{-\rmT} = c_\rmi \hbar / 2 \mathbb{I} $, while the final one becomes $ S_\rmi^{-1} \Sigma^\rmf S_\rmi^{-\rmT} = c_\rmf \hbar / 2 S_\rmi^{-1} S_\rmf S_\rmf^\rmT S_\rmi^{-\rmT} $, which is proportional to a symmetric and symplectic matrix, because transposition, inversion, and composition all preserve symplecticity.
Therefore the latter can be diagonalized by a matrix $ O $ that is  symplectic, i.e., $O^\rmT J O  = J$ with the block matrix
$J = \left( \begin{smallmatrix}
	0 & \mathbb{I} \\
	-\mathbb{I} & 0
\end{smallmatrix} \right) $, and simultaneously orthogonal ($O^\rmT O = \mathbb{I}$).
The composed symplectic matrix $ \tilde{S} = S_\rmi O $ transforms the final covariance matrix into diagonal form
$ \tilde{S}^{-1} \Sigma^\rmf \tilde{S}^{-\rmT} = c_\rmf \hbar / 2
	\left( \begin{smallmatrix}
		D_\rmf & 0 \\
		0 & D_\rmf^{-1}
	\end{smallmatrix} \right) $
while, due to orthogonality, the initial one stays proportional to the identity
$ \tilde{S}^{-1} \Sigma^\rmi \tilde{S}^{-\rmT} = c_\rmi \hbar / 2 \mathbb{I} $.
Nevertheless, we will stick to the form of \eref{eq:SigmaDiag} where both the initial and the final covariance matrices are non-trivially diagonal with $ D_\alpha \neq \mathbb{I} $ in general.
This has the advantage that we can restrict the symplectic matrix $ \tilde{S} $ to be a purely \textit{classically} defined canonical transformation that is not a function of $\hbar$.
Thus we ensure that the used phase-space coordinate system obeys generic rules of classical chaotic dynamics such as considerations on exponential separation of trajectories at Ehrenfest time scales.

\section{Approximate substitution of momenta}
\label{app:ApproxSubMomenta}
As discussed in the main text, \sref{sec:TWAsym}, the discrepancy between the full classical dynamics and its reduction to the immediate proximity of ${\SP}$ utilized to select symmetric trajectory families (see \sref{sec:extractsym}) inhibits the direct replacement of the initial and final momenta, $\vpig$ and $\vpfg$, in expression~\eref{eq:OmegatgsymWOPi} of the main text.
We derive here the corresponding approximate substitution rules given in\eqs\eref{eq:pfgparaNoshift}--\eref{eq:pigNoshift} and moreover their generalization to \qp-asymmetric uncertainties (see \ref{app:asym}).

\subsection{Initial-value determination of $\gamma$}
We start from writing expression~\eref{eq:OmegatgsymWOPi} of the main text as
\begin{equation} \label{eq:OmegatgsymFPi}
	\eqalign{
		\fl
		\langle \hat{\Omega} \rangle_t^{\rm sym} \simeq{}
		\int \!\rmd^L \vRi \int \!\rmd^L \vRf \int \!\rmd^L \vPi
		\sum_{\gamma\ \text{sym.}}
		\left| \det \!\left(
			\frac{\partial \sap{p}_{\gamma, l}^{\rmi}}{\partial R_{l^\prime}^\rmf}(\vRf, \vRi,t)
		\right)_{l,l^\prime} \right|
	\\ \times
		\prod_l \delta \!\left[ P^\rmi_l - \sap{p}^\rmi_{\gamma, l}( \vRf, \vRi, t) \right]
		F_\gamma
		(\vRf, \vRi, t) \psep,
	}
\end{equation}
where $ F_\gamma $ is short-hand for the product
\begin{equation} \label{eq:Fg}
	F_{\gamma} \!\left( \vRf, \vRi, t \right)
		= \Wi \!\left( \vRi, \vpig \bigr\vert_R \right)
		\Ow\!\left( \vRf, \vpfg \bigr\vert_R\right)
\end{equation}
of $ \Wi $ and $ \Ow $, which are evaluated at $ \vp^{\rmi,\rmf}_\gamma \bigr\vert_R = \vp^{\rmi,\rmf}_\gamma( \vRf, \vRi, t )$ and hence are functions of the summation index $ \gamma $.
The explicit dependence on the index $\gamma$ can be relaxed, because the Dirac delta in~\eref{eq:OmegatgsymFPi} makes it uniquely determined by the integration variables.
Indeed, owing to point-wise distinction~\eref{eq:bneqbprime}, two families $\gamma$ and $\gamma^\prime$ that have identical approximate momentum roots $\vpsapig ( \vRf, \vRi, t ) = \vPi = \vpsap^\rmi_{\gamma^\prime} ( \vRf, \vRi, t )$ for given $\vRi, \vRf$ have to be identical: $\gamma = \gamma^\prime$.
The Dirac delta~\eref{eq:deltaRdeltaPsym} for near-${\SP}$ evolution can thus be applied to replace the integration over $\vRf$ by the substitution $\vRf = \vqsap( \vRi, \vPi, t )$, giving
\begin{eqnarray}
	\fl
	\langle \hat{\Omega} \rangle_t^{\rm sym} \simeq
		\int \!\rmd^L \vRi  \int \!\rmd^L \vPi \int \!\rmd^L \vRf \,
		\prod_l \delta \!\left[ R^\rmf_l - \sap{q}_{l}( \vRi, \vPi, t) \right]
		F_{\sap{\gamma}(\vRi, \vPi, t)}(\vRf, \vRi, t) \nonumber \\
\label{eq:OmegatsymRPF}
	\fl
	\hphantom{\langle \hat{\Omega} \rangle_t^{\rm sym}} = \int \!\rmd^L \vRi  \int \!\rmd^L \vPi \,
		F_{\sap{\gamma}(\vRi, \vPi, t)} \!\left( \vqsap( \vRi, \vPi, t), \vRi, t \right) \psep,
\end{eqnarray}
where the symmetric trajectory family $ \gamma = \sap{\gamma}(\vRi, \vPi, t) $ is now uniquely determined by only the initial phase-space coordinates and the time $t$ via the definition
\begin{equation} \label{eq:gammaRP}
	\gamma  = \sap{\gamma}(\vRi, \vPi, t)
		\qquad \Leftrightarrow \qquad
		\vPi = \vpsapig \! \left( \vqsap(\vRi, \vPi, t), \vRi, t \right)
		\psep.
\end{equation}

\subsection{Initial-value substitution rules for $\vpig$ and $\vpfg$}
In the following, we derive the replacement rules~\eref{eq:pfgparaNoshift}, \eref{eq:pfgperpNoshift}, and~\eref{eq:pigNoshift} generalized to \qp-asymmetric uncertainties~\eref{eq:perpshbar2}.
This will in particular alter the order of the correction terms in\eqs\eref{eq:pfgparaNoshift}, \eref{eq:pfgperpNoshift}, and~\eref{eq:pigNoshift}.
We will then discuss in the subsequent \ref{app:neglectCorr} under which circumstances these corrections can safely be neglected. 

For the derivation we will make explicit use of the linear approximation~\eref{eq:qapproxlin} for $ \vqsap $.
Also, in many places we will truncate expansions in perpendicular components and neglect certain terms due to considerations on their scaling with $ \hbar $ induced by \eref{eq:perpshbar2}.
There is one crucial point about these considerations, though, to be clarified upfront.
We have to keep in mind that we want to maintain the validity of our approach and results for propagation times $ t $ as long as the Ehrenfest time 
\begin{equation}
	\tE = \lL^{-1} \log \hbar^{-1}
\end{equation}
and multiples thereof, where $ \lL $ is the Lyapunov exponent of the classically chaotic dynamics.
For such long times, exponential sensitivity to initial conditions renders expansions of unique classical evolution in terms of small variations in the initial conditions very delicate.
In a chaotic setup the long-time evolution of a phase-space coordinate $ x_l( \vx^\rmi, t ) $ generically depends exponentially on the initial phase-space coordinates, $ \partial x_l / \partial x^\rmi_{l^\prime} \sim \rme^{\lL t} $, where here we use the symbol $\vx$ to denote all phase-space variables and $l,l^\prime \in \{ 1, \ldots, 2L \}$.
This means that for times of the order of the Ehrenfest time, say
\begin{equation}
	t = \tau\, \tE \qquad \tau = \Ord(1) \psep,
\end{equation}
initial value derivatives have to be generically considered to be of order $ \partial x_l / \partial x^\rmi_{l^\prime} = \Ord( \hbar^{-\tau} ) $.
Even for initial value variations as small as $\Ord( \hbar )$ this inhibits the use of truncated expansions for our purposes.
Only if one hits the derivative in a stable direction $ \vek{n}^{\rm s}(\vx^\rmi) $ of the linearized dynamics, a set of zero measure, one gets $\sum_{l^\prime}n^{\rm s}_{l^\prime} \partial x_l / \partial x^\rmi_{l^\prime} \sim \rme^{ - \lL t} = \Ord(\hbar^\tau) $ corresponding to converging trajectories.

Note that the linearized classical evolution~\eref{eq:qapproxlin} of the position coordinates close to ${\SP}$ and its generalization~\eref{eq:qapprox1}--\eref{eq:qapprox3} are artificially introduced auxiliary objects to select only symmetric trajectory families.
They are not intended as good approximations to the full dynamics.
On the contrary, the fact that using the local approximation throws away all trajectories that do not belong to symmetric families shows the severity of such an invasive operation on the dynamics, here utilized in a specifically tailored way to reach our goal of selecting only the symmetric part.

The variation of boundary values as contrasted to initial values, i.e., derivatives of $\vpig( \vRf, \vRi, t )$ and $\vpig( \vRf, \vRi, t )$ with respect to $\vRi$ and $\vRf$, are generically not problematic because one of the boundary points as well as the trajectory family stay always fixed.
Since one can construct those derivatives from pairs of infinitesimally close
trajectories, this results in \textit{convergent} trajectories.
In particular, the derivatives of the functions $ \vp^\rmi $ and $ \vp^\rmf $ provide the stable directions mentioned above, either in forward or in time-reversed direction.
Consider for example a trajectory from $\vRi$ to $\vRf$, keep the final point fixed and slightly move the initial point by $ \delta \vRi $.
Then a simultaneous shift in the initial momentum by $ \delta \vPi = \sum_l \delta R^\rmi_l \, \partial \vpig / \partial R^\rmi_l $ yields a trajectory of the same family as the reference trajectory, meaning it stays close at all times and will never depart enough to accumulate additional conjugate points (or leave out conjugate points that it was passing through before).
Moreover, by definition it ends up at the same point $ \vRf $.
Thus, the shifted trajectory closes in on the reference trajectory, at least up to the time $t$.
In other words, the initial conditions have been varied in a direction sufficiently close to the stable ones that exponential departure will only be recognizable for times larger than $t$.
If instead one keeps the initial position fixed and varies the final one, this results in a tiny change $ \delta \vPi = \sum_l \delta R^\rmf_l \, \partial \vpig / \partial R^\rmf_l $ of the initial momentum which will produce an exponential departure from the reference trajectory over time to finally result in the separation $ \delta \vRf $.
Only the exceptional case that a stable direction exists very close to $\delta \vRi = \veknull$ is excluded from this generic picture.
Similar considerations on $ \vpfg $, directly related to the considerations on $ \vpig $ by looking at the time-reversed dynamics, lead to the generic picture that
\begin{equation} \label{eq:pifgderRif}
	\eqalign{
		\frac{\partial p^\rmi_{\gamma, l}}{\partial R^\rmi_{l^\prime}} = \Ord(1) \psep, \\
		\frac{\partial p^\rmf_{\gamma, l}}{\partial R^\rmf_{l^\prime}} = \Ord(1) \psep,
	}
\end{equation}
while the ``cross derivatives'' generically scale like
\begin{equation} \label{eq:pifgderRfi}
	\eqalign{
		\frac{\partial p^\rmi_{\gamma, l}}{\partial R^\rmf_{l^\prime}}
			\sim \rme^{-\lL t} = \Ord( \hbar^\tau )\psep, \\
		\frac{\partial p^\rmf_{\gamma, l}}{\partial R^\rmi_{l^\prime}}
			\sim \rme^{-\lL t} = \Ord( \hbar^\tau )  \psep,
	}
\end{equation}
assuming fully chaotic behavior.

As a reference we take again the explicitly used linear approximation~\eref{eq:qapproxlin} for the near-${\SP}$ dynamics $ \vqsap $, i.e.,
\begin{eqnarray} 
	\vqsappara( \vXi, t ) = \vqpara( \vXin, t ) \psep,\label{eq:qapproxlin2para}\\
	\vqsapperp( \vXi, t ) = \sum_{\xi=1}^{2(L-S)} \left.\frac{\partial \vqperp}{\partial X^\rmi_{\perp,\xi}}\right\rvert_{\SP} X^\rmi_{\perp,\xi} \psep,\label{eq:qapproxlin2perp}
\end{eqnarray}
using the simplified notation~\eref{eq:defX}--\eref{eq:defXperp} for phase-space coordinates.
This approximation induces also a specific form of the functions $ \vpsapig(\vRf, \vRi, t) $, i.e., the solutions for the initial momentum for which the linearized evolution $ \vqsap $, given an initial position $\vRi$, will result in $\vRf$, uniquely selected by a given symmetric trajectory family $\gamma$ [see \sref{sec:extractsym} and\eq\eref{eq:pgammasym} in particular].
The first defining condition $ \vRf = \vqsap\!\left(\vRi, \vpsapig(\vRf,\vRi,t), t \right) $ demands in particular that the parallel components fulfil
\begin{equation}
	\vRfpara = \vqpara\!\left(\vRin, \vpsapign(\vRf,\vRi,t), t \right) \psep,
\end{equation}
where the projected parallel dynamics~\eref{eq:qapproxlin2para} have already been used.
This condition is fulfilled for
\begin{equation} \label{eq:papproxlinearpara}
	\vpsapigpara(\vRf,\vRi,t) = \vpigpara(\vRfn, \vRin,t) \psep,
\end{equation}
which also satisfies $ \vpsapigpara \big\rvert_\Sq = \vpigpara \big\rvert_\Sq $.
Trivially, also the perpendicular components $ \vpsapigperp \big\rvert_\Sq = \vpigperp \big\rvert_\Sq = \veknull $ are equal on $\Sq$, which shows the validity of\eq\eref{eq:papproxlinearpara}.
For the perpendicular components the linearization of $ \vq $ implies
\begin{eqnarray}
	\vRfperp &{}= \vqsapperp\!\left( \vRi, \vpsapig( \vRf, \vRi, t ), t \right) \\
	&{}= \sum_\lambda \left.
			\frac{\partial \vqperp}{\partial R^\rmi_\lambda}
		\right\rvert_{\vRin, \vpsapign} R^\rmi_\lambda
		+ \sum_\lambda \left.
			\frac{\partial \vqperp}{\partial P^\rmi_\lambda}
		\right\rvert_{\vRin, \vpsapign} \sap{p}^\rmi_{\gamma, \lambda}( \vRf, \vRi, t ) \psep.
\end{eqnarray}
The derivatives do not depend on $\vRiperp$ and $\vRfperp$, since they are, in view of\eq\eref{eq:papproxlinearpara}, evaluated at $\vRin$ and $\vRfn$.
The perpendicular components $ \vpsapigperp $ are thus a linear function of $\vRiperp$ and $\vRfperp$.
Because in addition the functions $ \vpig $ and $ \vpsapig $ coincide in linear order around $ \vRi, \vRf \in \Sq $ [see \ref{app:linequivp}] we find the exact identity
\begin{equation}
	\vpsapigperp( \vRf, \vRi, t ) =
		\sum_\lambda
			\frac{\partial \vpigperp}{\partial R^\rmi_\lambda}
		\biggr\rvert_\Sq R^\rmi_\lambda
		+ \sum_\lambda 
			\frac{\partial \vpigperp}{\partial R^\rmf_\lambda}
		\biggr\rvert_\Sq R^\rmf_\lambda \psep.
\end{equation}
The momentum roots $ \vpsapig $ of the linearized dynamics $\vqsap$ are hence also a linearized version of the momentum roots $ \vpig $ of the full evolution $\vq$.
Since $ \vpsapigperp $ as well as the perpendicular derivatives of $ \vpsapigpara $ vanish on $\Sq$ (see \ref{app:locsepdyn}), one can write more compactly
\begin{equation}
	\vpsapig( \vRf, \vRi, t ) = \vpig( \vRfn, \vRin, t ) + 
		\sum_\lambda
				\frac{\partial \vpig}{\partial R^\rmi_\lambda}
			\biggr\rvert_\Sq R^\rmi_\lambda
		+ \sum_\lambda
				\frac{\partial \vpig}{\partial R^\rmf_\lambda}
			\biggr\rvert_\Sq R^\rmf_\lambda \psep.
\end{equation}
In view of\eqs\eref{eq:pifgderRif} and~\eref{eq:pifgderRfi} this fixes the accuracy with which $ \vpsapig $ and $ \vpig $ coincide:
\begin{equation} \label{eq:pigsappig}
	\vpsapig(\vRf, \vRi, t) = \vpig(\vRf, \vRi, t)
		+ \left[ \Ord( {\vRiperp} )
		+ \Ord( {\vRfperp} \hbar^\tau ) \right]^2 \psep.
%
\end{equation}

We recall that for the evaluation of $\vpig$ and $\vpfg$ [see equations~\eref{eq:OmegatgsymFPi}--\eref{eq:OmegatsymRPF}] we have to set $ \gamma = \sap{\gamma}(\vXi, t) $, defined in terms of initial phase-space coordinates by \eref{eq:gammaRP}.
Thus, with\eq\eref{eq:pigsappig} we can relate the momentum root to the initial momentum variable according to
\begin{equation}
	\vpig\!\left( \vqsap(\vXi), \vRi, t \right) = \vPi + \Ord(\hbar^{\kappa_\rmi})
\end{equation}
with
\begin{equation} \label{eq:kappai}
	\kappa_\rmi = \min \{ 2 \nu_\rmi, 2 \nu_\rmf + 2 \tau \} \psep,
\end{equation}
meaning the minimum of all entries if $\nu_{\rmi,\rmf}$ are multi-indexes.
The equivalence~\eref{eq:papproxlinearpara} of the parallel linearized momentum roots with the full momentum roots projected to $\Sq$ also implies that the trajectory starting at the projected initial phase-space point $\vXin$ belongs to the same family $ \gamma = \sap{\gamma}(\vXi, t) = \sap{\gamma}(\vXin, t) $.
To see this, consider the parallel components of\eq\eref{eq:gammaRP},
\begin{equation}
	\fl
	\vPipara = \vpsapigpara\!\left( \vqsap( \vXi, t ), \vRi, t \right)
	= \vpigpara\!\left( \vqsapn(\vXi, t), \vRin, t \right)
	= \vpigpara\!\left( \vq(\vXin, t), \vRin, t \right) \psep,
\end{equation}
complemented trivially by the perpendicular components
\begin{equation}
	\veknull = \vpigperp\!\left( \vq(\vXin, t), \vRin, t \right)
\end{equation}
to give
\begin{equation} \label{eq:Pinpig}
	\vPin = \vpig\!\left( \vq(\vXin, t), \vRin, t \right) \psep,
\end{equation}
a fact which, in analogy with\eq\eref{eq:gammaRP}, we could also express as 
\begin{equation} \label{eq:gsapg}
	\sap{\gamma}(\vXi, t) = \gamma(\vXin, t) \psep,
\end{equation}
where the function $\gamma$, as opposed to $ \sap{\gamma} $, is defined using the full dynamics $\vq, \vpig$ instead of $\vqsap, \vpsapig$:
\begin{equation} \label{eq:gammaXidef}
	\gamma = \gamma(\vXi, t)
		\qquad \Leftrightarrow \qquad
		\vPi = \vpig\!\left( \vq(\vXi, t), \vRi, t \right) \psep.
\end{equation}

To evaluate the final momentum root we use the local separation of dynamics and the considerations on stable and unstable directions~\eref{eq:pifgderRif}, \eref{eq:pifgderRfi} to get
\begin{equation}
	\vpfgpara ( \vRf, \vRi, t ) = \vpfgpara ( \vRfn, \vRin, t ) 
		+ \left[ \Ord( {\vRfperp} ) + \Ord( {\vRiperp} \hbar^\tau ) \right]^2 \psep.
\end{equation}
With the equivalence~\eref{eq:qapproxlin2para} of linearized and full dynamics in $ {\SP} $ as well as the equivalence~\eref{eq:gsapg} of linearized and projected trajectory families, this gives
\begin{equation}
	\vpfgpara\!\left( \vqsap( \vXi, t ), \vRi, t \right) = \vppara ( \vXin, t ) + \Ord( \hbar^{ \kappa_\rmf } )
\end{equation}
where $\gamma = \sap{\gamma}(\vXi, t)$ as before and
\begin{equation} \label{eq:kappaf}
	\kappa_\rmf = \min \{ 2 \nu_\rmf, 2 \nu_\rmi + 2 \tau \} \psep.
\end{equation}
For the perpendicular component we expand to linear order in the final and initial perpendicular position to write the momentum root $ \vpfgperp\!\left( \vqsap(\vXi,t), \vRi, t \right) $ of the linearized dynamics as the linearized version of the momentum root of the full dynamics:
\begin{equation} \label{eq:pfgperpeval1}
	\fl
	\eqalign{
		\vpfgperp\!\left( \vqsap(\vXi,t), \vRi, t \right) &{}=
			\sum_{\lambda}
				\frac{\partial \vpfgperp}{\partial R^\rmf_\lambda}
			\biggr\rvert_{ \vqsapn, \vRin } \sap{q}_\lambda( \vXi, t )
			+ \sum_{\lambda}
				\frac{\partial \vpfgperp}{\partial R^\rmi_\lambda}
			\biggr\rvert_{ \vqsapn, \vRin } R^\rmi_\lambda
			+ \Ord( \hbar^{ \kappa_\rmf } )
	\\
	& {}=
			\sum_{\lambda, \lambda^\prime}
				\frac{\partial \vpfgperp}{\partial R^\rmf_\lambda}
			\biggr\rvert_{ \vqsapn, \vRin }
			\left(
				\frac{\partial q_\lambda}{\partial R^\rmi_{\lambda^\prime}}
						\biggr\rvert_{ \vXin } R^\rmi_{\lambda^\prime}
				+ \frac{\partial q_\lambda}{\partial P^\rmi_{\lambda^\prime}}
						\biggr\rvert_{ \vXin } P^\rmi_{\lambda^\prime}
			\right)
	\\
		&{}\quad + \sum_{\lambda}
				\frac{\partial \vpfgperp}{\partial R^\rmi_\lambda}
			\biggr\rvert_{ \vqsapn, \vRin } R^\rmi_\lambda
			+ \Ord( \hbar^{ \kappa_\rmf } )
	\\
		&{}= \sum_{\xi= 1}^{2(L-S)}
			\frac{\partial}{\partial X^\rmi_{\perp,\xi}}
			\left[
				\vpfgperp\!\left( \vq(\vXi,t), \vRi, t \right)
			\right] \Bigr\rvert_{\vXin}  X^\rmi_{\perp,\xi}
			+ \Ord( \hbar^{ \kappa_\rmf } ) \psep,
	}
\end{equation}
where $\gamma$ is fixed and considered as constant under the derivative, only afterwards evaluated to $\gamma = \sap{\gamma}(\vXi,t) = \gamma(\vXin, t)$ [see equations~\eref{eq:Pinpig} and~\eref{eq:gsapg}].
The final key step is to relax this fixation of $\gamma$ in order to allow us to evaluate the momentum root before taking the derivative in the last line of\eq\eref{eq:pfgperpeval1}.
For any finite propagation time, an \textit{infinitesimal} change of the initial conditions does not change the trajectory family.
The index $\gamma$, as a discrete quantity, e.g. represented by natural numbers, cannot change continuously with $\vXi$.
Instead, it would have to undergo a \textit{jump} from one family to another when one passes through a point where $\vRi$ and $\vq(\vXi,t)$ are conjugate to each other, i.e., a caustic or even a focal point.
As this only applies to a set of zero measure in all of phase space, we consider it as exceptional and correspondingly use the generic fact that
\begin{equation}
	\frac{\partial \gamma(\vXi,t)}{\partial X^\rmi_{\perp, \xi}}  \biggr\rvert_{\vXin} = 0
\end{equation}
to replace the fixed $\gamma = \gamma(\vXin,t)$ in the last line of~\eref{eq:pfgperpeval1} by $\gamma(\vXi,t)$.
This allows us to identify the final momentum
\begin{equation}
	\vp^\rmf_{\gamma(\vXi,t)}\!\left( \vq(\vXi,t), \vRi, t \right) = \vp(\vXi, t)
\end{equation}
with the exact unique evolution $ \vp(\vXi, t) $ of the momentum prior to taking the derivative and finally evaluate
\begin{equation}
	\vpfgperp\!\left( \vqsap(\vXi,t), \vRi, t \right) = \sum_{\xi= 1}^{2(L-S)}
		\left. \frac{\partial \vpperp}{\partial X^\rmi_{\perp,\xi}}
		 \right\rvert_{\vXin}  X^\rmi_{\perp,\xi}
				+ \Ord( \hbar^{ \kappa_\rmf } ) \psep.
\end{equation}

To summarize, the arguments to the Wigner transforms $ \Wi $ and $\Ow$ in\eq\eref{eq:Fg}, respectively~\eref{eq:OmegatgsymWOPi} of the main text, are
\begin{eqnarray}
\label{eq:evalcorrRi}
	\vRi = \vRi \psep, \\
\label{eq:evalcorrpi}
	\vpig = \vPi + \Ord( \hbar^{\kappa_\rmi} ) \psep, \\
\label{eq:evalcorrRf}
	\vRf = \vq(\vXin,t) + \sum_{\xi=1}^{2(L-S)}
		\left. \frac{\partial \vq}{\partial X^\rmi_{\perp,\xi}} \right\rvert_{\vXin} X^\rmi_{\perp, \xi} \psep, \\
\label{eq:evalcorrpf}
	\vpfg = \vp(\vXin,t) + \sum_{\xi=1}^{2(L-S)} 
		\left. \frac{\partial \vp}{\partial X^\rmi_{\perp,\xi}} \right\rvert_{\vXin} X^\rmi_{\perp, \xi}
		+ \Ord( \hbar^{\kappa_\rmf} ) \psep,
\end{eqnarray}
where the order $\Ord(\hbar^{\kappa_\alpha})$ of the corrections is given by\eqs\eref{eq:kappai} and~\eref{eq:kappaf}.
In the special case of \qp-symmetric uncertainties $\nu_\alpha = 1/2$, e.g., for coherent states, this gives $\kappa_\alpha = 1$, resulting in the substitution rules~\eref{eq:pfgparaNoshift}, \eref{eq:pfgperpNoshift}, and~\eref{eq:pigNoshift} given in the main text.

\subsection{Neglecting corrections to momentum arguments}
\label{app:neglectCorr}
We analyze under which circumstances the correction terms in\eqs\eref{eq:evalcorrRi}--\eref{eq:evalcorrpf} can be safely ignored when used as the arguments of the Wigner transforms $\Wi$ and $\Ow$ in\eq\eref{eq:OmegatgsymWOPi} of the main text, leading to the results~\eref{eq:Omegatsymparaperp} and~\eref{eq:OmegatsymDet}.
For coherent states~\eref{eq:Wcoh}, the Wigner transforms do not exhibit peaks sharper than $\Ord(\sqrt{\hbar})$ in any phase-space coordinate, such that those corrections, then of $\Ord(\hbar)$, can indeed be neglected.
The situation is a bit more delicate if one admits arbitrary $ \Wi $ and $\Ow$, especially when they exhibit uncertainties asymmetric in $q$ and $p$ (see \ref{app:asym}).
In this more generic case the momenta could possibly be too sharply defined to neglect the corrections when evaluating $ \vpig $ and $ \vpfg $.
This is not only an issue in the perpendicular coordinates.
Also the parallel components could be peaked at certain values.
For the perpendicular components we have considered tight localization on $ {\SP} $ as a necessary prerequisite such that a saturation of minimum uncertainty (up to a constant) is present, as expressed by\eq\eref{eq:perpshbar2}.
In contrast to that, arbitrarily broad distributions in the parallel components are unproblematic, while difficulties arise in the opposite case of a sharp definition.
For example $ \Ow(\vq, \vp) $ could be centered very sharply around a finite value of $ \vppara $, such that a small correction, say, of order $ \Ord(\hbar) $ to $ \vpfgpara $ can make a significant difference.
This problem would definitely occur when the corresponding peak of $ \Ow $ in $ \vppara $ has a width that scales smaller than $ \Ord(\hbar) $.

Here we will analyse the requirements on the phase-space distributions to give negligible corrections.
For this purpose we will assume a ``worst case'' point of view in that we consider the parallel components being peaked around some value with saturation (up to a dimensionless constant) of minimum uncertainty.
The minimum uncertainty principle applies in any case to the Wigner function $\Wi$ and is as well a reasonable presumption for the observable $ \Ow $.
Evidently, minimum uncertainty of the latter is fulfilled in the case of transition probabilities, where $ \Ow $ is the Wigner function of the final state (up to normalization).
Moreover, operators whose Wigner transform $ \Ow $ have finer structures that fall below minimum uncertainty would in general not be sufficiently well described within the TWA-like approach in the first place.
Instead, this would require quantum corrections, e.g., through implementing stochastic quantum jumps \cite{Polkovnikov2010}.

Similar to the considerations in \ref{app:asym} we will therefore assume
\begin{equation} \label{eq:parashbar2}
	\eqalign{
		\Delta \vq^\alpha_\para = \Ord(\hbar^{\mu_\alpha}) \psep, \\
		\Delta \vp^\alpha_\para = \Ord(\hbar^{1-\mu_\alpha}) \psep,
	}
\end{equation}
for the widths of a certain peak in the marginal distributions $W_\para$ or $\Omega_\para$, see\eqs\eref{eq:Wpara} and~\eref{eq:Opara} in the main text.
Again, the asymmetry parameters $ \mu_\alpha $ have to be generally understood as multi-indexes.
Also the local ``frame'' of parallel phase-space coordinates (by canonical transformation) complies with the main axes of the covariance matrices corresponding to the peaks of both, $W_\para$ and $\Omega_\para$, simultaneously.
This is in full analogy to the analysis of perpendicular components (see \ref{app:asym}).

If $\kappa_\alpha > 1 - \nu_\alpha$ and $\kappa_\alpha > 1 - \mu_\alpha$ for $\alpha \in \{ \rmi, \rmf \}$, then the corrections of order $\Ord( \hbar^{\kappa_\alpha} ) $ cannot become comparable to the corresponding uncertainty given by\eqs\eref{eq:perpshbar2} and~\eref{eq:parashbar2}, as long as $\hbar$ is sufficiently small.
According to that, the phase-space distributions have to be sufficiently broad in the momenta (parallel as well as perpendicular) to neglect these corrections.
For \qp-symmetric uncertainties $\nu_\alpha = \mu_\alpha = 1/2$, as is the case for transition probabilities between coherent states (see \sref{sec:transprobcs}), these conditions are easily met since then $\kappa_\alpha = 1$ for any propagation time $\tau > 0$.
A less restrictive overall sufficient (but not necessary) condition is given by
\begin{equation}
	\frac{1}{3} \leq \nu_\alpha \leq \frac{2}{3} \quad \wedge \quad \frac{1}{3} \leq \mu_\alpha \leq \frac{2}{3} \psep,
\end{equation}
because then the exponents $\kappa_\alpha$~\eref{eq:kappai}, \eref{eq:kappaf} of all correction terms are $ \kappa_\alpha > 2 / 3 $.

In \ref{app:swapandshift} we show how these restrictions on the sharpness of $\Wi$ and $\Ow$ can be further relaxed.

\section{Relaxing sharpness conditions by swapping and shifting}
\label{app:swapandshift}
In \ref{app:neglectCorr} we have given lower bound conditions on the sharpness of peaks in the phase-space distributions of $\Wi$ and $\Ow$.
In the first instance these come as requirements on the sharpness in momentum coordinates, since, by construction of the method, those are accompanied by correction terms~\eref{eq:evalcorrpi} and~\eref{eq:evalcorrpf}, while position coordinates~\eref{eq:evalcorrRi} and~\eref{eq:evalcorrRf} are evaluated directly as the desired linearization around ${\SP}$.

\subsection{Swapping of momentum and position}
\label{app:qpswap}
Firstly we can therefore exploit canonical invariance to loosen up the requirements.
For each individual canonical pair of conjugate phase-space coordinates we have the option to swap the role of position and momentum (up to a minus sign) in advance.
With this freedom we can for example choose to always guarantee that the final phase-space distribution $ \Ow $ is sharper in positions than in momenta, i.e., that $ 1 - \nu_\rmf \leq 1 / 2 $ and $ 1 - \mu_\rmf \leq 1 / 2 $.
If the propagation time in units of the Ehrenfest time is additionally at least $ \tau > 1/4 - \nu_\rmi $, where $ \nu_\rmi $ is the (possibly \qp-swapped) scale of perpendicular position uncertainty, then the correction to the final momenta~\eref{eq:evalcorrpf} is surely negligible.
Consequently, the latter are always negligible if the \textit{sufficient} condition $ \frac{1}{4} < \nu_\rmi < \frac{3}{4} $ on $\Wi$ is met in any given form before a possible \qp-swap.

\subsection{Shifting in the parallel initial values}
Secondly, we can use the freedom to change the integration variables.
The idea is to slightly shift the initial values $ \vRi, \vPi$ in the stable (unstable) directions, to exactly cancel the corrections to the initial (final) momenta in trade for newly introduced corrections to the initial (final) position coordinates.
This can be done in a quite symmetric way for the initial and the final variables.
But instead of demonstrating all possibilities, we rely on the preceding \qp-swap (see above), such that we assume already a situation in which the corrections to final momenta are negligible.

We assume the notation of subsumed phase-space coordinates~\eref{eq:defX} and start with shifting the initial coordinates $ \vXi \mapsto \vXbi = \vXi + \delta \vx^\rmi $ to compensate the corrections to the parallel momentum components $ \vpigpara $ of order $ \Ord(\hbar^{\kappa_\rmi}) $, which fixes $ \delta \vp^\rmi_\para = \Ord(\hbar^{\kappa_\rmi}) $ to fulfil $ \vpigpara = \vPbi_\para $.
We demand that the parallel initial variables get shifted in a direction very close to one of the \textit{stable} directions of the original (projected) trajectory in $ {\SP} $, starting at $ \vXin $, in order to minimize the effect on the final variables.
In particular, the trajectory starting at $ \vXbin $ shall belong to the same family as the one starting at $ \vXin $.
This determines the position shift $ \delta \vek{r}^\rmi_\para $ quite precisely in terms of the momentum shift $ \delta \vp^\rmi_\para $.
A precise prescription is to define the $S$ components of the former implicitly by demanding the $S$ constraints
\begin{equation} \label{eq:shiftparaconstraint}
	\vpigpara\!\left( \vq( \vXin, t ), \vRbin, t \right) = \vPbi_\para \psep,
\end{equation} 
which expresses that the final position of the shifted trajectory remains constant, i.e., $ \vq( \vXbin, t ) = \vq( \vXin, t ) $ or equivalently
\begin{equation} \label{eq:shiftqparaequal}
	\vqsappara( \vXbi, t ) = \vqsappara( \vXi, t ) \psep,
\end{equation}
and that the trajectory family is unchanged, $ \gamma( \vXbin, t ) = \gamma( \vXin, t ) $, corresponding to\eq\eref{eq:gammaXidef}.
Expanding\eq\eref{eq:shiftparaconstraint} in the shift $ \delta \vek{r}^\rmi_\para $ and using\eq\eref{eq:Pinpig} fixes it to
\begin{equation}
	\delta \vek{r}^\rmi_\para =
	\underbrace{
		\left[ \left( 
			\frac{\partial p^\rmi_{\gamma, \sigma}}{\partial R^\rmi_{\sigma^\prime}}
			\biggr\rvert_{ \vq(\vXin), \vRin }		
		\right)_{\sigma, \sigma^\prime}
		\right]^{-1} 
	}_{\Ord( 1 )} \delta \vp^\rmi_\para  + \Ord( \hbar^{2 \kappa_\rmi} ) \psep.
\end{equation}
In particular, the shift in the parallel position is of the same order $ \delta \vek{r}^\rmi_\para = \Ord( \hbar^{\kappa_\rmi} ) $ as the parallel momentum shift $\delta \vp^\rmi_\para$.

In contrast to the parallel final position, the influence on the perpendicular final position does not vanish per se.
The linearized perpendicular dynamics depend on the parallel position, in which they have to be expanded:
\begin{equation} \label{eq:qsapperpshiftpara}
	\vqsapperp( \vXbi, t ) =
		\biggl( \frac{\partial q_\lambda}{\partial X^\rmi_{\xi}} \biggr\rvert_{\vXbin} \biggr)_{\lambda, \xi}  \vXbi_\perp
		= \biggl( \frac{\partial q_\lambda}{\partial X^\rmi_{\xi}} \biggr\rvert_{\vXin} \biggr)_{\lambda, \xi}
		\underbrace{
			\left( \mathbb{I} + \Ord( \delta \vx^\rmi_\para ) \right) \vXbi_\perp
		}_{ \displaystyle \equiv \vXiperp } \psep,
\end{equation}  
where the expression in parentheses is a $2(L-S) \times 2(L-S)$ matrix that expresses the influence on the linearized perpendicular dynamics by a variation in the parallel directions.
Here, we are allowed to use expansion and truncation of the unique time evolution in the initial coordinates since we already have ensured by\eq\eref{eq:shiftparaconstraint} that the variation is in a stable direction regarding the projected in-plane dynamics and that the trajectory family is unchanged.
The stated order of this deviation can be derived from considering infinitesimally separated partner trajectories.
By shifting the perpendicular initial phase-space coordinates corresponding to this matrix, i.e., according to the implicit definition given in\eq\eref{eq:qsapperpshiftpara}, we can compensate this deviation to fulfil
\begin{equation}
	\vqsapperp( \vXbi, t ) = \vqsapperp( \vXi, t ) \psep.
\end{equation}
Thus, together with\eq\eref{eq:shiftqparaequal}, all components of the final linearized position remain unchanged:
\begin{equation} \label{eq:shift1qequal}
	 \vqsap( \vXbi, t ) = \vqsap( \vXi, t ) \psep.
\end{equation}
The price is that we introduce a new correction to the perpendicular position $ \vRiperp = \vRbi_\perp + \Ord( \hbar^{\kappa_\rmi} ) $, while the shift in the perpendicular momentum remains of the same order as the existing correction [see equation~\eref{eq:evalcorrpi}] and hence does not add any new difficulty.

The influence on the final momenta is easily estimated by expanding $ \vpfg $ in the initial position shift $ \delta \vek{r}^\rmi $, which is exponentially suppressed by the generic stability considerations~\eref{eq:pifgderRfi}:
\begin{equation} \label{eq:pfshiftpara}
	\vpfg\!\left( \vqsap( \vXbi, t ), \vRbi, t \right) = \vpfg\!\left( \vqsap( \vXi, t ), \vRi, t \right)
		+ \underbrace{\Ord( \hbar^\tau ) \delta \vek{r}^\rmi}_{ {\mathclap{ \displaystyle = \Ord( \hbar^{\kappa_\rmi + \tau} ) }} } \psep.
\end{equation}
In total, the situation after the shift as compared to\eqs\eref{eq:evalcorrRi}--\eref{eq:evalcorrpf} is that the initial positions $ \vRi $ get a correction of order $ \Ord( \hbar^{\kappa_\rmi} ) $, the correction to the parallel initial momenta $ \vPipara $ is exactly cancelled, the final positions $\vqsap$ remain without correction, and the final momenta $ \vpfg $ get an additional correction of order $ \Ord( \hbar^{\kappa_\rmi + \tau} ) $.

\subsection{Shifting in the perpendicular initial values}
From this, one can perform an additional shift in only the perpendicular initial coordinates to fully absorb the correction of the perpendicular initial position into the one of the perpendicular initial momentum, or vice versa.
The linearization of the perpendicular dynamics thereby simplifies this shift in comparison with the one just used to compensate parallel momentum corrections.
To simplify notation, we use now $ \vXi $ as the variables after the (possibly applied) first shift, whereas $ \vXbi = \vXi + \delta \vx^\rmi $ denotes the variables after the second, solely perpendicular shift ($ \delta \vx^\rmi_\para = \veknull $).  
In order to cancel the $\Ord( \hbar^{\kappa_\rmi} )$-correction to $ \vRiperp $ or $ \vPiperp $ one has to fix $ \delta \vek{r}^\rmi_\perp = \Ord( \hbar^{\kappa_\rmi} ) $ or $ \delta \vp^\rmi_\perp = \Ord( \hbar^{\kappa_\rmi} ) $, respectively.
Conversely, $\delta \vp^\rmi_\perp$ or $\delta \vek{r}^\rmi_\perp$ are fixed by demanding that the shift be close to a stable direction.
In particular we can choose to keep the final perpendicular position $ \vqsapperp $ invariant by relating
\begin{equation}
	\delta \vp^\rmi_\perp =
		\underbrace{
			\biggl( \frac{\partial p^\rmi_{\gamma, \lambda}}{\partial R^\rmi_{\lambda^\prime}} \biggr\rvert_{\vq(\vXin,t), \vRin} \biggr)_{\lambda, \lambda^\prime} 
		}_{ \displaystyle = \Ord(1) } \delta \vek{r}^\rmi_\perp \psep, 
\end{equation}
such that all components of the shift are of order $ \delta \vx^\rmi_\perp = \Ord( \hbar^{\kappa_\rmi} ) $.
The parallel position trivially remains constant under the shift and thus we guarantee again that $ \vqsap( \vXbi, t ) = \vqsap( \vXi, t ) $, as we did when shifting of parallel components~\eref{eq:shift1qequal}.

In a similar manner to the parallel shift~\eref{eq:pfshiftpara}, the effect on the final perpendicular momentum is estimated by expanding $ \vpfgperp $ in $ \delta \vx^\rmi_\perp $.
As before, this adds a correction of order $ \Ord( \hbar^{\kappa_\rmi + \tau} ) $, while the parallel components of the final momentum actually stay constant, since $ \vppara( \vXbi, t ) = \vppara( \vXbin, t ) = \vppara( \vXin, t ) $.
Due to the linearization of the local dynamics a purely perpendicular shift is here sufficient, leaving also $ \vRipara $ and $ \vPipara $ invariant.

Let us finally note that the new integration variables after the demonstrated shifts are not \textit{constant} shifts, but functions of $ \vXi $, which demands to account for the corresponding Jacobian when changing the integration variables from $ \vXi $ to $ \vXbi $.
Nevertheless, the Jacobian is unity plus a correction that vanishes for $ \hbar \to 0 $ and thus irrelevant since the overall approach amounts to describe $ \langle \hat{\Omega} \rangle_t $ to leading order in $\hbar$ only.

\subsection{Relaxed sharpness conditions}
The combined flexibility of possible preceding canonical \qp-swaps, parallel shifts and perpendicular shifts covers a large parameter space.
Moreover we took a somewhat ``worst case'' perspective in many places, such that all given conditions should be understood as sufficient rather than necessary.
In the majority of ``natural'' scenarios the result will thus be valid up to corrections of order $ \Ord( \hbar^\epsilon ) $ with some $ \epsilon > 0 $.

More precisely, all corrections are negligible if
\begin{equation} \label{eq:combcondkf}
	(\kappa_\rmf > 1 - \nu_\rmf) \quad \wedge \quad (\kappa_\rmf > 1 - \mu_\rmf)
\end{equation}
and \textit{one} of the four conditions
\begin{eqnarray}
	\fl
		\text{i)} \qquad &\Big[ \kappa_\rmi > 1 - \nu_\rmi \quad \wedge \quad \kappa_\rmi > 1 - \mu_\rmi \Big] \psep, \label{eq:combcondki1}
	\\\fl
		\text{ii)} \qquad &\Big[ \kappa_\rmi > 1 - \nu_\rmi \quad \wedge \quad \kappa_\rmi > \mu_\rmi
			\quad \wedge \quad (\kappa_\rmi + \tau > 1 - \mu_\rmf) \Big] \psep,\label{eq:combcondki2}
	\\\fl
		\text{iii)} \qquad &\Big[ \kappa_\rmi > \nu_\rmi \quad \wedge \quad \kappa_\rmi > 1 - \mu_\rmi
			\quad \wedge \quad (\kappa_\rmi + \tau > 1 - \nu_\rmf) \Big] \psep, \label{eq:combcondki3}
	\\\fl
		\text{iv)} \qquad &\Big[ \kappa_\rmi > \nu_\rmi \quad \wedge \quad \kappa_\rmi > \mu_\rmi
			\quad \wedge \quad (\kappa_\rmi + \tau > 1 - \nu_\rmf)
			\quad \wedge \quad (\kappa_\rmi + \tau > 1 - \mu_\rmf) \Big] \label{eq:combcondki4}
\end{eqnarray}
is met, where the conditions have to be fulfilled for all entries of the multi-indexes and where we assume the preceding \qp-swap as described above, i.e., $\nu_\rmf \geq 1/2, \mu_\rmf \geq 1/2 $, denoting by $\nu_\alpha$ and $\mu_\alpha$ the parameters associated with the (possibly already swapped) positions.
In the case of only moderately asymmetric initial perpendicular uncertainties (independent of the \qp-swap)
\begin{equation} \label{eq:nuimoderate}
	\frac{1}{4} < \nu_\rmi < \frac{3}{4} \psep,
\end{equation}
already mentioned in \ref{app:qpswap}, the conditions~\eref{eq:combcondkf} are automatically met, and, since then $ \kappa_\rmi > 1 / 2 $, also all conditions in parentheses in \eref{eq:combcondki1}--\eref{eq:combcondki4} are fulfilled.
It is then very likely that the given parameters fulfil at least one of the remaining conditions.
In fact, if (after the \qp-swap) the entries of $ \nu_\rmi $ are either all $ \leq 1 / 2 $ or all $ \geq 1 / 2 $ and also the entries of $ \mu_\rmi $ are either all $ \leq 1 / 2 $ or all $ \geq 1 / 2 $, one of the four conditions is always fulfilled.
Only if uncertainties are squeezed in a pathologically awkward manner that treats initial and final positions and momenta in a very uneven way, the validity of the method is not assured and has to be checked.

\section{Gaussian approximation}
\label{app:GaussApprox}
We discuss the approximation of perpendicular phase-space distributions $W_\perp$ and $\Omega_\perp$.
As the separation of marginal distributions given by\eqs\eref{eq:Wpara} and~\eref{eq:Opara} in the main text implies the normalization conditions
\begin{eqnarray}
	\eqalign{	
		\int \!\rmd \vX_\perp \, W_\perp( \vX_\perp ; \vX_\para ) = 1 \psep, \\
		\int \!\rmd \vX_\perp \, \Omega_\perp( \vX_\perp ; \vX_\para ) = 1 \psep,
	}
\end{eqnarray}
one could write
\begin{equation} \label{eq:WOGauss}
	\eqalign{
		W_\perp(\vXiperp ; \vXipara) \simeq (\pi \hbar)^{S-L} \sqrt{\det A^\rmi}
			\exp \!\left( - \frac{1}{\hbar} (\vXiperp)^\rmT A^\rmi \vXiperp \right) \psep, \\
		\Omega_\perp(\vXfperp ; \vXfpara) \simeq (\pi \hbar)^{S-L} \sqrt{\det A^\rmf}
			\exp \!\left( - \frac{1}{\hbar} (\vXfperp)^\rmT A^\rmf \vXfperp \right) \psep,
	}
\end{equation}
and match the symmetric $ 2 (L-S) \times 2 (L-S) $ matrices $ A^\alpha = A^\alpha(\vX^\alpha_\para) = (A^\alpha)^\rmT $ to the actual perpendicular shape of the distributions $\Wi$ and $\Ow$ under consideration.
For instance, matching the covariances
\begin{equation} \label{eq:covariances}
	\eqalign{
		\Sigma^\rmi_{\xi, \xi^\prime}
			= \left\langle X_{\perp, \xi} X_{\perp, \xi^\prime} \right\rangle_{W_\perp}
			= \left( W_\para\!\left( \vX_\para \right) \right)^{-1} \int \!\rmd \vX_\perp \,
				X_{\perp, \xi} X_{\perp, \xi^\prime} \, \Wi(\vX) \psep, \\			
		\Sigma^\rmf_{\xi, \xi^\prime}
			= \left\langle X_{\perp, \xi} X_{\perp, \xi^\prime} \right\rangle_{\Omega_\perp}
			= \left( \Omega_\para\!\left( \vX_\para \right) \right)^{-1} \int \!\rmd \vX_\perp \,
				X_{\perp, \xi} X_{\perp, \xi^\prime} \, \Ow(\vX) \psep,
	}
\end{equation}
where $ \xi, \xi^\prime \in \{ 1, \ldots, 2( L - S ) \} $, yields the matrices
\begin{equation}
	A^\alpha = \frac{\hbar}{2} (\Sigma^\alpha)^{-1}
\end{equation}
together with the formula
\begin{eqnarray}
	\fl
	\langle \hat{\Omega} \rangle_t^{\rm sym} &{}\simeq
			( \pi \hbar )^{2(S-L)}
			\int \!\rmd \vXipara \;
				W_\para\!\left( \vXipara \right)
				\, \Omega_\para\!\left( \vx_\para( \vXin, t ) \right)
				\sqrt{ \det ( A^\rmi A^\rmf ) }
	\nonumber\\\fl & \qquad \qquad \qquad \times
		\int \!\rmd \vXiperp \;
			\exp \!\left[ - \frac{1}{\hbar}
				( \vXiperp )^\rmT
					\left( A^\rmi + \Mstab^\rmT A^\rmf \Mstab \right)
				\vXiperp
			\right]
	\nonumber\\\fl
\label{eq:OmegatsymparaDet}
	&{}=
		( \pi \hbar )^{S-L}
		\int \!\rmd \vXipara \,
			W_\para\!\left( \vXipara \right)
			\, \Omega_\para\!\left( \vx_\para( \vXin, t ) \right)
			\frac{\sqrt{ \det ( A^\rmi A^\rmf ) }}
				{\sqrt{\det\!\left( A^\rmi + \Mstab^\rmT A^\rmf \Mstab \right)}} \psep,
\end{eqnarray}
which is the counterpart of\eq\eref{eq:OmegatsymDet} in the main text without explicitly invoking Gaussian shapes of $W_\perp$ and $\Omega_\perp$.
Instead, in the expression~\eref{eq:OmegatsymparaDet} the marginal distributions would have to be calculated by integrating $\Wi$ and $\Ow$ in $ \vX^{\rmi, \rmf}_\perp$ in addition to matching the covariances~\eref{eq:covariances}.
In that respect, equation~\eref{eq:OmegatsymparaDet} is a slightly more general version of the result~\eref{eq:OmegatsymDet} given in the main text.

If $W_\perp$ and $\Omega_\perp$ have profiles that are indeed very close to Gaussian shapes, the exactly normalized distributions~\eref{eq:WOGauss} coincide with~\eref{eq:WOGauss0} in the main text, where the inverse covariance matrices are then simply given by the second derivatives of $\Wi$ and $\Ow$ with respect to $\vX^\alpha_\perp$, evaluated on ${\SP}$, i.e., for $\vX^\alpha_\perp = 0$.
The precise prescription is
\begin{eqnarray}
\label{eq:Ai}
	A^\rmi_{\xi, \xi^\prime}( \vXipara )
		= - \frac{\hbar}{2 \Wi( \vXin )}
				\frac{\partial^2 \Wi}{\partial X^\rmi_{\perp, \xi} \partial X^\rmi_{\perp, \xi^\prime}}
			\biggr\rvert_{ \vXin } \psep,
	\\
\label{eq:Af}
	A^\rmf_{\xi, \xi^\prime}( \vXfpara )
		= - \frac{\hbar}{2 \Ow( \vXfn )}
				\frac{\partial^2 \Ow}{\partial X^\rmf_{\perp, \xi} \partial X^\rmf_{\perp, \xi^\prime}}
			\biggr\rvert_{ \vXfn } \psep,
\end{eqnarray}
where $ \xi, \xi^\prime \in \{ 1, \ldots, 2( L - S ) \} $.
The explicit calculation of the marginal distributions is then obsolete, as they are determined by the parameters of the Gaussians:
\begin{eqnarray}
	W_\para( \vXipara )
		\simeq (\pi \hbar)^{L-S} \frac{\Wi( \vXin )}{\sqrt{\det A^\rmi}} \psep, 
	\\	
	\Omega_\para( \vXfpara )
		\simeq (\pi \hbar)^{L-S} \frac{\Ow( \vXfn )}{\sqrt{\det A^\rmf}} \psep.
\end{eqnarray}
This yields the final result as stated in the main text, equation~\eref{eq:OmegatsymDet}.

Note that the obtained results~\eref{eq:Omegatsymparaperp} and~\eref{eq:OmegatsymDet} of the main text and\eq\eref{eq:OmegatsymparaDet} are invariant with respect to canonical transformations in the perpendicular phase-space coordinates, as long as the Wigner distributions are transformed correspondingly.
At the level of the Gaussian approximation, e.g., this means that the inverse covariance matrices transform as $ A^\alpha \mapsto \tilde{A}^\alpha $ with $ A^\alpha = S^\rmT \tilde{A}^\alpha S $, 
where we specify here the canonical transformation $ \vx_\perp \mapsto \tilde{\vx}_\perp $ in its differential form $ S_{ \xi, \xi^\prime } = {\partial \tilde{x}_{\perp, \xi}} / {\partial x_{\perp, \xi^\prime}} $.
On the other hand, the marginal distributions $ W_\para $ and $ \Omega_\para $ as well as the evaluation of the distributions at $ {\SP} $, $ \Wi(\vX_0) $ and $ \Ow(\vX_0) $, are invariant.
The stability matrix transforms as $ \Mstab \mapsto \Mcan $ with $ \Mstab = S^{-1} \Mcan S $.
Because $ \det S = 1 $ for any canonical transformation, it follows that the determinants are also invariant, i.e.,
$ \det ( A^\rmi A^\rmf ) = \det ( S^\rmT \tilde{A}^\rmi S S^\rmT \tilde{A}^\rmf S ) = \det ( \tilde{A}^\rmi \tilde{A}^\rmf ) $ and
$ \det ( A^\rmi + \Mstab^\rmT A^\rmf \Mstab ) = \det ( S^\rmT (\tilde{A}^\rmi + \McanT \tilde{A}^\rmf \Mcan) S ) = \det ( \tilde{A}^\rmi + \McanT \tilde{A}^\rmf \Mcan ) $.\\
\\

\bibliographystyle{my-iopart-num}
\bibliography{twastab_arxiv_submission}

\end{document}